\documentclass[a4paper,10pt]{article}
\usepackage{setspace}
\usepackage[explicit]{titlesec}
\usepackage[utf8]{inputenc}
\usepackage{amsfonts,amsmath,amssymb,mathtools}
\usepackage[left=2cm,right=2cm,top=2cm,bottom=2cm,bindingoffset=0cm]{geometry}
\usepackage{graphicx}
\usepackage{authblk}
\usepackage{xcolor}
\usepackage{hyperref}
\usepackage{authblk}
\usepackage{slashed}
\usepackage{braket}
\usepackage{comment}
\usepackage{pb-diagram}
\usepackage{wasysym}
\usepackage{amsthm}
\usepackage{cancel}
\usepackage{tikz}\usetikzlibrary{tikzmark}
\usetikzlibrary{fadings}
\usetikzlibrary{positioning}
\usetikzlibrary{backgrounds} 
\usetikzlibrary{decorations.pathreplacing}
\usetikzlibrary{arrows}
\definecolor{airforceblue}{rgb}{0.36, 0.54, 0.66}	
\definecolor{beige}{rgb}{0.96, 0.96, 0.86}
\definecolor{bittersweet}{rgb}{1.0, 0.44, 0.37}
\definecolor{melon}{rgb}{0.99, 0.74, 0.71}
\definecolor{mustard}{rgb}{1.0, 0.86, 0.35}
\definecolor{lava}{rgb}{0.81, 0.06, 0.13}
\definecolor{magnolia}{rgb}{0.97, 0.96, 1.0}
\definecolor{lavendermist}{rgb}{0.9, 0.9, 0.98}
\definecolor{lavendergray}{rgb}{0.77, 0.76, 0.82}
\definecolor{palepink}{rgb}{0.98, 0.85, 0.87}
\definecolor{palesilver}{rgb}{0.79, 0.75, 0.73}
\definecolor{cadetgrey}{rgb}{0.57, 0.64, 0.69}
\definecolor{anti-flashwhite}{rgb}{0.95, 0.95, 0.96}
\colorlet{Light0anti-flashwhite}{anti-flashwhite!70!white}
\colorlet{Lightanti-flashwhite}{anti-flashwhite!50!white}
\colorlet{Light2anti-flashwhite}{anti-flashwhite!30!white}
\definecolor{linkcolor}{rgb}{0,0,1}
\definecolor{urlcolor}{rgb}{0,0,1}

\usepackage{tikz-cd}
\usetikzlibrary{decorations.markings}
\usetikzlibrary{positioning}
\usepackage{braids}
\usepackage{array}
\usepackage{ytableau}
\usepackage{longtable}
\usepackage{empheq}
\usepackage{multicol}
\usepackage{mathrsfs}
\usepackage{diagbox}
\usepackage[vcentermath]{youngtab}
\usepackage[natbib=true, backend = bibtex, style=numeric-comp, sorting=none,maxbibnames=99]{biblatex}
\hypersetup{pdfstartview=FitH,  linkcolor=linkcolor,urlcolor=urlcolor,citecolor=urlcolor, colorlinks=true}

\newcommand\bem{\begin{pmatrix}}
\newcommand\eem{\end{pmatrix}}
\newcommand\beq{\begin{equation}}
\newcommand\eeq{\end{equation}}
\newcommand\beqs{\begin{equation*}}
\newcommand\eeqs{\end{equation*}}

\newcommand{\sgn}{\,\text{sgn}}

\date{}

\def\be{\begin{eqnarray}}
\def\ee{\end{eqnarray}}
\def\nn{\nonumber}

\definecolor{red}{rgb}{1,0,0}
\definecolor{orange}{rgb}{1,0.5,0}
\definecolor{violet}{rgb}{0.7,0,1}

\usepackage{tikz}\usetikzlibrary{tikzmark}
\usetikzlibrary{fadings}
\usetikzlibrary{positioning}
\usetikzlibrary{backgrounds} 
\usetikzlibrary[backgrounds]
\usetikzlibrary{decorations.pathreplacing}
\usetikzlibrary{arrows}
\definecolor{airforceblue}{rgb}{0.36, 0.54, 0.66}	
\definecolor{beige}{rgb}{0.96, 0.96, 0.86}
\definecolor{bittersweet}{rgb}{1.0, 0.44, 0.37}
\definecolor{melon}{rgb}{0.99, 0.74, 0.71}
\definecolor{mustard}{rgb}{1.0, 0.86, 0.35}
\definecolor{lava}{rgb}{0.81, 0.06, 0.13}
\definecolor{magnolia}{rgb}{0.97, 0.96, 1.0}
\definecolor{lavendermist}{rgb}{0.9, 0.9, 0.98}
\definecolor{lavendergray}{rgb}{0.77, 0.76, 0.82}
\definecolor{palepink}{rgb}{0.98, 0.85, 0.87}
\definecolor{palesilver}{rgb}{0.79, 0.75, 0.73}
\definecolor{cadetgrey}{rgb}{0.57, 0.64, 0.69}
\definecolor{anti-flashwhite}{rgb}{0.95, 0.95, 0.96}
\colorlet{Light0anti-flashwhite}{anti-flashwhite!70!white}
\colorlet{Lightanti-flashwhite}{anti-flashwhite!50!white}
\colorlet{Light2anti-flashwhite}{anti-flashwhite!30!white}
\definecolor{linkcolor}{rgb}{0,0,1}
\definecolor{urlcolor}{rgb}{0,0,1}

\usepackage{tikz-cd}
\usetikzlibrary{decorations.markings}
\usetikzlibrary{positioning}
\usepackage{array}
\usepackage{ytableau}
\usepackage{longtable}
\usepackage{empheq}
\usepackage{multicol}
\usepackage{mathrsfs}
\usepackage{diagbox}
\usepackage[vcentermath]{youngtab}

\hypersetup{pdfstartview=FitH,  linkcolor=linkcolor,urlcolor=urlcolor,citecolor=urlcolor, colorlinks=true}

\newcommand\ST{{\bf t}}
\newcommand\Kh{{\rm Kh}}
\newcommand\n{n}

\begin{document}

\title{\bf Towards tangle calculus for Khovanov polynomials
}

\author[2,3]{{\bf A. Anokhina}\thanks{\href{mailto:anokhina@itep.ru}{anokhina@itep.ru}}}
\author[1,3]{{\bf E. Lanina}\thanks{\href{mailto:lanina.en@phystech.edu}{lanina.en@phystech.edu}}}
\author[1,2,3]{{\bf A. Morozov}\thanks{\href{mailto:morozov@itep.ru}{morozov@itep.ru}}}

\vspace{5cm}

\affil[1]{Moscow Institute of Physics and Technology, 141700, Dolgoprudny, Russia}
\affil[2]{Institute for Information Transmission Problems, 127051, Moscow, Russia}
\affil[3]{NRC "Kurchatov Institute", 123182, Moscow, Russia\footnote{former Institute for Theoretical and Experimental Physics, 117218, Moscow, Russia}}
\renewcommand\Affilfont{\itshape\small}

\maketitle

\vspace{-7cm}

\begin{center}
	\hfill MIPT/TH-17/23\\
	\hfill ITEP/TH-23/23\\
	\hfill IITP/TH-18/23
\end{center}

\vspace{4.5cm}

\begin{abstract}

{
We provide new evidence that the  tangle calculus and ``evolution'' are applicable to the Khovanov polynomials
for families of long braids inside the knot diagram. We show that jumps in evolution, peculiar for superpolynomials,
are much less abundant than it was originally expected.
Namely, for torus and twist satellites of a fixed companion knot, the main (most complicated) contribution
does not jump, all jumps are concentrated in the torus and twist part correspondingly,
where these jumps are necessary to make the Khovanov polynomial positive.
Among other things, this opens a way to define a jump-free part of the colored
Khovanov polynomials, which differs from the naive colored polynomial
just ``infinitesimally''.
The separation between jumping and smooth parts involves a combination of Rasmussen index 
and a new knot invariant, which we call ``Thickness''.
}
\end{abstract}

\tableofcontents

\section{Introduction}
In this paper, we investigate whether and how the tangle calculus is applicable to  the Khovanov polynomials. In {\it 2d} conformal field theory, non-perturbative correlation functions are efficiently calculated with the use of holomorphic factorization. Namely, correlation functions are expressed through bilinear combinations of conformal blocks summed over just a few internal states, but not over the entire huge Hilbert space of states. In {\it 3d} topological field theory similar story takes place. Wilson loops in three-dimensional Chern--Simons theory~\cite{CS,Witten} with a gauge field taken in an arbitrary $\mathfrak{sl}_N$ representation $R$ and integrated over any link -- the colored (by $R$) HOMFLY polynomials~\cite{alexander1928topological,leech1970computational,jones1983invent,freyd1985new,kauffman1987state,przytycki1987kobe,Guadagnini:1989kr,GUADAGNINI1990575,Alvarez_1993,Alvarez1994tt,Alvarez_1997,Labastida1997uw}, obey the so called {\it tangle calculus}~\cite{kaul1998chern,mironov2018tangle}.

Tangle calculus is targeted at making knots and links by gluing tangles.
With each tangle $T$ we associate a tensor with $\otimes_i^{n_T} R_i$,
where $n_T$ is the number of free legs in tangle $T$, and $R_i$ are the corresponding
representations. With each tangle, one can associate some rational function or a matrix
which enters a resulting HOMFLY polynomial as a factor.
Gluing of tangles causes summation over irreducible representations of
intermediate states in a link polynomial. The Reshetikhin--Turaev formalism~\cite{Reshetikhin,guadagnini1990chern2,reshetikhin1990ribbon,turaev1990yang,reshetikhin1991invariants} explains how to glue the HOMFLY polynomials
from tangles, associated with elementary vertices (${\cal R}$-matrices) and with lines
($\cal U$-matrices).
A more interesting calculus involves {\it evolution}~\cite{mironov2013evolution,dunin2013superpolynomials,mironov2016racah,mironov2015colored2,mironov2014colored,dunin2022evolution,anokhina2019nimble,willis2021khovanov,nakagane2019action,anokhina2018khovanov}, i.e. tangles,
made from braids of arbitrary length (i.e. numbers of crossings inside tubes),
which become {\it parameters} of the evolution families.
In this case, we represent knots/links as tubes of arbitrary lengths
connected by complicated but ``small'' vertices with no parameters.
The question then is what kind  of {\it connectors} is needed,
and what are explicit expressions for them. First example of the last mentioned tangle calculus was considered in~\cite{mironov2015colored} for arborescent knots. Another
example was provided for satellite knots~\cite{morozov2018knot}. Tangle calculus for the colored HOMFLY polynomials in
a more general form was investigated in~\cite{mironov2018tangle}.

The colored HOMFLY polynomials can be $\ST$-deformed to the Khovanov--Rozansky polynomials~\cite{khovanov2000categorification,bar2002khovanov,khovanov2004sl,khovanov2007virtual,khovanov2010categorifications,dolotin2014introduction} and superpolynomials~\cite{dunin2013superpolynomials,dunfield2006superpolynomial,gukov2005khovanov}. The original definition of the Khovanov--Rozansky polynomials~\cite{khovanov2000categorification,khovanov2007virtual} is formulated in terms of complexes
constructed from each link separately, and there is no straightforward evidence for this definition to provide
continuous evolution or some sort of tangle calculus. Extension of tangle calculus to $\ST$-deformed polynomials is still a question. The analogue of Reshetikhin--Turaev formalism
with elementary vertices, which also guarantees the existence of tangle calculus for the HOMFLY polynomials, does not look available.

However, there is certain evidence, that the evolution approach can be applicable,
with a single correction that dependence on evolution parameters is
not always smooth -- there can be jumps at particular values of the parameters around zero.
For Reshetikhin--Turaev approach, all braids are of unit length, and we are fully in the region of jumps in all directions. This fact explains the problems with a direct generalisation of this approach to the Khovanov case.
But when tubes are long, they do not produce jumps, and the only question is whether
the ``connectors'' break smooth evolution. In other words, unlike the case of HOMFLY polynomials, ``connectors'' for superpolynomials inherit information of tangles to be connected.
Another question is a more careful control of jumps, and attempts to make
combinations of tangles which do not have them.
These are difficult questions -- primarily because of the lack of knowledge
about superpolynomials, only few examples are available in literature (for example, in~\cite{fuji2012volume,nawata2012super}).

Evolution for the Khovanov--Rozansky polynomials does exists at least in some regions of parameters
and some families of knots. This was observed by examples of the Khovanov--Rozansky polynomials for torus and twist
knots in~\cite{anokhina2018khovanov,dolotin2014introduction}, for the Khovanov polynomials ($\ST$-deformation of $\mathfrak{sl}_2$ HOMFLY polynomials) for
pretzel knots in~\cite{anokhina2019nimble}. An attempt to introduce tangle calculus for the Khovanov polynomials for
torus and twist satellite knots was made in~\cite{anokhina2021khovanov}. It also seems to be possible to analyse the
existence of continuous evolution and tangle calculus for the Khovanov polynomials using the fact that complexes can be
constructed for separate tangles~\cite{krasner2008computation,bar2005khovanov}. Then, one performs a sort of tensor product of such complexes
in order to obtain a complex for a whole knot which in turn gives an expression for the Khovanov
polynomial~\cite{bar2007fast}.

In this paper, we consider a particular case of the uncolored Khovanov
polynomials (i.e. the Khovanov polynomials colored by fundamental representation) which are much better studied
and can even be practically calculated with the help of commonly available computer program from~\cite{lewark,katlas}. Although we concentrate on the uncolored Khovanov case, a minimal extension from fundamental representation $R=[1]$ to the first symmetric representation $R=[2]$
is needed -- as usual in the evolution method.
Restriction to $R=[2]$ implies restriction to 2-strand braids, and for $N=2$
there is no difference between parallel and antiparallel braids
(representation $[1,1]=\varnothing$ and $[2]={\rm adj}$ is both the first symmetric and adjoint).
We are concentrated on satellite knots, distinguishing two families:
torus and twist satellites, and \textbf{our claim} is that
\begin{eqnarray}
\boxed{
	{\color{red}{\Kh}^{{\rm {\cal S}}_{T_\n }({\cal K})}}
	=
	{\color{red} {\Kh}^{T_{\n+2\,{\rm Sh}_{\cal K}}}} + \mathfrak{c}_T \cdot \lambda_{\rm adj}^{-\n}
	\cdot \mathfrak{Kh}_{\rm adj}^{ {\cal K} }\,.
}
\label{maindeco}
\end{eqnarray}
The formula is the same for torus and twist satellites of ${\cal K}$ shown in Fig.\ref{fig:1} below.
Thus, we use the subscript $T$ to refer both cases.
The only difference is in the coefficient $\mathfrak{c}_T\,$.
Namely, $\mathfrak{c}_{tor}=\mu_{\rm adj}$, and $\mathfrak{c}_{tw}=\tau_{\rm adj}$, see~\eqref{GenMainClaim}.
The {\bf second term} at the r.h.s. is naturally associated with representation $[2]={\rm adj}$.
It is rather complicated, but
{\bf is everywhere smooth} (does not have jumps).
The first term is almost independent of the {\it companion} knot ${\cal K}$ and has jumps --
but it is a well known and simple {\it fundamental} Khovanov polynomial for torus or twist knots,
with a slightly shifted parameter $n$
(the shift ${\rm Sh}_{\cal K}$ is the only thing which depends on ${\cal K}$ in the first term).
To emphasize which quantities are non-analytic in $n$ (i.e. suffer from jumps),
we write them in red.
The truly interesting ${\cal K}$-dependent contribution  $\mathfrak{Kh}_{\rm adj}^{ {\cal K} }$ is black.
It is the same for torus and twist satellites
and it differs only slightly from adjoint Khovanov polynomial for ${\cal K}$.
Additional property of decomposition formula (\ref{maindeco}) is that both terms at the r.h.s.,
not only their sum,  are positive polynomials.

We have begun our presentation by making a precise statement (\ref{maindeco}) about {\it evolution}
of Khovanov polynomials -- a particular case of tangle calculus, when some tangles are long braids.
The simplest class of this type is provided by torus and twist knots,
the next level of complexity are torus and twist  {\it satellites} --
and they are the main subject of this paper. The rest part of \textbf{our text is organised as follows}. In Section~\ref{sec:picture}, we provide pictures of torus and twist satellite knots and clarify their notations.
We explain the statement~\eqref{maindeco} in detail for the simplest satellites in Section~\ref{sec:HOMFLY} at the level of HOMFLY and Jones polynomials. In Section~\ref{sec:KhRed}, we claim the existence of tangle calculus for the Khovanov polynomials for torus and twist satellite knots.
We state how to tame jumps of vertices in the satellite Khovanov polynomials and present our main result~\eqref{maindeco} with examples in Section~\ref{sec:main}.
After that in Section~\ref{sec:checks}, we provide a list of knots for which we have checked our main claim and the existence of tangle calculus for the Khovanov polynomials. Jumps of the reduced satellite Khovanov polynomials depend on a new knot invariant, which we call the shift ${\rm Sh}_{\cal K}$. We describe it in details in Section~\ref{sec:shift}. In Section~\ref{sec:double-evo} we make the first step towards generalising (\ref{maindeco}) to multi-evolution,
namely consider satellites {\it of} torus and twist knots. Finally, in Section~\ref{sec:conclusion} we conclude with the claim that tangle calculus has unexpectedly nice perspectives
for application to Khovanov polynomials: the main problem is attributed to {\it jumps} --
the violation of smooth evolution at the low values of evolution parameters,
and these jumps can be fully controlled by the {\it skeleton} of the knot,
in spirit of the newly discovered formula \eqref{maindeco}. We provide evidence that the statement (\ref{maindeco}) is true for arbitrary torus and twist satellites in Appendix~\ref{App} --
though the general proof remains to be found.

\section{Picture}\label{sec:picture}

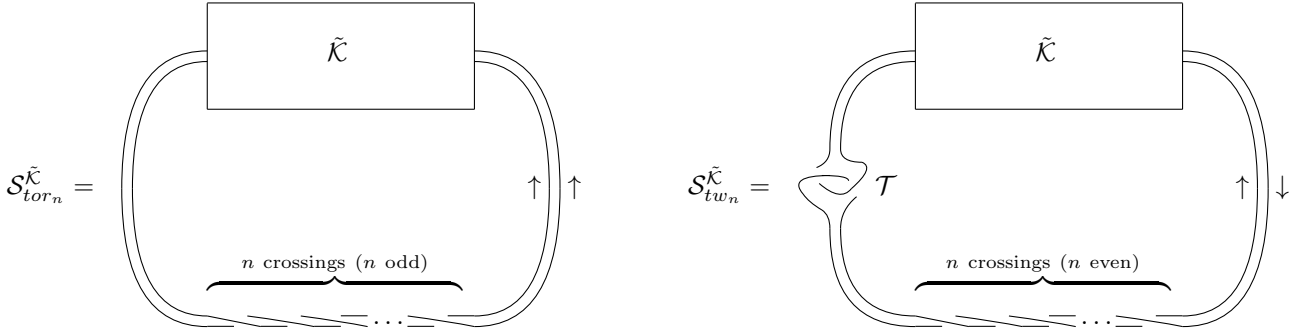
\begin{figure}[h]  \label{gluing}
\begin{picture}(300,150)(-70,-90)

\put(0,0){\line(1,0){100}}
\put(0,40){\line(1,0){100}}
\put(0,0){\line(0,1){40}}
\put(100,0){\line(0,1){40}}


\qbezier(100,18)(128,18)(128,-30)
\qbezier(100,-78)(128,-78)(128,-30)
\qbezier(100,22)(132,22)(132,-30)
\qbezier(100,-82)(132,-82)(132,-30)



\qbezier(0,18)(-28,18)(-28,-30)
\qbezier(0,-78)(-28,-78)(-28,-30)
\qbezier(0,22)(-32,22)(-32,-30)
\qbezier(0,-82)(-32,-82)(-32,-30)

\put(45,18){\mbox{$\tilde{\cal K}$}}

\qbezier(0,-78)(10,-80)(20,-82) \put(0,-82){\line(1,0){10}}
\qbezier(15,-78)(30,-80)(40,-82) \put(20,-82){\line(1,0){10}}
\qbezier(35,-78)(50,-80)(60,-82) \put(40,-82){\line(1,0){10}} \put(50,-78){\line(1,0){10}}
\put(62,-81){\mbox{$\ldots$}}
\qbezier(75,-78)(90,-80)(100,-82) \put(75,-82){\line(1,0){10}} \put(90,-78){\line(1,0){10}}
\put(0,-75){\mbox{$\overbrace{\phantom{----------}}^{\n\ {\rm crossings} \ (\n\ {\rm odd})}$}}

\put(120,-32){\mbox{$\uparrow$}}
\put(135,-32){\mbox{$\uparrow$}}

\put(-75,-32){\mbox{${\cal S}_{tor_{\n}}^{\tilde{\cal K}}  = $}}

\put(265,0){

\put(0,0){\line(1,0){100}}
\put(0,40){\line(1,0){100}}
\put(0,0){\line(0,1){40}}
\put(100,0){\line(0,1){40}}


\qbezier(100,18)(128,18)(128,-30)
\qbezier(100,-78)(128,-78)(128,-30)
\qbezier(100,22)(132,22)(132,-30)
\qbezier(100,-82)(132,-82)(132,-30)

\qbezier(0,18)(-28,18)(-28,-15)
\qbezier(0,-78)(-28,-78)(-28,-45)
\qbezier(0,22)(-32,22)(-32,-15)
\qbezier(0,-82)(-32,-82)(-32,-45)

\qbezier(-28,-15)(-28,-20)(-20,-20)
\qbezier(-20,-20)(-15,-22)(-24,-30)
\qbezier(-36,-28)(-28,-35)(-24,-30)
\qbezier(-32,-15)(-32,-20)(-37,-22)

\qbezier(-28,-45)(-28,-38)(-22,-33)
\qbezier(-25,-27)(-25,-25)(-35,-25)
\qbezier(-32,-25)(-48,-25)(-42,-32)
\qbezier(-35,-38)(-40,-35)(-42,-32)
\qbezier(-32,-45)(-32,-40)(-35,-38)

\put(45,18){\mbox{$\tilde{\cal K}$}}

\qbezier(0,-78)(10,-80)(20,-82) \put(0,-82){\line(1,0){10}}
\qbezier(15,-78)(30,-80)(40,-82) \put(20,-82){\line(1,0){10}}
\qbezier(35,-78)(50,-80)(60,-82) \put(40,-82){\line(1,0){10}} \put(50,-78){\line(1,0){10}}
\put(62,-81){\mbox{$\ldots$}}
\qbezier(75,-78)(90,-80)(100,-82) \put(75,-82){\line(1,0){10}} \put(90,-78){\line(1,0){10}}
\put(0,-75){\mbox{$\overbrace{\phantom{----------}}^{\n\ {\rm crossings} \ (\n\ {\rm even)}}$}}

\put(-15,-32){\mbox{$\cal T$}}

\put(120,-32){\mbox{$\uparrow$}}
\put(135,-32){\mbox{$\downarrow$}}

\put(-85,-32){\mbox{${\cal S}_{tw_{\n}}^{\tilde{\cal K}} = $}}
}

\end{picture}
\caption{\footnotesize
Two 2-cabled knot diagrams $\tilde{\cal K}$, cut at arbitrary place,   with the free ends glued together
in two different ways to form a new knot.
Position of cut and new $n$ crossings does not matter
(all choices are topologically/Reidemeister equivalent).
Notations are adjusted to allow a unified notation ${\cal S}_{T_\n}^{\tilde{\cal K}}$ for both cases. Note that we fix the parity of $n$ crossings to be odd in the case of torus satellites and even in the case of twist satellites to choose formulas for $\mu$ and $\tau$ elements, see subsequent sections. Arrows show orientation of a resulting knot. Twisting tangle is denoted by $\cal T$.
A possible content of the box is shown in Fig. \ref{fig:2}.
Note that this definition involves not just the knot ${\cal K}$,
but a particular knot diagram $\tilde{\cal K}$.
However, dependence on the diagram is simple:
it can be eliminated by s shift of $\n$ by $2w_{\tilde{\cal K}}$, where $w_{\tilde{\cal K}}$ is
the {\it writhe number} of the diagram (the difference between the numbers
 of positive and negative crossings in $\tilde{\cal K}$):
${\cal S}_{T_\n}({\cal K}):={\cal S}_{T_{\n-2w_{\tilde{\cal K}}}}^{\tilde{\cal K}}$
does not depend on the diagram.
}\label{fig:1}
\end{figure}

\begin{figure}[h] \label{contenttref}
\begin{picture}(300,130)(-80,-60)

\put(150,0){

\qbezier(-44,12)(-30,-65)(50,5)     \qbezier(-48,9)(-30,-69)(50,0)
\qbezier(20,42)(110,51)(50,0)      \qbezier(24,39)(100,48)(50,5)
\qbezier(0,40)(-20,38)(-60,12)      \qbezier(6,37)(-20,34)(-60,8)
\qbezier(-49,22)(-48,90)(52,16)    \qbezier(-45,24)(-45,84)(50,12)
\qbezier(-60,8)(-122,-35)(-145,-30) \qbezier(-60,12)(-122,-31)(-145,-26)
\qbezier(61,5)(125,-30)(155,-30)    \qbezier(64,8)(125,-26)(155,-26)

\put(-130,-40){\line(1,0){270}}
\put(-130,60){\line(1,0){270}}
\put(-130,-40){\line(0,1){100}}
\put(140,-40){\line(0,1){100}}

}

\end{picture}
\caption{\footnotesize  2-cabled trefoil as an example of possible content of the box
in Fig. \ref{fig:1}.
See Fig. \ref{fig:3} below for possible 2-strand and 3-strand realizations of this trefoil,
which are convenient for calculations in the case of HOMFLY polynomials and clarify the structure of the answer
in the Khovanov case.
}
\label{fig:2}
\end{figure}
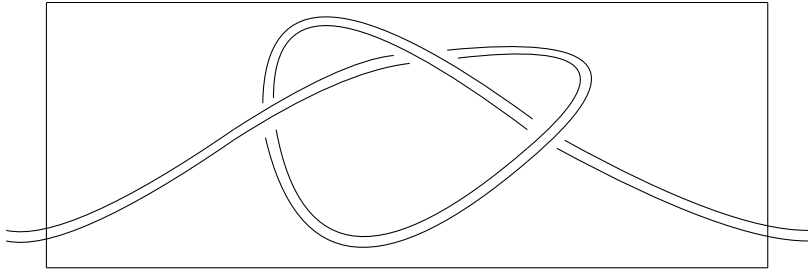

In this paper, we study knot polynomials for the peculiar class of {\it satellite} knots
${\cal S}^{\tilde{{\cal K}}}_{{\cal K}'}$. We take a two-strand tangle of a companion knot diagram $\tilde{\cal K}$ and double its strand (see Fig.~\ref{fig:2} as an example). This operation is called 2-cabling and described in detail in~\cite{anokhina2014cabling}. Then, we glue together the ends of the resulting tangle.
In the case of 2-cabling there are just two simple ways to make such a gluing,
these are known as torus ${\cal K}'=tor_n$ and twist ${\cal K}'=tw_{n}$ satellites of ${\tilde{\cal K}}$, see Fig.~\ref{fig:1}.
Generalisations of tangle calculus to $n$-cables are straightforward, but the variety of possible gluings,
making a {\it knot} (rather than a link) from $n$-cables increases.

\newpage

\setcounter{equation}{0}
\section{HOMFLY polynomials for satellite knots}\label{sec:HOMFLY}
We begin our study of polynomial knot invariants for torus and twist satellites by considering the HOMFLY case. In this paper, we consider only the case of fundamental representation, as our goal is to describe the Khovanov polynomials for torus and twist satellite knots. Nowadays, big enough data for the Khovanov polynomials can be obtained only for the uncolored case.

For the colored HOMFLY polynomials, tangle calculus can be derived analytically with the use of Reshetikhin--Turaev approach and provides some intuition on how tangle calculus for the Khovanov polynomials can be realised.
\subsection{Twist satellites}
According to equation (8) of \cite{morozov2018knot},
tangle calculus implies for the right picture in Fig. \ref{fig:1} that
{\it unreduced} HOMFLY polynomial in the fundamental representation and in the topological framing is
\be\label{twistgluing}
{\cal H}^{{\cal S}_{tw_\n}^{\tilde{\cal K}}}_\Box = {\cal T}_{\varnothing}+ \overbrace{(-A)^{-\n}}^{\lambda_{\rm adj}^{-\n}}\cdot \overbrace{\frac{\{Aq\}\{A/q\}}{\{q\}^2}}^{\rm qdim_{adj}}
\cdot\,{\cal T}_{\rm adj} \
\cdot (-A)^{-2w_{\tilde{{\cal K}}}}{H}^{{\cal K}}_{\rm adj}\,,
\ee
where ${\rm qdim}_R$ is the quantum dimension of representation $R$ and
\begin{equation}
    {\cal T}_\varnothing = A^{-2}\{A\}\left(-\frac{q^2}{A}-\frac{1}{A q^2}+A+\frac{1}{A}\right)\quad \text{and}\quad {\cal T}_{\rm adj} =-A\,\frac{\{q\}^2}{\{A\}}\,.
\end{equation}
{\it Reduced}  HOMFLY is obtained by division of the unreduced one over ${\rm qdim}_\Box = \frac{\{A\}}{\{q\}}$.
We also absorb all quantum dimensions and in what follows use the notation
$\tau_R:={\cal T}_R\cdot\frac{ {\rm qdim}_R}{{\rm qdim}_\Box}$.
The quantities ${H}_R^{{\cal K}}$ at the r.h.s. are actually the reduced HOMFLY polynomials in appropriate
representations.
There also appears the framing factor $\lambda_{\rm adj}^{-2w_{\tilde{\cal K}}}$, see details in~\cite{anokhina2014cabling}.
Abbreviated notation $\{x\}:=x-x^{-1}$  is used in formulas for quantum dimensions.

Formula~\eqref{twistgluing} provides a good illustration of tangle calculus for the HOMFLY polynomial. Knot ${\cal S}_{tw_{\n}}^{\tilde{\cal K}}$ is split into three tangles with four free legs. Twisting tangle, see Fig.~\ref{fig:1}, gives contribution ${\cal T}_Y$ to the satellite HOMFLY polynomial. Tangle containing $n$ crossings corresponds to $\lambda_Y^{-n}$, and tangle of ${\cal K}$ gives $\lambda_Y^{-2w_{\tilde{\cal K}}} H_Y^{\cal K}$. Gluing together all the tangles causes the appearance of quantum dimension factor. Each tangle carries representation $\Box\otimes \overline{\Box}=\varnothing+{\rm adj}$. Thus, the summation over $Y=\varnothing,\,{\rm adj}$ is implied\footnote{Note that $\lambda_\varnothing={\rm qdim}_\varnothing=H^{\cal K}_\varnothing=1$, so that they do not contribute to~\eqref{twistgluing}.}.

Let us also explain the notions used in Introduction. Here braid tangle with $n$ crossings provides a {\it tube} of length $n$. The role of a {\it connector} is played by the twisting tangle. Gluing tangle of $\tilde{\cal K}$ and $n$-tube with the use of the connector causes contribution ${\cal T}_Y\cdot {\rm qdim}_Y$ to~\eqref{twistgluing}.

One can rewrite formula~\eqref{twistgluing} for the reduced case in a knot diagram independent form:
\begin{equation}
H^{{\cal S}_{tw_{\n}}({\cal K})}_\Box
:=H^{{\cal S}_{tw_{\n-2w_{\tilde{\cal K}}}}^{\tilde{\cal K}}}_\Box
=\overbrace{ {\tau}_{\varnothing}
	+ {\tau}_{\rm adj} \cdot (-A)^{-\n -2\,{\rm Sh}}\cdot
}^{  H_{\Box}^{{\rm Tw}_{\n+2\,{\rm Sh}}}}
+\, {\tau}_{\rm adj} \cdot (-A)^{-\n}\cdot
\Big(\overbrace{{H}^{{\cal K}}_{\rm adj}-
	(-A)^{-2\,{\rm Sh}}
}^{\mathfrak{H}_{\rm adj}^{\cal K}}\Big)\,,
\label{eqH}
\end{equation}
where ${\rm Sh}$ is an arbitrary number.
This can look like a superficial complication, but
actually gives a good starting point for consideration of superpolynomials.
Unfortunately,  the lack of experimental information suggests that we better proceed in steps
and begin with Khovanov polynomials (i.e. from $\mathfrak{sl}_2$), where experimental data is easily available.
It is at that stage that the role of the factor
$(-A)^{-2\,{\rm Sh}}=\lambda_{\rm adj}^{-2\,{\rm Sh}}$ will get clarified.

\subsection{Torus satellites}
For the left picture in Fig.~\ref{fig:1}, we have
\be \label{HSatPar}
{\cal H}^{{\cal S}_{tor_\n}^{\tilde{\cal K}} }_\Box = \left( \frac{q}{A} \right)^{-\n}
\cdot \overbrace{\frac{\{A\}\{Aq\}}{\{q\}\{q^2\}}}^{{\rm qdim}_{[2]}} \cdot  \left(\frac{q}{A}\right)^{-2w_{\tilde{{\cal K}}}}{H}^{{\cal K}}_{[2]}
+ \left(-\frac{1}{qA}\right)^{-\n}
\cdot \overbrace{\frac{\{A\}\{A/q\}}{\{q\}\{q^2\}}}^{{\rm qdim}_{[1,1]}} \cdot \left(-\frac{1}{qA}\right)^{-2w_{\tilde{{\cal K}}}}{H}^{{\cal K}}_{[1,1]}\,.
\ee
Here tangle calculus also manifests. The only differences are the absence of twisting tangle and the orientation of a torus satellite knot -- now strands are parallel and the summation is made over the representation $Y=[1,1],\,[2]\in \Box\otimes \Box$. In this case, the contribution of connecting {\it vertex} is ${\rm qdim}_Y$.

Formula~\eqref{HSatPar} can be rewritten for the reduced case in a knot diagram independent form:
{\footnotesize \begin{equation}\label{HTorSat}
		\begin{aligned}
   H^{{\cal S}_{tor_{\n-2w_{\tilde{\cal K}}}}^{\tilde{\cal K}} }_\Box =
   (-q A)^{-2\,{\rm Sh}}\cdot \Big\{\overbrace{\left(-\frac{1}{qA}\right)^{-(\n+2\,{\rm Sh})}\mu_{[1,1]}+\left( \frac{q}{A} \right)^{-(\n+2\,{\rm Sh})}\mu_{[2]}}^{H_\Box^{tor_{\n+2\,{\rm Sh}}}}\Big\} \cdot {H}^{{\cal K}}_{[1,1]}+\left( \frac{q}{A} \right)^{-\n} \mu_{[2]}\cdot \Big({H}^{{\cal K}}_{[2]}-(-q^2)^{-2\,{\rm Sh}}{H}^{{\cal K}}_{[1,1]}\Big)\,,
		\end{aligned}
\end{equation}}
where Sh is an arbitrary number.
Here we use the notation
\begin{equation}
	\mu_R :=\frac{{\rm qdim}_R}{{\rm qdim}_\Box}\,.
\end{equation}
In what follows, we study the Khovanov polynomials for torus satellite knots. Khovanov polynomial is the $\ST$-deformation of Jones polynomial. Thus, consider formula~\eqref{HTorSat} in the case of Jones polynomials $A=q^2$ (i.e. for $\mathfrak{sl}_2$), in which it simplifies and possesses additional properties:
\begin{equation}\label{JonesTorSat}
	J^{{\cal S}_{tor_{\n}}
		({\cal K})}_\Box
	:=(-q^3)^{-\n}J^{{\cal S}_{tor_{\n-2w_{\tilde{\cal K}}}}^{\tilde{\cal K}}}_\Box
	=\overbrace{\left\{\mu_{\rm adj}\cdot(-q^2)^{-(\n+2\,{\rm Sh})}+\mu_{[1,1]}\right\}
	}^{J_\Box^{tor_{\n+2\,{\rm Sh}}}}
	+\mu_{\rm adj}\cdot(-q^2)^{-\n}\cdot\Big(\overbrace{J_{\rm adj}^{\cal K}-(-q^2)^{-2\,{\rm Sh}}}^{\mathfrak{H}_{\rm adj}^{\cal K}\big|_{A=q^2}}\Big)\,.
\end{equation}
For $\mathfrak{sl}_2$, representations $\Box$ and $\overline{\Box}$ are isomorphic. This property is inherited by Jones polynomials in vertical framing. In order to provide explicit isomorphism $\varnothing=[1,1]$ and ${\rm adj}=[2]$, we separated framing factor $(-q^3)^{\n}$ from Jones polynomial in the topological framing. Note that in this expression the same $\mathfrak{H}_{\rm adj}^{\cal K}$ as in~\eqref{eqH} is present, and we expect this fact to hold in the Khovanov case too.

\subsection{Unified formula for torus and twist satellites}
Formulas~\eqref{eqH},~\eqref{JonesTorSat} become almost the same for the Jones polynomials:
\begin{equation}\label{JonesUniSat}
    J^{{\cal S}_{T_n}({\cal K})}_\Box = J_\Box^{T_{\n + 2\,{\rm Sh}}}+c_T\cdot \lambda_{\rm adj}^{-n}\cdot \mathfrak{H}_{\rm adj}^{\cal K}\big|_{A=q^2}\,,
\end{equation}
where $\lambda_{\rm adj}=-q^2$ for $A=q^2$ and $T=tor,\, tw$ refers to both torus and twist satellite knots, correspondingly. In this formula, only $c_T$ multiplier is different for torus and twist cases, $c_{tor}=\mu_{\rm adj}$ and $c_{tw}=\tau_{\rm adj}$. We emphasise that formula~\eqref{JonesUniSat} looks exactly like~\eqref{maindeco}, but is everywhere smooth (i.e. all terms take the same form for any value of the {\it evolution parameter} $n$). An important peculiarity of formula~\eqref{maindeco} is that shift ${\rm Sh}$ becomes dependent on a companion knot $\cal K$ to provide quantities free of jumps. This fact is disscussed in details in subsequent sections.

%
%
%


%


\setcounter{equation}{0}
\section{Reduced Khovanov polynomials for satellite knots}\label{sec:KhRed}

In this section, we describe a generalisation of the above results for the HOMFLY tangle calculus to the {\it reduced} Khovanov polynomials for torus and twist satellites. As it happens for the uncolored Khovanov polynomials for torus and twist knots, the satellite one-parametric Khovanov polynomials have two branches with an ``infinitesimal" jump. In Subsection~\ref{sec:gen-claim-KTC}, we state how tangle calculus for the satellite Khovanov polynomials works for {\it any} companion knot $\cal K$. We demonstrate how the Khovanov polynomials for the satellite knots split for the simplest satellites of ${\cal K}=3_1$ in Subsections~\ref{sec:diag-dep},~\ref{sec:diag-dep-tor}.
In Appendix~\ref{App} we provide evidence that claims of this section remain true for satellites of other
knots ${\cal K}$ -- and that the jumping term is actually almost independent of ${\cal K}$.
We also address the question of dependence on the knot diagram and explain how to handle it properly in Section~\ref{sec:main}.


\subsection{General claim}\label{sec:gen-claim-KTC}
The {\bf claim of this section} is that the Khovanov polynomials for twist and torus satellites of an {\it arbitrary} companion knot $\cal K$ are consistent with tangle calculus. Namely, the following splits hold:
\begin{equation}\label{GenKhTwSat}
\begin{aligned}
    {\color{red}\Kh^{{\cal S}^{\tilde{\cal K}}_{tw_{\n-2w_{\tilde{\cal K}}}}}}&=\overbrace{{\color{red} (-\ST)^{\theta(\n+2\,{\rm Sh}_{\cal K})}}\cdot q^2\,
\frac{1+q^4\ST ^2}{1-q^2\ST }}^{{\color{red}\tau_{\varnothing}}
}
&\overbrace{ -\,\ST ^{-1}\frac{1+q^6 \ST ^3}{1-q^2\ST }}^{\tau_{\rm adj}}\cdot \overbrace{(q^2\ST )^{-\n}}^{\lambda_{\rm adj}^{-\n}}\cdot\, {\color{red} \Kh_{\rm adj}^{\cal K}}\,, \\
{\color{red}\Kh ^{{\cal S}_{tor_{\n-2w_{\tilde{\cal K}}}}^{\tilde{\cal K}}}}
	&=\overbrace{{\color{red} (-\ST)^{\theta(\n+2\,{\rm Sh}_{\cal K})}}\cdot \frac{q}{1-q^2 \ST}}^{{\color{red} \mu_{[1,1]}}}\cdot\overbrace{(q^3\ST)^{\n}}^{\lambda_{[1,1]}^{\n}}
	&+\overbrace{ \frac{q^4 \ST^2-q^2 \ST+1}{q \ST \left(q^2 \ST-1\right)}}^{\mu_{\rm adj}}\cdot\overbrace{q^{\n}}^{\lambda_{[2]}^{\n}}\cdot\,{\color{red}\Kh _{\rm adj}^{\cal K}}\,,
\end{aligned}
\end{equation}
where $\theta(x)$ is the Heaviside step function:
\begin{equation}
    \theta(x)=\begin{cases}
        1,\quad x\geq 0\,, \\
        0,\quad x < 0\,,
    \end{cases}
\end{equation}
and we remind that we mark in red terms having jumps in {\it evolution parameter} $\n$. Here we have shifted the parameter $\n$ to obtain knot diagram independent quantities (see Subsections~\ref{sec:diag-dep}–\ref{sec:diag-dep-tor} for discussion). We emphasise that these formulas have two smooth branches with a jump at the point $n+2\,{\rm Sh}_{\cal K}=0$. The adjoint Khovanov polynomial ${\color{red} \Kh_{\rm adj}^{\cal K}}$ possess a non-multiplicative jump\footnote{By the notion of ``multiplicative jump'' we mean that the Khovanov polynomials on two branches differ by just a single multiplier, as it happens for example for Khovanov polynomials for torus and twist knots~\eqref{Tor&TwKh}, where the jump is concentrated in a multiplier ${\color{red} (-\ST)^{\theta(n)}}$. On the contrary, the adjoint Khovanov polynomials have non-mutiplicative jump which is concentrated in the only additive term, as it is shown in~\eqref{KhTw31}.} at this point, but for {\it any} knot $\cal K$ the difference is concentrated in one summand {\it only}, as it is illustrated in the example of trefoil in the subsequent sections\footnote{In fact, there exists even simpler example of ${\cal K}=\,$unknot. In this case, ${\cal S}^{\rm unknot}_{tw_{\n-2w_{\rm unknot}}}=tw_\n\,,\; {\cal S}^{\rm unknot}_{tor_{\n-2w_{\rm unknot}}}=tor_\n$ and
\begin{equation}\label{Tor&TwKh}
\begin{aligned}
{\color{red} \Kh_\Box^{tw_\n}}&=\overbrace{{\color{red} (-\ST)^{\theta(\n)}}\cdot q^2\,
\frac{1+q^4\ST ^2}{1-q^2\ST }}^{{\color{red}\tau_{\varnothing}}
}
&\overbrace{ -\,\ST ^{-1}\frac{1+q^6 \ST ^3}{1-q^2\ST }}^{\tau_{\rm adj}}\cdot \overbrace{(q^2\ST )^{-\n}}^{\lambda_{\rm adj}^{-\n}}\cdot\, \overbrace{{\color{red} (-\ST)^{\theta(\n)}}}^{{\color{red} \Kh_{\rm adj}^{\rm unknot}}}\,, \\
{\color{red}\Kh_\Box^{tor_{\n}}}
	&=\overbrace{{\color{red} (-\ST)^{\theta(\n)}}\cdot \frac{q}{1-q^2 \ST}}^{{\color{red} \mu_{[1,1]}}}\cdot\overbrace{(q^3\ST)^{\n}}^{\lambda_{[1,1]}^{\n}}
	&+\overbrace{ \frac{q^4 \ST^2-q^2 \ST+1}{q \ST \left(q^2 \ST-1\right)}}^{\mu_{\rm adj}}\cdot\overbrace{q^{\n}}^{\lambda_{[2]}^{\n}}\cdot\,\overbrace{{\color{red} (-\ST)^{\theta(\n)}}}^{{\color{red}\Kh _{\rm adj}^{\rm unknot}}}\,,
\end{aligned}
\end{equation}
what corresponds to~\eqref{GenKhTwSat} for ${\rm Sh}_{\rm unknot}=0$.}. An important {\it feature} is that ${\rm Sh}_{\cal K}$ dependent of ${\cal K}$ is present in~\eqref{GenKhTwSat}. Its connection with $\rm Sh$ which we have implemented in~\eqref{eqH} and~\eqref{HTorSat} will get clarified in Section~\ref{sec:main}.

As we expect equations~\eqref{GenKhTwSat} to hold for any knot $\cal K$, we claim that {\bf tangle calculus for the Khovanov polynomials for twist satellite knots does hold}, but on two branches unlike the HOMFLY case~\eqref{twistgluing},~\eqref{HSatPar}. Here we use the same notations $\lambda_{\rm adj}\,,\;\lambda_{[1,1]}\,,\;\lambda_{[2]}$ and $\tau_\varnothing\,,\;\tau_{\rm adj}\,,\;\mu_{[1,1]}\,,\;\mu_{\rm adj}$ for $\ST$-deformed elements, and one can easily check that they become the same elements for the Jones case by setting $\ST=-1$. In addition, the Khovanov polynomials give the Jones polynomials by the same substitution:
\begin{equation}
    {\color{red}\Kh^{{\cal S}_{T_{\n-2w_{\tilde{\cal K}}}}^{\tilde{\cal K}}}(\ST=-1,q)}=J_\Box^{{\cal S}_{T_{\n-2w_{\tilde{\cal K}}}}^{\tilde{\cal K}}}\quad \text{and}\quad {\color{red}\Kh_{\rm adj}^{\cal K}(\ST=-1,q)}=J_{\rm adj}^{\cal K}\,,
\end{equation}
and this is valid on both branches. Here we use the notation $T=tw,\,tor$ to refer to both twist and torus satellite knots.

As we have declaired in Introduction, for the Khovanov polynomials the ``connectors'' ${\color{red} \tau_\varnothing}\,,\;{\color{red} \mu_{[1,1]}}$ contain information about glued parts. In other words, it is different on two branches and its jump depends on the tube through the evolution parameter $\n$ and on the companion knot through the shift ${\rm Sh}_{\cal K}$. The adjoint Khovanov polynomial for a knot $\cal K$ also ``feels'' that it is connected to $n$-tube -- it has the jump at the same point $n+2\,{\rm Sh}_{\cal K}=0$.

\subsection{Example of twist satellites of the trefoil. Dependence on a diagram}\label{sec:diag-dep}

Let us consider the simplest non-trivial example of twist satellite -- with ${\cal K}={\rm trefoil}$ given by diagrams in Fig.~\ref{fig:3}.

{\begin{figure}[h] \label{contenttref}
\resizebox{.35\textwidth}{!}{%
\begin{picture}(200,110)(-145, -10)

\put(-100,  10){
\put(-20,-20){\line(0,1){40}}  \put(-15,-12){\mbox{{\footnotesize$Y\in \Box\otimes \overline{\Box}$ or}}}
\put(-15,-20){\mbox{{\footnotesize$Y\in \Box\otimes \Box$}}}
\put(60,-20){\line(0,1){80}}  \put(65,-20){\mbox{{\footnotesize$Q\in Y^{\otimes 2}$}}}
\put(-30,-1){\line(-1,0){10}} \put(-30,1){\line(-1,0){10}}
\put(230,-1){\line(1,0){10}} \put(230,1){\line(1,0){10}}
}

\put(200, 10){
\put(-20,-20){\line(0,1){40}}  \put(-15,-12){\mbox{{\footnotesize$Y\in \Box\otimes \overline{\Box}$ or}}}
\put(-15,-20){\mbox{{\footnotesize$Y\in \Box\otimes \Box$ }}}
\put(90,-20){\line(0,1){90}}  \put(95,-20){\mbox{{\footnotesize$Q\in Y^{\otimes 3}$}}}
\put(-30,-1){\line(-1,0){10}} \put(-30,1){\line(-1,0){10}}
\put(210,-1){\line(1,0){10}} \put(210,1){\line(1,0){10}}
}

 \linethickness{1mm}
\put(-100, 10){

\qbezier(-30,0)(30,0)(40,20)
\qbezier(40,20)(50,40)(70,40)
\qbezier(70,40)(90,40)(100,20)
\qbezier(100,20)(110,0)(130,0)
\qbezier(130,0)(150,0)(160,20)
\qbezier(160,20)(170,40)(190,40)

\qbezier(0,40)(30,40)(40,20)
\qbezier(40,20)(50,0)(70,0)
\qbezier(70,0)(90,0)(100,20)
\qbezier(100,20)(110,40)(130,40)
\qbezier(130,40)(150,40)(160,20)
\qbezier(160,20)(170,0)(230,0)

\qbezier(0,40)(-30,40)(-30,60)
\qbezier(-30,60)(-30,80)(100,80)
\qbezier(190,40)(220,40)(220,60)
\qbezier(220,60)(220,80)(100,80)

}

\put(200, 10){

\qbezier(-30,0)(30,0)(50,30)
\qbezier(50,30)(70,60)(90,60)
\qbezier(90,60)(110,60)(120,50)
\qbezier(120,50)(130,40)(150,40)

\qbezier(0,40)(30,40)(40,20)
\qbezier(40,20)(50,0)(70,0)
\qbezier(70,0)(90,0)(110,30)
\qbezier(110,30)(130,60)(150,60)

\qbezier(30,60)(60,60)(80,30)
\qbezier(80,30)(100,0)(210,0)

\qbezier(30,60)(10,60)(10,70)
\qbezier(10,70)(10,80)(80,80)
\qbezier(150,60)(170,60)(170,70)
\qbezier(170,70)(170,80)(80,80)

\qbezier(0,40)(-20,40)(-20,65)
\qbezier(-20,65)(-20,90)(80,90)
\qbezier(150,40)(200,40)(200,65)
\qbezier(200,65)(200,90)(80,90)

}

\end{picture}%
}
\caption{\footnotesize
Two realizations of the trefoil from Fig. \ref{fig:2}.
The trefoil itself can be represented as a 2-strand braid with each strand
in representation $[2]\oplus [1,1]$ or $\varnothing\oplus {\rm adj}$.
The corresponding knot polynomial is then decomposed in contributions of irreducible
representations from $[2]\otimes[2]$ and $[1,1]\otimes[1,1]$
or $\varnothing\otimes\varnothing=\varnothing$ and a triple from ${\rm adj}\otimes {\rm adj}$.
Alternative representation of the same trefoil is 3-strand. Then cubes of representations
appear instead of squares.
If the trefoil is substituted by an arbitrary $m$-strand knot, representations contributing to a knot polynomial are
irreducible representations from the $m$-th powers $[2]^{\otimes m}$, $[1,1]^{\otimes m}$, $\varnothing^{\otimes m}=\varnothing$, ${\rm adj}^{\otimes m}$.
In the case of Jones and Khovanov polynomials ($N=2$) for satellite {\it knots} (not links) the only non-trivial contribution is
from ${\rm adj}^{\otimes m}=[2]^{\otimes m}$.
}
\label{fig:3}
\end{figure}
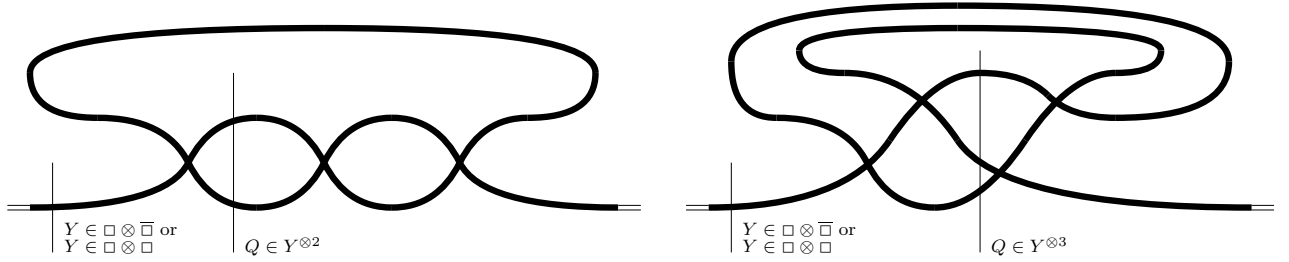}
In this section, we consider two diagrams of the trefoil in order to illustrate how the dependence of ${\cal S}^{\tilde{\cal K}}_{tw_n}$ on a knot diagram $\tilde{\cal K}$ influences on satellite Khovanov polynomials, as predicted by appearance of $\lambda_Y^{-2 w_{\tilde{\cal K}}}$ factors in~\eqref{twistgluing}.
As usual with evolution of Khovanov polynomials \cite{dunin2013superpolynomials}, the answer has two branches,
for positive and negative values of the evolution parameter $\n$.
In fact, the boundary in this case is at $\n=0$ and the difference between the two formulas is shown in red:
{\scriptsize \begin{equation}\label{KhTw31}
\begin{aligned}
{\color{red}\Kh ^{{\cal S}_{tw_\n}^{3_1^{(2)}}}}\Big|_{\n\geq 0}
& =\overbrace{{\color{red} (-\ST)}\cdot q^2\,
\frac{1+q^4\ST ^2}{1-q^2\ST }}^{\tau_{\varnothing}
}
& \overbrace{ -\,\ST ^{-1}\frac{1+q^6 \ST ^3}{1-q^2\ST }}^{\tau_{\rm adj}
}\overbrace{(q^2\ST )^{-(\n-6)}}^{\lambda_{\rm adj}^{-(\n\overbrace{-6}^{2w_{3_1}^{(2)}})}}
\overbrace{\left\{   (q^2\ST)^{-11}
(q^{18}\ST ^{11}+q^{12} \ST ^7-q^{10} \ST ^7-{\color{red} q^{10}\ST ^6} -q^8 \ST ^5+q^6 \ST ^3+q^4 \ST ^2-q^2 \ST-1)
\right\}
}^{{\color{red}\Kh _{\rm adj}^{3_1}}}\,,
\\
{\color{red}\Kh ^{{\cal S}_{tw_\n}^{3_1^{(2)}}}}\Big|_{\n<0}
&=\overbrace{q^2\,\frac{1+q^4
\ST ^2}{1-q^2\ST }}^{\tau_{\varnothing}}
&\overbrace{-\,\ST ^{-1}\frac{1+q^6 \ST ^3}{1-q^2\ST }}^{\tau_{\rm adj}}
\overbrace{(q^2\ST )^{-(\n-6)}}^{\lambda_{\rm adj}^{-(\n-6)}}\overbrace{
\left\{  (q^2\ST)^{-11}
(q^{18}\ST ^{11}+q^{12} \ST ^7-q^{10} \ST ^7 +{\color{red} q^{10}\ST ^5}-q^8 \ST ^5+q^6 \ST ^3+q^4 \ST ^2-q^2 \ST-1)
\right\}}^{{\color{red}\Kh _{\rm adj}^{3_1}}}\,.
\end{aligned}
\end{equation}}

\noindent
Superscript $(2)$ refers to a 2-strand realisation of the trefoil $3_1$.
To compare, in the 3-strand realisation we get instead
{\scriptsize \begin{equation}\label{KhTw31Str3}
\begin{aligned}
{\color{red}\Kh ^{{\cal S}_{tw_\n}^{3_1^{(3)}}}}\Big|_{\n\geq2}
& =\overbrace{{\color{red} (-\ST)}\cdot q^2\,
\frac{1+q^4\ST ^2}{1-q^2\ST }}^{{\color{red}\tau_{\varnothing}}
}
& \overbrace{ -\,\ST ^{-1}\frac{1+q^6 \ST ^3}{1-q^2\ST }}^{\tau_{\rm adj}}\cdot \overbrace{(q^2\ST )^{-(\n-8)}}^{\lambda_{\rm adj}^{-(\n\overbrace{-8}^{2w_{3_1}^{(3)}})}}
\overbrace{\left\{   (q^2\ST)^{-11}
(q^{18}\ST ^{11}+q^{12} \ST ^7-q^{10} \ST ^7-{\color{red} q^{10}\ST ^6} -q^8 \ST ^5+q^6 \ST ^3+q^4 \ST ^2-q^2 \ST-1)
\right\}
}^{{\color{red} \Kh _{\rm adj}^{3_1}}}\,,
\\
{\color{red}\Kh ^{{\cal S}_{tw_\n}^{3_1^{(3)}}}}\Big|_{\n<2}
&=\overbrace{q^2\,\frac{1+q^4
\ST ^2}{1-q^2\ST }}^{{\color{red} \tau_{\varnothing}}}
&\overbrace{-\,\ST ^{-1}\frac{1+q^6 \ST ^3}{1-q^2\ST }}^{\tau_{\rm adj}}
\cdot \overbrace{(q^2\ST )^{-(\n-8)}}^{\lambda_{\rm adj}^{-(\n-8)}}\overbrace{
\left\{  (q^2\ST)^{-11}
(q^{18}\ST ^{11}+q^{12} \ST ^7-q^{10} \ST ^7 +{\color{red} q^{10}\ST ^5}-q^8 \ST ^5+q^6 \ST ^3+q^4 \ST ^2-q^2 \ST-1)
\right\}}^{{\color {red}\Kh _{\rm adj}^{3_1}}}\,.
\end{aligned}
\end{equation}}

\noindent
Clearly, dependence on the number of strands (i.e. on the choice of a knot diagram of $3_1$)
disappears from the shifted quantity
${\color{red}\Kh ^{{\cal S}_{tw_{\n}}(3_1)}}:={\color{red} \Kh ^{{\cal S}^{\tilde{3}_1}_{tw_{\n-2w_{\tilde{3}_1}}}}}$,
where $w_{\tilde{3}_1}$ is the writhe number of a diagram of $3_1$. This dependence on a knot diagram can be seen already at the level of the HOMFLY and Jones polynomials, but above examples are used to illustrate
that it works just in the same way after the ${\ST}$-deformation to the Khovanov polynomials.

Note that these examples are in accordance with the general claim~\eqref{GenKhTwSat} for ${\rm Sh}_{3_1}=3$. The adjoint Khovanov polynomials for $3_1$ differ on two branches only ``infinitesimally'', as we have mentioned in Subsection~\ref{sec:gen-claim-KTC}.


\subsection{Example of torus satellites of the trefoil. Dependence on a diagram}\label{sec:diag-dep-tor}
We illustrate our consideration of the Khovanov polynomials for torus satellite knots with the simplest example of the trefoil given by the left diagram in Fig.~\ref{fig:3}\,:
\begin{equation}\label{KhTor31}
	\begin{aligned}
	{\color{red}\Kh ^{{\cal S}_{tor_\n}^{3_1^{(2)}}}}\Big|_{\n\geq 0}
	& =\overbrace{{\color{red} (-\ST)}\cdot \frac{q}{1-q^2 \ST}}^{{\color{red} \mu_{[1,1]}}}\cdot\overbrace{(q^3\ST)^{\n-6}}^{\lambda_{[1,1]}^{\n+2w_{3_1}^{(2)}}}
	& +\overbrace{ \frac{q^4 \ST^2-q^2 \ST+1}{q \ST \left(q^2 \ST-1\right)}}^{\mu_{\rm adj}}\cdot\overbrace{q^{\n-6}}^{\lambda_{[2]}^{\n+2w_{3_1}^{(2)}}}\cdot\,{\color{red}\Kh _{\rm adj}^{3_1}}\,,
	\\
	{\color{red}\Kh ^{{\cal S}_{tor_\n}^{3_1^{(2)}}}}\Big|_{\n<0}
	&=\overbrace{\frac{q}{1-q^2 \ST}}^{{\color{red} \mu_{[1,1]}}}\cdot\overbrace{(q^3\ST)^{\n-6}}^{\lambda_{[1,1]}^{\n+2w_{3_1}^{(2)}}}
	& +\overbrace{ \frac{q^4 \ST^2-q^2 \ST+1}{q \ST \left(q^2 \ST-1\right)}}^{\mu_{\rm adj}}\cdot\overbrace{q^{\n-6}}^{\lambda_{[2]}^{\n+2w_{3_1}^{(2)}}}\color{red}\cdot\,\Kh _{\rm adj}^{3_1}
	\end{aligned}
	\end{equation}
with the same $\color{red}\Kh _{\rm adj}^{3_1}$ as in Subsection~\ref{sec:diag-dep}. In the case of torus satellites, the Khovanov polynomial also has two branches with the boundary at $\n=0$. A dependence on a particular diagram of the knot $3_1$ appears through $2w_{\tilde{3}_1}$ shifts of evolution parameters, compare with the HOMFLY case~\eqref{HSatPar}. We illustrate this phenomenon by consideration of three-strand realisation of the trefoil (right picture in Fig.~\ref{fig:3}):
\begin{equation}
	\begin{aligned}
	{\color{red}\Kh ^{{\cal S}_{tor_\n}^{3_1^{(3)}}}}\Big|_{\n\geq 2}
	& =\overbrace{{\color{red} (-\ST)}\cdot \frac{q}{1-q^2 \ST}}^{{\color{red} \mu_{[1,1]}}}\cdot\overbrace{(q^3\ST)^{\n-8}}^{\lambda_{[1,1]}^{\n+2w_{3_1}^{(3)}}}
	& +\overbrace{ \frac{q^4 \ST^2-q^2 \ST+1}{q \ST \left(q^2 \ST-1\right)}}^{\mu_{\rm adj}}\cdot\overbrace{q^{\n-8}}^{\lambda_{[2]}^{\n+2w_{3_1}^{(3)}}}\cdot \,{\color{red}\Kh _{\rm adj}^{3_1}}\,,
	\\
	{\color{red}\Kh ^{{\cal S}_{tor_\n}^{3_1^{(3)}}}}\Big|_{\n<2}
	&=\overbrace{\frac{q}{1-q^2 \ST}}^{{\color{red}\mu_{[1,1]}}}\cdot\overbrace{(q^3\ST)^{\n-8}}^{\lambda_{[1,1]}^{\n+2w_{3_1}^{(3)}}}
	& +\overbrace{ \frac{q^4 \ST^2-q^2 \ST+1}{q \ST \left(q^2 \ST-1\right)}}^{\mu_{\rm adj}}\cdot\overbrace{q^{\n-8}}^{\lambda_{[2]}^{\n+2w_{3_1}^{(3)}}}\cdot\,{\color{red}\Kh _{\rm adj}^{3_1}}\,.
	\end{aligned}
	\end{equation}
Dependence on a certain knot diagram disappears if one considers the quantity with shifted by $2w_{\tilde{3}_1}$ evolution parameter: ${\color{red}\Kh ^{{\cal S}^{\tilde{3}_1}_{tor_{\n-2w_{\tilde{3}_1}}}}}$. The HOMFLY and Jones polynomials possess the same diagram dependence, and we see that this property conserves after $\ST$-deformation. These examples are also in accordance with the general claim~\eqref{GenKhTwSat} for ${\rm Sh}_{3_1}=3$.


\setcounter{equation}{0}
\section{Taming the jumps in satellite Khovanov polynomials}\label{sec:main}
In this section, our main concern is the problem of smoothness violation (jumps) in $\n$-evolution. We explain that it is fully contained in the first term at the r.h.s. of (\ref{maindeco}).
In the simplest example of satellites of the knot $3_1$ we demonstrate the (actually, minor) difference between (\ref{maindeco}) for twist and torus satellites -- and this again models the situation for arbitrary ${\cal K}$. For further examples demonstrating that result~\eqref{GenMainClaim} is true far beyond
$3_1$, see Appendix~\ref{App}.

We have already discussed knot diagram dependence in Section~\ref{sec:KhRed}. Thus, in this section we consider shifted Khovanov polynomials ${\color{red} \Kh^{{\cal S}^{\tilde{\cal K}}_{T_{\n-2w_{\tilde{\cal K}}}}}}$, where we remind that $\tilde{\cal K}$ is a knot diagram of a knot $\cal K$.

\subsection{Main claim}
It turns out that for an {\it arbitrary} companion knot ${\cal K}$ the adjoint Khovanov polynomials on two braches extracted from~\eqref{GenKhTwSat} differ by the only summand of ${\color{red}(-\ST)^{\theta(\n+2\,{\rm Sh}_{\cal K})}}\cdot \lambda_{\rm adj}^{-2\,{\rm Sh}_{\cal K}}$ with $\lambda_{\rm adj}=q^2\ST$. In other words, the quantity
\begin{equation}\label{GenKhAdjWithoutJumps}
\mathfrak{Kh}_{{\rm adj}}^{\cal K}\ := \
{\color{red}\Kh_{\rm adj}^{\cal K}}\Big|_{\n\geq\,-2\,{\rm Sh}_{\cal K}}-{\color{red}     (-\ST)}\cdot\lambda^{-2\,{\rm Sh}_{\cal K}}_{\text{adj}}
   \ =\ {\color{red}\Kh _{\rm adj}^{\cal K}}\Big|_{\n<\, -2\,{\rm Sh}_{\cal K}}-
    \lambda^{-2\,{\rm Sh}_{\cal K}}_{\text{adj}}
\end{equation}
is free of jumps. Moreover, if we add and subtract ${\color{red}(-\ST)^{\theta(\n+2\,{\rm Sh}_{\cal K})}}\cdot \lambda_{\rm adj}^{-2\,{\rm Sh}_{\cal K}}$ from the adjoint Khovanov polynomials in~\eqref{GenKhAdjWithoutJumps}, we arrive to a fascinating expression which is the {\bf main result} of our research:
\begin{equation}\label{GenMainClaim}
\begin{aligned}
    {\color{red} \Kh^{{\cal S}_{tw_\n}({\cal K})}}&:=
{\color{red} \Kh ^{{\cal S}_{tw_{\n-2w_{\tilde{\cal K}}}}^{\tilde{\cal K}}}}
=&{\color{red} \Kh ^{tw_{\n+2\,{\rm Sh}_{\cal K}}}}
+ \tau_{\rm adj}\cdot \lambda_{\rm adj}^{-\n}\cdot  {\mathfrak{Kh}_{{\rm adj}}^{\cal K}}\,, \\
{\color{red} \Kh^{{\cal S}_{tor_\n}({\cal K})}}&:=
(q^3\ST)^{-\n}\cdot {\color{red}\Kh^{{\cal S}_{tor_{\n-2w_{\tilde{\cal K}}}}^{\tilde{\cal K}}}}
=&{\color{red} \Kh^{tor_{\n+2\,{\rm Sh}_{\cal K}}}}
+ \mu_{\rm adj}\cdot \lambda_{\rm adj}^{-\n}\cdot {\mathfrak{Kh}_{{\rm adj}}^{\cal K}}\,,
\end{aligned}
\end{equation}
which we illustrate in subsections below by an example of satellites of the trefoil knot. These two formulas can be unified by introduction of $\mathfrak{c}_T$ with $T$ referring to torus and twist satellites -- $\mathfrak{c}_{tor}=\mu_{\rm adj}$, $\mathfrak{c}_{tw}=\tau_{\rm adj}$. Using this notation we straightforwardly arrive to the already stated main claim~\eqref{maindeco}. Note that formula~\eqref{GenMainClaim} is a direct $\ST$-deformation of corresponding expression for the Jones polynomial~\eqref{JonesUniSat}. Unlike the case of Jones polynomials, the shift ${\rm Sh}$ depends on a knot $\cal K$. This shift depends on the companion knot ${\cal K}$, but {\it not} on the diagram $\tilde{\cal K}$. First of all, this dependence is needed to provide quantities without jumps. Second, the jump of the fundamental Khovanov polynomial for torus and twist satellite at the special point $\n+2\,{\rm Sh}_{\cal K}=0$ demands one of the allocated quantities possess a jump at the same point. This fact provides a hint that ${\color{red} \Kh^{T_{\n+2\,{\rm Sh}_{\cal K}}}}$ are the proper objects to be extracted from the satellite polynomials.

Another meaningful feature is that in split~\eqref{GenMainClaim} both two terms in the r.h.s. are positive polynomials. This fact also means that ${\mathfrak{Kh}_{{\rm adj}}^{\cal K}}$ provides a good substitute to the adjoint Khovanov polynomial ${\color{red} \Kh^{\cal K}_{\rm adj}}$, which is actually not positive and unnaturally depends on braids inserted into satellites, but not only on the knot $\cal K$.

\bigskip

{\footnotesize
{\bf Additional comment.}
 Note that formulas~\eqref{GenMainClaim} differ by the only factor $\mathfrak{c}_T$ (see~\eqref{maindeco}):
\begin{equation}\label{tau-mu}
    \tau_{\rm adj}=q^3 \ST (1+q^2\ST)\cdot \mu_{\rm adj}\lambda_{\rm adj}^{-1}\,,
\end{equation}
what can be used to make the difference between the Khovanov polynomials of torus and twist satellite knots by the only additive term. With the use of relation~\eqref{tau-mu} we obtain:
\begin{equation}
    {\color{red} \Kh^{{\cal S}_{tw_{2\n}}({\cal K})}}=q^3 \ST(1+q^2\ST)\cdot{\color{red} \Kh^{{\cal S}_{tor_{2\n+1}}({\cal K})}}+{\color{red} \Kh^{tw_{2\n+2\,{\rm Sh}_{\cal K}}}}-q^3\ST(1+q^2\ST)\cdot {\color{red}\Kh^{tor_{2\n+1+2\,{\rm Sh}_{\cal K}}}}\,.
\end{equation}
Here we use our choice of satellite knots with even number of crossings in twist satellite knots and odd number of crossings in torus satellite knots, see Fig.~\ref{fig:1}. This formula completely simplifies after the substitution of the Khovanov polynomials for torus and twist knots:
\begin{equation}
    {\color{red} \Kh^{{\cal S}_{tw_{2\n}}({\cal K})}}=q^3 \ST(1+q^2\ST)\cdot{\color{red} \Kh^{{\cal S}_{tor_{2\n+1}}({\cal K})}}+{\color{red} (-\ST)^{\theta(\n+{\rm Sh}_{\cal K})}}\cdot q^2\,.
\end{equation}
}

\subsection{Example of twist satellite of the trefoil}
Terms denoted in red in formulas~\eqref{KhTw31},~\eqref{KhTw31Str3} are quantities which are not fully analytic in evolution parameter
and exhibit jumps at certain value of $\n=-6$ for the diagram independent quantity ${\color{red} \Kh^{{\cal S}^{\tilde{3}_1}_{tw_{\n-2w_{\tilde{3}_1}}}}}$, where $\tilde{3}_1$ denotes a knot diagram of the trefoil. The adjoint Khovanov polynomials ${\color{red} \Kh_{\rm adj}^{3_1}}$ on different branches differ by just a single term,
i.e. can be expressed via a single quantity, which does not jump at $\n=-6\,$:
\begin{equation}\label{KhAdjWithoutJumps}
\mathfrak{Kh}_{{\rm adj}}^{3_1}\ := \
{\color{red}\Kh_{\rm adj}^{3_1}}\Big|_{\n\geq -6}-{\color{red}     (-\ST)}\cdot\lambda^{-6}_{\text{adj}}
   \ =\ {\color{red}\Kh _{\rm adj}^{3_1}}\Big|_{\n<-6}-
    \lambda^{-6}_{\text{adj}}\,,
\end{equation}
where $\lambda_{\rm adj} = q^2\ST$ and  $(q^2 \ST)^{-11}\cdot{\color{red} q^{10} \ST^5} = q^{-12} \ST^{-6} = \lambda_{\rm adj}^{-6}$.
This corresponds to the fact that in the case of Khovanov polynomials ${\rm Sh}_{3_1}=3$, see~\eqref{GenKhAdjWithoutJumps}.
We can add and subtract $\lambda^{-6}_{\text{adj}}$ from ${\color{red} \Kh _{\rm adj}^{3_1}}$, and get:
\be
\label{Cont31}
   {\color{red} \Kh ^{{\cal S}_{tw_{\n-2w_{\tilde{3}_1}}}^{\tilde{3}_1}}}\
    -\ \overbrace{{\color{red} (-\ST )^{\theta(\n+6)}}\cdot q^2\left\{\frac{1+q^4
    \ST ^2}{1-q^2\ST }-q^{-2}\ST ^{-1}\frac{1+q^6 \ST ^3}{1-q^2\ST }\lambda_{\rm adj}^{-(\n+6)}
    \right\}}^{{\color{red} \Kh ^{tw_{\n+6}}}}
    =\nn \\
    = \overbrace{(q^2\ST )^{-\n}}^{\lambda_{\rm adj}^{-\n}}\cdot\overbrace{\ST^{-1}\frac{(1+q^6 \ST ^3)
     }{q^2\ST-1}}^{{\tau}_{\rm adj}}
    \cdot \overbrace{(q^2\ST-1)
    (q^2\ST)^{-11} (1+q^2\ST)
     (q^{14}\ST ^9+q^{10} \ST ^7+q^8 \ST ^5+q^2 \ST +1)}^{\mathfrak{Kh}_{{\rm adj}}^{3_1}}\,.
\ee
In this formula, there are jumps at the l.h.s. at $\n=-6\,$:
a non-multiplicative jump in ${\color{red} \Kh^{{\cal S}_{tw_{\n-2w_{\tilde{3}_1}}}^{\tilde{3}_1}}}$
and a multiplicative jump in ${\color{red} \Kh^{tw_{\n+6}}}$.
They compensate each other so that the most interesting remaining piece of ${\color{red} \Kh^{{\cal S}_{tw_{\n-2w_{\tilde{3}_1}}}^{\tilde{3}_1}}}$
at the r.h.s. of~\eqref{Cont31} is free of jumps. In particular, $\mathfrak{Kh}_{{\rm adj}}^{3_1}$ does not have jumps.
Note also that $(q^2\ST-1)$ terms cancel, what provides an explicitly positive polynomial at the r.h.s. --
like the nonanalytic (jumping) ${\color{red} \Kh^{{tw_{\n+6}}}}$ ($tw_{-2}$ corresponds to knot $3_1$). The structure of formula~\eqref{Cont31} is in correspondence with our general claim~\eqref{GenMainClaim}.


\bigskip

{\footnotesize
{\bf Additional comment.}
We see that while the colored ${\color{red} \Kh_{\rm adj} ^{3_1}}$
is not analytic in evolution parameter,
the non-analyticity is actually "small" --
concentrated in a single term of a complicated polynomial
(see further examples in Appendix~\ref{App} for illustration of the relative complexity).
A possible interpretation of our result here is that the minor modification
produces an analytical (jump-free) quantity ${\mathfrak{Kh}_{{\rm adj}}^{3_1}}$,
which can serve as a nice/practically relevant substitute of the colored Khovanov polynomial \cite{anokhina2021khovanov}.
From this perspective it is a natural question to ask for generalisations to higher representations
$[r]$ instead of ${\rm adj}=[2]$ -- for this purpose one should study $r$-cablings,
what remains for the future work beyond the scope of the present paper.
}

\subsection{Example of torus satellite of the trefoil}
We tame the jumps by introduction of the same $\mathfrak{Kh}_{\rm adj}^{3_1}$~\eqref{KhAdjWithoutJumps}. In the case of torus satellites, we also add and subtract $\lambda_{\rm adj}^{-6}$ from the adjoint Khovanov polynomial to eliminate the ``infinitesimal'' difference on different branches:
\be
{\color{red} \Kh ^{{S_{tor_{\n-2w_{\tilde{3}_1}}}^{\tilde{3}_1}}}}\
-\ (q^3\ST)^{\n}\cdot\overbrace{{\color{red} (-\ST )^{\theta(\n+6)}}\left\{\frac{q}{1-q^2 \ST}+\frac{q^4 \ST^2-q^2 \ST+1}{q t \left(q^2 \ST-1\right)}(q^2\ST)^{-\n}\cdot(q^2\ST)^{-6}
	\right\}}^{{\color{red} \Kh ^{tor_{\n+6}}}}
=\nn \\
= \overbrace{q^{\n}}^{\lambda_{[2]}^{\n}}\cdot \overbrace{
	\frac{(q^4 \ST^2-q^2 \ST+1)}{q \ST(q^2\ST-1)}
}^{{\mu}_{\rm adj}}\cdot
\overbrace{(q^2\ST-1)
	(q^2\ST)^{-11} (1+q^2\ST)
	(q^{14}\ST ^9+q^{10} \ST ^7+q^8 \ST ^5+q^2 \ST +1)}^{\mathfrak{Kh}_{{\rm adj}}^{3_1}}\,.
\ee
Multiplying this equation by $(q^3\ST)^{-\n}$ corresponding to the $\ST$-deformed framing factor (see equation~\eqref{JonesTorSat}), we arrive to a particular case of formula~\eqref{GenMainClaim}. Here we also eliminate the diagram dependence by the shift of evolution parameter $\n$.



\section{The actual checks}\label{sec:checks}

In the next two sections
and in Appendix~\ref{App} we provide additional comments on different aspects of tangle calculus
and further examples illustrating the validity of equation (\ref{maindeco}).
These examples include:\\ \\
$\bullet$ torus knots $tor_k\,$: \ ${\rm Sh}_{tor_{k>0}}=\frac{3}{2}(k-1)$ and ${\rm Sh}_{tor_{k<0}}=\frac{3}{2}(k+1)\,$;\\ \\
$\bullet$ twist knots $tw_k\,$: \ ${\rm Sh}_{tw_{k>0}}=0$ and ${\rm Sh}_{tw_{k<0}}=3$; \\ \\
$\bullet$ knots from the beginning of the Rolfsen table:
\begin{table}[h!]
	\centering
	\begin{tabular}{|c|c|c|c|c|c|c|c|c|c|c|c|c|c|c|c||c|c|c|c|}
		\hline
		${\cal K}$ & $3_1$ & $4_1$ & $5_1$ & $5_2$ & $6_1$ & $6_2$ & $6_3$
		& $7_1$ & $7_2$ & $7_3$ & $7_4$ & $7_5$ & $7_6$ & $7_7$ & $8_1$ &   ${\color{blue} \mathbf{8_{19}}}$ & $8_{20}$ & $8_{21}$ & $9_1$ 
		\\
		& {\scriptsize\!\!\!$tor_3=tw_{-2}$\!\!\!}&{\footnotesize\!\!\!$tw_{2}$\!\!\!}& {\footnotesize\!\!$tor_5$\!\!} & {\footnotesize\!\!$tw_{-4}$\!\!} &  {\small\!\!$tw_{4}$\!\!} & &
		&{\footnotesize $tor_7$ }&{\footnotesize\!\!$tw_{-6}$\!\!} & & & & & & {\footnotesize\!\!$tw_{6}$\!\!}& {\footnotesize\!\! $T[3,4]$\!\!} & & & {\footnotesize\!\!$tor_9$\!\! }
		\\
		\hline
		${\rm Sh}_{\cal K}$ & 3 & 0 & 6 & 3 & 0 & 3 & 0 & 9 & 3 & $6$ & $3$ & 6 & 3 & 0 & 0 & ${\color{blue}\mathbf{8}}$ & 0 & 3 & 12  
		\\
		\hline 
		${\rm RSI}_{\cal K}$ & 2 & 0 & 4 & 2 & 0 & 2 & 0 & 6 & 2 & $4$ & $2$ & 4 & 2 & 0 & 0 & $6$ & 0 & 2 & 8
		\\
		\hline
	\end{tabular}
\end{table}
\begin{table}[h!]
	\centering
	\begin{tabular}{|c|c||c|c||c||c||c||c||c||c||c||c|c||c|c|c|c|}
		\hline
		${\cal K}$ & $9_2$ & ${\color{blue}\mathbf{9_{42}}}$ & $9_{43}$ & $10_1$ & ${\color{blue}\mathbf{10_{124}}}$ & ${\color{blue}\mathbf{10_{128}}}$ & ${\color{blue}\mathbf{10_{132}}}$ & ${\color{blue}\mathbf{10_{136}}}$ & ${\color{blue}\mathbf{10_{139}}}$ & ${\color{blue}\mathbf{10_{145}}}$ & ${\color{blue}\mathbf{10_{152}}}$ & ${\color{blue}\mathbf{10_{153}}}$ & ${\color{blue} \mathbf{14n_{21881}}}$ 
		\\
		&{\footnotesize\!\!$tw_{-8}$\!\!} & & & & {\footnotesize $tw_8$} & {\footnotesize $T[3,5]$} & & & & & & & {\footnotesize $T[3,7]$} 
		\\
		\hline
		${\rm Sh}_{\cal K}$ & 3 & 0 & $6$ & 0 & ${\color{blue}\mathbf{11}}$ & ${\color{blue}\mathbf{8}}$ & 3 & 0 & ${\color{blue}\mathbf{11}}$ & 6 & ${\color{blue}\mathbf{11}}$ & 0 & {\color{blue} \bf 16} 
		\\
		\hline 
		${\rm RSI}_{\cal K}$ & 2 & 0 & $4$ & 0 & $8$ & $6$ & 2 & 0 & $8$ & 4 & $8$ & 0 & 12 
		\\
		\hline
	\end{tabular}
	\caption{Knots and their Rasmussen $s$-invariants ${\rm RSI}_{\cal K}$ and invariants ${\rm Sh}_{\cal K}$. Thick knots and deviations from the empirical rule ${\rm Sh}_{\cal K}=\frac{3}{2}{\rm RSI}_{\cal K}$ are colored in blue. Torus knots are denoted as $T[l,m]$. For short, $T[2,k]$ are denoted as $tor_k$ throughout the text.}
	\label{tab:knots-list}
\end{table}



\section{On the knot invariant ${\rm Sh}_{\cal K}$}\label{sec:shift}
In this section, we provide the description of the quantity ${\rm Sh}_{\cal K}$ and its properties. We have separated this information from the explanation of the main result in Sections~\ref{sec:KhRed} and~\ref{sec:main} in order to make the statements clearer.

\subsection{Definition of ${\rm Sh}_{\cal K}$}
Reduced HOMFLY polynomials  possess  differential expansion \cite{itoyama2012homfly,morozov2016factorization,kononov2016rectangular,morozov2018factorization,kameyama2020cyclotomic,morozov2019extension,morozov2020kntz,bishler2020perspectives,morozov2022differential}
which for the fundamental representation states that
\be
H_\Box^{\cal K} -1 \ \vdots \ \{Aq\}\{A/q\}\ \stackrel{ N=2}{ \longrightarrow} \{q^3\}\{q\}\,,
\ee
where $A:=q^N$ and $\{x\}:=x-x^{-1}$.
It can be generalised to superpolynomials with ${\bf t}\neq -1$ \cite{dunin2013superpolynomials},
but breaks after reduction to Khovanov polynomials.

Still, they satisfy  a similarly-looking (though very different) relation,
which can help to understand the origins and the meaning of our parameter ${\rm Sh}_{\cal K}$. We start with the description of Rasmussen $s$-invariant ${\rm RSI}_{\cal K}$ \cite{rasmussen2010khovanov,lewark2022rasmussen}. Rasmussen invariant can be defined for all explicitly known cases from the {\it unreduced} Khovanov polynomial~\cite{rasmussen2010khovanov}:
\begin{equation}\label{RSIKh}
	Kh_\Box^{\cal K} = q^{-{\rm RSI}_{\cal K}}(q+q^{-1})+(1+q^4\ST)\cdot Kh'^{\cal K}\,.
\end{equation}
For first several decades of knots in the Rolfsen table
a similar  relation holds for the {\it reduced} Khovanov polynomials as well:
\begin{equation}\label{RedRSIdef}
	\Kh_\Box^{\cal K}\stackrel{?}{=}q^{-{\rm RSI}_{\cal K}}+q\,(1+q^2 \ST)\cdot Kh'^{\cal K}\,,
\end{equation}
Here we denote the remaining part of the Khovanov polynomial $\Kh^{\cal K}$ as $Kh'^{\cal K}$, which is a positive integer polynomial.
Note that it is the same in \eqref{RedRSIdef} as in \eqref{RSIKh}.
The question mark in~\eqref{RedRSIdef} emphasises that this relation is not always true.
For {\it some} of the {\it thick} knots \eqref{RedRSIdef} breaks.
In  Table~\ref{tab:knots-list} we mark thick knots by blue,
and for those which violate \eqref{RedRSIdef} blue is also the entry in the line ${\rm Sh}_{\cal K}$.
Reduced Khovanov polynomials in this case satisfy
\begin{equation} \label{RedShdef}
    \Kh_\Box^{\cal K}=q^{-2\,({\rm Sh}_{\cal K}-{\rm RSI}_{\cal K})}+q\,(1+q^2 \ST)\cdot Kh''^{\cal K}
\end{equation}
which can be treated as a {\it definition} of the knot invariant ${\rm Sh}_{\cal K}$, as this expansion seems to hold for {\it all} knots.
One of our claims in this paper is that {\bf it is exactly the same quantity that enters (\ref{maindeco})}.

Two differently-looking formulas (\ref{RedRSIdef}) and (\ref{RedShdef}) can be made
more similar by introduction of some measure of knot thickness\footnote{We emphasise that our ``Thickness'' is not homological thickness~\cite{rasmussen2010khovanov}.}:
\be\label{Sh}
{\rm Sh}_{\cal K} + \text{Thickness}_{\cal K} = \frac{3}{2}{\rm RSI}_{\cal K}\,.
\ee
All the above examples of knots with ${\rm Sh}_{\cal K}$ marked in blue provide $\text{Thinkness}_{\cal K}=1$, except the knot $T[3,7]$ with $\text{Thinkness}_{T[3,7]}=2$. All other knots (and even some thick knots) from Table~\ref{tab:knots-list} have $\text{Thinkness}_{\cal K}=0$. It remains for a future research to find knots of greater ``Thickness'' (if any) exist.
In particular, it seems to stay unity or zero even when Turaev genus \cite{dasbach2008jones} grows from 1 to 2. We have checked this property for all such knots with up to 12 crossings.

If one uses the above defined ``Thickness'' of a knot $\cal K$, the deviation of formula~\eqref{RedShdef} from~\eqref{RedRSIdef} becomes clearer:
\begin{equation} \label{Khred1}
    \Kh_\Box^{\cal K}=q^{-{\rm RSI}_{\cal K}+2\,{\rm Thickness}_{\cal K}}+q\,(1+q^2 \ST)\cdot Kh''^{\cal K}\,.
\end{equation}
Note that for ${\rm Thickness}_{\cal K}\neq 0$ there is a stronger difference between (\ref{Khred1}) and (\ref{RSIKh})
for reduced and unreduced polynomials: $Kh''^{\cal K}\neq Kh'^{\cal K}$. Moreover, $Kh''^{\cal K}$ is non-positive, although still integer polynomial.

\subsection{Examples}
In this section we provide three examples to demonstrate the validity of equations~\eqref{RSIKh}–\eqref{Khred1}. First, consider the simplest example of the trefoil. For the unreduced Khovanov polynomial we get:
\begin{equation}
Kh^{3_1}=q^{-2}(q+q^{-1})+(1+q^4 \ST)\cdot q^{-9} \ST^{-3}\,,
\end{equation}
and we have ${\rm RSI}_{3_1}=2$, what coincides with the value in Table~\ref{tab:knots-list}. For the reduced Khovanov polynomial,~\eqref{RedRSIdef} holds:
\begin{equation}
\Kh^{3_1}=q^{-2}+q\,(1+q^2\ST)\cdot q^{-9} \ST^{-3}\,,
\end{equation}
and~\eqref{RedShdef}–\eqref{Khred1} with ${\rm Sh}_{3_1}=3$ and ${\rm Thickness}_{3_1}=0$ are also valid. 

Second, consider the simplest {\color{blue} \bf thick} knot ${\color{blue} \bf 8_{19}}$. One can extract RSI from the unreduced Khovanov polynomial:
\begin{equation}
Kh^{8_{19}}=q^{-6}(q+q^{-1})+\left(1+q^4 \ST\right)\cdot q^{-17} \ST^{-5}\left(q^4 \ST^2+q^2+1\right)\,.
\end{equation}
However, for the reduced Khovanov polynomial~\eqref{RedRSIdef} breaks:
\begin{equation}
\Kh^{8_{19}}=q^{-6}+q^{-16} \ST^{-5}(q^6 \ST^3+q^4 \ST^2+q^4 \ST+1)\,,
\end{equation}
but~\eqref{RedShdef}–\eqref{Khred1} still hold:
\begin{equation}
\Kh^{8_{19}}=q^{-2\,({\color{blue} \bf 8}-6)}+\left(1+q^2 \ST\right)\cdot q^{-16} \ST^{-5}\left(q^4 \ST^2+1\right) \left(-q^6 \ST^2+q^4 \ST^2+q^4 \ST-q^2 \ST+1\right)
\end{equation}
with ${\rm Sh}_{8_{19}}={\color{blue} \bf 8}$ and ${\rm Thickness}_{8_{19}}=1$. We have also found a unique knot of ${\rm Thickness}=2$ -- the knot ${\color{blue} \mathbf{14n_{21881}}}=T[3,7]$. In this case, the unreduced Khovanov polynomial
\begin{equation}
    Kh^{14n_{21881}}=q^{-12}(q+q^{-1})+q^{-29} \ST^{-9} \left(q^4 \ST+1\right) \left(q^4 \ST^2+q^2+1\right) \left(q^6 \ST^4+1\right)
\end{equation}
with ${\rm RSI}_{14n_{21881}}=12$. And for the reduced Khovanov polynomial~\eqref{RedRSIdef} breaks, but~\eqref{RedShdef}–\eqref{Khred1} hold with ${\rm Thickness}_{14n_{21881}}=2$ and ${\rm Sh}_{14n_{21881}}={\color{blue} \mathbf{16}}$:
{\small \begin{equation}
    \Kh^{14n_{21881}}=q^{-2\,({\color{blue} \mathbf{16}}-12)}q^{-28} \ST^{-9}\left(q^2 \ST+1\right) \left(q^4 \ST^2+1\right) \left(-q^{14} \ST^6+q^{12} \ST^5+q^{10} \ST^6-q^8 \ST^5+q^6 \ST^4-q^6 \ST^2+q^4 \ST^2+q^4 \ST-q^2 \ST+1\right)\,.
\end{equation}}

\subsection{Properties of ${\rm Sh}_{\cal K}$}
Let us discuss some important properties of the parameter ${\rm Sh}_{\cal K}$. First of all, as we have already mentioned, ${\rm Sh}_{\cal K}$ is a {\it knot invariant}. Throughout the text we choose knots $\cal K$ to have ${\rm Sh}_{\cal K}\geq 0$ and ${\rm RSI}_{\cal K}\geq 0$ and their mirror images $\overline{\cal K}$ have ${\rm Sh}_{\overline{\cal K}}=-{\rm Sh}_{\cal K}$ and ${\rm RSI}_{\overline{\cal K}}=-{\rm RSI}_{\cal K}$. This fact also implies that for amphichiral knots, for which ${\cal K}=\overline{\cal K}$, ${\rm Sh}_{\cal K}=0$ and ${\rm RSI}_{\cal K}=0$ what can be seen by the examples in Table~\ref{tab:knots-list}.

The last statement is also in correspondence with the fact that the uncolored Khovanov polynomials inherit the duality between a knot and its mirror image after the change of variables $\ST\rightarrow \ST^{-1}$, $q\rightarrow q^{-1}$, see equations~\eqref{RSIKh}–\eqref{RedShdef}. The sign change of ${\rm Sh}_{\cal K}$ under inversion of a knot also serves the Khovanov polynomials to respect the duality:
\begin{equation}
    {\cal S}_{T_\n}({\cal K})\, \longleftrightarrow \, {\cal S}_{T_{-\n}}(\ \overline{{\cal K}} \ )\,,
\end{equation}
see Fig.~\ref{fig:1}. We recall that the subscript $T$ refers to both torus and twist satellites, $T=tor$ or $T=tw$. One can check that
\begin{equation}
    \Kh^{{\cal S}_{T_\n}(\cal K)}(q,\ST)=\Kh^{{\cal S}_{T_{-\n}}( \overline{{\cal K}}  )}(q^{-1},\ST^{-1})\,.
\end{equation}
Moreover, this property is also respected by the adjoint Khovanov polynomials:
\begin{equation}\label{MirrKhAdj}
    \Kh_{\rm adj}^{\cal K}(q,\ST)\Big|_{n}=\Kh_{\rm adj}^{\overline{\cal K}}(q^{-1},\ST^{-1})\Big|_{-n}\quad {\rm and} \quad \mathfrak{Kh}_{\rm adj}^{\cal K}(q,\ST)=(-\ST)\cdot \mathfrak{Kh}_{\rm adj}^{\overline{\cal K}}(q^{-1},\ST^{-1})\,,
\end{equation}
see Appendix~\ref{App} for examples.

\setcounter{equation}{0}
\section{Internal structure of adjoint Khovanov polynomials}\label{sec:double-evo}
In this section, we use some material from Appendix~\ref{App} to discuss the next important issue --
what happens to the structure of~\eqref{maindeco} for a double evolution.
We consider just the simplest cases, when the companion knots ${\cal K}$ are themselves twist and torus knots, so there is a new evolution parameter $k$
in addition to our old $\n$.

\subsection{Twist knots ${\cal K}=tw_k$}
In this section, we discuss the internal structure of adjoint Khovanov polynomials for two-strand twist knots $tw_k$ with even $k$. In this case, two strands are antiparallel and carry representation $[2]={\rm adj}$.
Thus, we expect that $\mathfrak{Kh}_{\text{adj}}^{tw_k}$ (and thus, $\Kh_{\rm adj}^{tw_{k}}$) have spectral expansion over three
eigenvalues corresponding to $[2]\otimes \overline{[2]}=\varnothing\oplus \text{adj}\oplus [4]\,$. This property does hold:
\begin{equation}\label{KhAdjSpectr}
    \mathfrak{Kh}_{\text{adj}}^{tw_k}=\lambda_{\varnothing}^{k}\cdot \tau^{\rm adj}_{\varnothing}+\lambda_{\text{adj}}^{k}\cdot \tau^{\rm adj}_{\text{adj}}+\lambda_{[4]}^{k}\cdot \tau^{\rm adj}_{[4]}
\end{equation}
with eigenvalues
\begin{equation}\label{EigenVal}
    \lambda_{\varnothing}=1,\quad \lambda_{\text{adj}}=q^2 \ST,\quad \lambda_{[4]}=q^6 \ST^4\,,
\end{equation}
and $\tau$-elements for $k<0\,$ ($k=-2$ corresponds to $3_1$):
{\footnotesize \begin{equation}\label{tau-neg}
\begin{aligned}
    \tau^{\rm adj}_{\varnothing}&=\frac{\left(q^4 \ST^3+1\right) \left(q^8 \ST^5+q^6 \ST^4+q^4 \ST^3+1\right)}{q^{10} \ST^4 \left(q^3 \ST^2-1\right) \left(q^3 \ST^2+1\right)}\,,\\
    \tau_{\rm adj}^{\text{adj}}&=-\frac{q^{18} \ST^{12}+q^{16} \ST^{11}+q^{16} \ST^{10}+q^{14} \ST^9+q^{12} \ST^9+3 q^{12} \ST^8+q^{12} \ST^7+2 q^{10} \ST^7+q^{10} \ST^6-q^8 \ST^4+q^6 \ST^4+q^4 \ST^3+q^2 \ST^2+q^2 \ST+1}{q^{12} \ST^5 \left(q^4 \ST^3-1\right) \left(q^4 \ST^3+1\right)}\,, \\
    \tau^{\rm adj}_{[4]}&=\frac{\left(q^2 \ST-1\right) \left(q^2 \ST+1\right)^2 \left(q^8 \ST^5+1\right) \left(q^{10} \ST^7+1\right)}{q^{10} \ST^3 \left(q^3 \ST^2-1\right) \left(q^3 \ST^2+1\right) \left(q^4 \ST^3-1\right) \left(q^4 \ST^3+1\right)}\,,
\end{aligned}
\end{equation}}
while for $k>0$ ($k=2$ corresponds to $4_1$):
{\footnotesize \begin{equation}\label{tau-pos}
\begin{aligned}
    \tau^{\rm adj}_{\varnothing}&=-\frac{q^{18} \ST^{11}+q^{14} \ST^{10}-q^{12} \ST^8-q^{12} \ST^7-q^{10} \ST^7-q^{10} \ST^6-q^8 \ST^6-2 q^8 \ST^5-q^6 \ST^4-q^2 \ST-1}{q^{12} \ST^7 \left(q^3 \ST^2-1\right) \left(q^3 \ST^2+1\right)}\,, \\
    \tau_{\rm adj}^{\text{adj}}&=-\frac{q^{18} \ST^{12}-q^{16} \ST^{12}-q^{16} \ST^{11}+q^{14} \ST^9+q^{12} \ST^9+3 q^{12} \ST^8+q^{12} \ST^7+2 q^{10} \ST^7+q^{10} \ST^6+q^8 \ST^6+2 q^8 \ST^5+q^6 \ST^4+q^4 \ST^3+q^2 \ST^2+q^2 \ST+1}{q^{12} \ST^7 \left(q^4 \ST^3-1\right) \left(q^4 \ST^3+1\right)}\,,\\
    \tau^{\rm adj}_{[4]}&=\frac{\left(q^2 \ST-1\right) \left(q^2 \ST+1\right)^2 \left(q^8 \ST^5+1\right) \left(q^{10} \ST^7+1\right)}{q^{10} \ST^5 \left(q^3 \ST^2-1\right) \left(q^3 \ST^2+1\right) \left(q^4 \ST^3-1\right) \left(q^4 \ST^3+1\right)}\,.
\end{aligned}
\end{equation}}
Here we use the same symbol $\tau$ as in formulas of previous sections (for example in equation~\eqref{GenKhTwSat}), because in fact it corresponds to the same tangle calculus element but carrying other representations instead of $\Box\otimes\overline{\Box}=\varnothing\oplus {\rm adj}$. For $\tau$-elements~\eqref{tau-neg},~\eqref{tau-pos} we have the following relation:
\begin{equation}\label{tauEv}
\begin{aligned}
    \left\{\tau^{\rm adj}_{\varnothing}+(q^2\ST)^{-6}\right\}\Big|_{k<0}-(-\ST)^2\left\{\tau^{\rm adj}_{\varnothing}+1\right\}\Big|_{k>0}&=\frac{(\ST+1) \left(q^8 \ST^6+1\right)}{q^{12} \ST^6}\,, \\
    \tau^{\rm adj}_{\rm adj}\Big|_{k<0}-(-\ST)^2\tau_{\rm adj}^{\rm adj}\Big|_{k>0}&=-\frac{(\ST+1)^2}{q^4 t(q^2\ST-1)}\,,\\
    \tau^{\rm adj}_{[4]}\Big|_{k<0}-(-\ST)^2\tau^{\rm adj}_{[4]}\Big|_{k>0}&=0\,.
\end{aligned}
\end{equation}
In the case of fundamental representation,
the Khovanov polynomial for twist knots possesses multiplicative jumps, see~\eqref{Tor&TwKh},
but in the case of adjoint Khovanov polynomials $\mathfrak{Kh}_{\rm adj}^{tw_k}$ and $\Kh_{\rm adj}^{tw_k}$ the evolution is defined by~\eqref{tauEv}. Note that terms that are added to $\tau_\varnothing^{\rm adj}$ are exactly $\lambda_{\rm adj}^{-2\,{\rm Sh}_{\cal K}}$. This fact returns us to $\tau$-elements of $\Kh_{\rm adj}^{tw_k}$. Formula~\eqref{KhAdjSpectr} together with~\eqref{EigenVal} and~\eqref{tau-neg},~\eqref{tau-pos} gives the adjoint Khovanov polynomials for all torus knots, which provide positive polynomials for all values of $k$.

\subsection{Torus knots ${\cal K}=tor_k$}
In this section, we discuss the internal structure of the adjoint Khovanov polynomials for two-strand torus knots $tor_k$ with odd $k$. In this case, two strands are parallel and carry representation $[2]={\rm adj}$. Thus, we expect the adjoint Khovanov polynomial $\mathfrak{Kh}_{\text{adj}}^{tor_k}$ (and thus, $\Kh_{\rm adj}^{tor_k}$) to have spectral expansion over three
eigenvalues corresponding to ${\rm adj}\otimes {\rm adj}=\varnothing\oplus\text{adj}\oplus[4]\,$. This is actually the case:
\begin{equation}\label{KhAdjSpectrTorus}
    \mathfrak{Kh}_{\text{adj}}^{tor_k}=\lambda^{-4k}_{\text{adj}}\left(\lambda_{\varnothing}^{k}\cdot \mu^{\rm adj}_{\varnothing}+\lambda_{\text{adj}}^{k}\cdot \mu_{\rm adj}^{\text{adj}}+\lambda_{[4]}^{k}\cdot \mu^{\rm adj}_{[4]}\right)
\end{equation}
with the same eigenvalues~\eqref{EigenVal}, and for $k>0\,$ ($k=3$ corresponds to the knot $3_1$):
\begin{equation}\label{muPos-k}
\begin{aligned}
    \mu^{\rm adj}_{\varnothing}&=\frac{q^2 \ST \left(q^2 \ST+1\right)}{\left(q^3 \ST^2-1\right) \left(q^3 \ST^2+1\right)}\,, \\
    \mu^{\rm adj}_{\text{adj}}&=-\frac{\left(q^2 \ST+1\right) \left(q^{12} \ST^8-q^{10} \ST^7+q^8 \ST^6-q^6 \ST^5+q^6 \ST^4-q^4 \ST^3+q^2 \ST^2+1\right)}{\left(q^4 \ST^3-1\right) \left(q^4 \ST^3+1\right)}\,, \\
    \mu^{\rm adj}_{[4]}&=\frac{\left(q^2 \ST-1\right) \left(q^2 \ST+1\right) \left(q^{14} \ST^9+q^{10} \ST^7-q^4 \ST^3+1\right)}{q^2 \ST \left(q^3 \ST^2-1\right) \left(q^3 \ST^2+1\right) \left(q^4 \ST^3-1\right) \left(q^4 \ST^3+1\right)}\,.
\end{aligned}
\end{equation}
Here we use the same symbol $\mu$ as in formulas of previous sections (for example in equation~\eqref{GenKhTwSat}), because in fact it corresponds to the same tangle calculus element but carrying other representations instead of $\Box\otimes\Box=\varnothing\oplus {\rm adj}$. Torus knots with negative crossings $k$ correspond to mirror knots, for which relations for the adjoint Khovanov polynomials~\eqref{MirrKhAdj} hold. Thus, for $k<0\,$ the corresponding $\mu$-elements can be obtained from~\eqref{muPos-k}\footnote{Their explicit expressions for $k<0$ are
\begin{equation}\label{mu2p<0}
\begin{aligned}
    \mu^{\rm adj}_{\varnothing}&=\frac{q^2 \ST^3 \left(q^2 \ST+1\right)}{\left(q^3 \ST^2-1\right) \left(q^3 \ST^2+1\right)}\,, \\
    \mu^{\rm adj}_{\text{adj}}&=-\frac{\left(q^2 \ST+1\right) \left(q^{12} \ST^8+q^{10} \ST^6-q^8 \ST^5+q^6 \ST^4-q^6 \ST^3+q^4 \ST^2-q^2 \ST+1\right)}{q^6 \ST^2 \left(q^4 \ST^3-1\right) \left(q^4 \ST^3+1\right)}\,, \\
    \mu^{\rm adj}_{[4]}&=\frac{\ST \left(q^2 \ST-1\right) \left(q^2 \ST+1\right) \left(q^{14} \ST^9-q^{10} \ST^6+q^4 \ST^2+1\right)}{q^2 \left(q^3 \ST^2-1\right) \left(q^3 \ST^2+1\right) \left(q^4 \ST^3-1\right) \left(q^4 \ST^3+1\right)}\,.
\end{aligned}
\end{equation}}:
\begin{equation}
    \mu^{\rm adj}_{Y}(q,\ST)\Big|_{k>0}=(-\ST)\cdot \mu^{\rm adj}_{Y}(q^{-1},\ST^{-1})\Big|_{k<0}\,\quad \text{with} \quad Y=\varnothing\,,\;{\rm adj}\,,\;[4]\,.
\end{equation}
The jumps of $\mu$-elements~\eqref{muPos-k},~\eqref{mu2p<0} of the adjoint Khovanov polynomials at the point $k=0$ are much more complicated and non-multiplicative compared to the simple multiplicative jumps of the fundamental Khovanov polynomials for torus knots~\eqref{Tor&TwKh}. 

Note that formula~\eqref{KhAdjSpectrTorus} together with~\eqref{EigenVal} and~\eqref{muPos-k},~\eqref{mu2p<0} gives the adjoint Khovanov polynomials for all torus knots, which provide positive polynomials for all values of $k$.

\setcounter{equation}{0}
\section{Conclusion}\label{sec:conclusion}

In a knot diagram $\tilde{\cal K}$, one can blow up an arbitrary vertex,
substituting it by a 2-strand braid of arbitrary length $n$.
As a slight generalisation, one can insert $m$-strand braids in place
of combination of vertices.
Dependence of a knot polynomial on the length $n$ is called {\it evolution} \cite{mironov2013evolution},
because it is described by a simple formula
\be
{\rm Pol}_R^{\tilde{\cal K}_n} = \sum_Y {\color{red}B_Y}\lambda_Y^{-n}\,,
\label{evof}
\ee
where the sum goes over representations $Y\in R^{\otimes m}$
(some $R$ in the tensor product can be conjugate, depending on the orientation
of strands).
This formula (\ref{evof}) is a simple statement in the tangle calculus \cite{mironov2018tangle},
inspired by the Reshetikhin--Turaev formalism for
the Jones and the HOMFLY polynomials \cite{reshetikhin1990ribbon,morozov2010chern,mironov2013evolution,artdiff},
but becomes rather non-trivial in the theory of complexes, leading to the Khovanov
and the Khovanov--Rozansky polynomials \cite{khovanov2000categorification,khovanov2007virtual,anokhina2018khovanov}.
Moreover, in this generalisation the coefficients $B_Y$ are not fully independent of $n$,
they jump at some value $n_0$, when the underlying complex changes drastically \cite{anokhina2021khovanov,dunin2022evolution} --
this is emphasised in (\ref{evof}) by writing $B_Y$ in red.

In this paper, we have studied the jumps in (\ref{evof}) for a peculiar case of evolution,
corresponding to 2-strand torus and twist satellites ${\cal S}_{T_n}^{\tilde{\cal K}}$ \cite{morozov2018knot,anokhina2021khovanov}.
Our claim is that the jump structure in Khovanov polynomials is much simpler than expected:
\be
{\color{red} \Kh_\Box^{{\cal S}_{T_{n-2w_{\tilde{\cal K}}}}^{{\cal K}}}} = {\color{red} \Kh_\Box^{T_{n+2\,{\rm Sh}_{\cal K}}}}
+ \mathfrak{c}_{T} \cdot\lambda_{\rm adj}^{-n}\cdot \mathfrak{Kh}_{\rm adj}^{ {\cal K} }\,,
\label{evoSf}
\ee
where only the first part at the r.h.s. is jumping, which depends on ${\cal K}$ only through the shift ${\rm Sh}_{\cal K}$
(which is, moreover, independent
on a choice of knot diagram $\tilde{\cal K}$).
The most interesting and complicated piece of the adjoint Khovanov polynomial of a knot ${\cal K}$ in the second term at the r.h.s. is free of jumps.

We have provided much evidence to support this claim, by analyzing a variety of knots ${\cal K}$,
including their additional evolutions along other directions.
A proof based on analysis of complexes would be desirable now. Further generalisations can be obtained in three directions:
\begin{itemize}

\item{}
to other evolutions and their interplay
(naively the jumping pattern of multi-evolution can be rich and somewhat involved \cite{anokhina2019nimble} --
but now we can expect separation of a jumping part into ${\cal K}$-independent multi-braid items),

\item{}
to other representations $R\neq \Box$,

\item{}
to the Khovanov--Rozansky polynomials.

\end{itemize}

\section*{Ackhowledgements}

We are grateful for enlightening discussions to D. Galakhov, A. Mironov, V. Mishnyakov, And. Morozov, A. Popolitov, A. Sleptsov, N. Tselousov.

This work is supported by the Russian Science Foundation (Grant No.21-12-00400).

\newpage





\printbibliography

@article{mironov2018tangle,
  title={Tangle blocks in the theory of link invariants},
  author={Mironov, A. and Morozov, A. and Morozov, And.},
  journal={Journal of High Energy Physics},
  volume={2018},
  number={9},
  pages={1--45},
  year={2018},
  publisher={Springer},
    eprint = {1804.07278},
    archivePrefix = {arXiv},
    primaryClass = {hep-th}
}

@article{mironov2015colored,
  title={Colored HOMFLY polynomials of knots presented as double fat diagrams},
  author={Mironov, A. and Morozov, A. and Morozov, And. and Ramadevi, P. and Singh, V.K.},
  journal={Journal of High Energy Physics},
  volume={2015},
  number={7},
  pages={1--70},
  year={2015},
  publisher={Springer},
    eprint = {1504.00371},
    archivePrefix = {arXiv},
    primaryClass = {hep-th}
}

@article{morozov2018knot,
  title={Knot polynomials for twist satellites},
  author={Morozov, A.},
  journal={Physics Letters B},
  volume={782},
  pages={104--111},
  year={2018},
  publisher={Elsevier},
    eprint = {1801.02407},
    archivePrefix = {arXiv},
    primaryClass = {hep-th}
}

@article{dolotin2014introduction,
  title={Introduction to Khovanov homologies. III. A new and simple tensor-algebra construction of Khovanov--Rozansky invariants},
  author={Dolotin, V. and Morozov, A.},
  journal={Nuclear Physics B},
  volume={878},
  pages={12--81},
  year={2014},
  publisher={Elsevier},
    eprint = {1308.5759},
    archivePrefix = {arXiv},
    primaryClass = {hep-th}
}

@article{anokhina2019nimble,
  title={Nimble evolution for pretzel Khovanov polynomials},
  author={Anokhina, A. and Morozov, A. and Popolitov, A.},
  journal={The European Physical Journal C},
  volume={79},
  number={10},
  pages={867},
  year={2019},
  publisher={Springer},
    eprint = {1904.10277},
    archivePrefix = {arXiv},
    primaryClass = {hep-th}
}

@article{morozov2022differential,
  title={Differential expansion for antiparallel triple pretzels: the way the factorization is deformed},
  author={Morozov, A. and Tselousov, N.},
  journal={The European Physical Journal C},
  volume={82},
  number={10},
  pages={912},
  year={2022},
  publisher={Springer},
    eprint = {2205.12238},
    archivePrefix = {arXiv},
    primaryClass = {hep-th}
}

@article{bishler2020perspectives,
  title={Perspectives of differential expansion},
  author={Bishler, L. and Morozov, A.},
  journal={Physics Letters B},
  volume={808},
  pages={135639},
  year={2020},
  publisher={Elsevier},
    eprint = {2006.01190},
    archivePrefix = {arXiv},
    primaryClass = {hep-th}
}

@article{morozov2020kntz,
  title={The KNTZ trick from arborescent calculus and the structure of the differential expansion},
  author={Morozov, A.},
  journal={Theoretical and Mathematical Physics},
  volume={204},
  number={2},
  pages={993--1019},
  year={2020},
  publisher={Springer},
    eprint = {2001.10254},
    archivePrefix = {arXiv},
    primaryClass = {hep-th}
}

@article{morozov2019extension,
  title={Extension of KNTZ trick to non-rectangular representations},
  author={Morozov, A.},
  journal={Physics Letters B},
  volume={793},
  pages={464--468},
  year={2019},
  publisher={Elsevier},
    eprint = {1903.00259},
    archivePrefix = {arXiv},
    primaryClass = {hep-th}
}

@article{kameyama2020cyclotomic,
  title={Cyclotomic expansions of HOMFLY-PT colored by rectangular Young diagrams},
  author={Kameyama, M. and Nawata, S. and Tao, R. and Zhang, H.D.},
  journal={Letters in Mathematical Physics},
  volume={110},
  pages={2573--2583},
  year={2020},
  publisher={Springer},
    eprint = {1902.02275},
    archivePrefix = {arXiv},
    primaryClass = {math.GT}
}

@article{morozov2018factorization,
  title={Factorization of differential expansion for non-rectangular representations},
  author={Morozov, A.},
  journal={Modern Physics Letters A},
  volume={33},
  number={12},
  pages={1850062},
  year={2018},
  publisher={World Scientific},
    eprint = {1606.06015},
    archivePrefix = {arXiv},
    primaryClass = {hep-th}
}

@article{kononov2016rectangular,
  title={On rectangular HOMFLY for twist knots},
  author={Kononov, Ya. and Morozov, A.},
  journal={Modern Physics Letters A},
  volume={31},
  number={38},
  pages={1650223},
  year={2016},
  publisher={World Scientific},
    eprint = {1610.04778},
    archivePrefix = {arXiv},
    primaryClass = {hep-th}
}

@article{morozov2016factorization,
  title={Factorization of differential expansion for antiparallel double-braid knots},
  author={Morozov, A},
  journal={Journal of High Energy Physics},
  volume={2016},
  number={9},
  pages={1--31},
  year={2016},
  publisher={Springer}
}

@article{itoyama2012homfly,
  title={HOMFLY and superpolynomials for figure eight knot in all symmetric and antisymmetric representations},
  author={Itoyama, H. and Mironov, A. and Morozov, A. and Morozov, And.},
  journal={Journal of High Energy Physics},
  volume={2012},
  number={7},
  pages={1--21},
  year={2012},
  publisher={Springer},
    eprint = {1203.5978},
    archivePrefix = {arXiv},
    primaryClass = {hep-th}
}

@article{anokhina2021khovanov,
  title={Khovanov polynomials for satellites and asymptotic adjoint polynomials},
  author={Anokhina, A. and Morozov, A. and Popolitov, A.},
  journal={International Journal of Modern Physics A},
  volume={36},
  number={34n35},
  pages={2150243},
  year={2021},
  publisher={World Scientific},
    eprint = {2104.14491},
    archivePrefix = {arXiv},
    primaryClass = {hep-th}
}

@article{bar2007fast,
  title={Fast Khovanov homology computations},
  author={Bar-Natan, D.},
  journal={Journal of Knot Theory and Its Ramifications},
  volume={16},
  number={03},
  pages={243--255},
  year={2007},
  publisher={World Scientific},
    eprint = {math/0606318},
    archivePrefix = {arXiv},
    primaryClass = {math.GT}
}

@article{bar2005khovanov,
  title={Khovanov’s homology for tangles and cobordisms},
  author={Bar-Natan, D.},
  journal={Geometry \& Topology},
  volume={9},
  number={3},
  pages={1443--1499},
  year={2005},
  publisher={Mathematical Sciences Publishers},
    eprint = {math/0410495},
    archivePrefix = {arXiv},
    primaryClass = {math.GT}
}

@article{CS,
    author = "Chern, S.-S. and Simons, J.",
    title = "{Characteristic forms and geometric invariants}",
    doi = "",
    journal = "Annals Math.",
    volume = "99",
    pages = "48--69",
    year = "1974"
}

@article{Witten,
  title={Quantum field theory and the Jones polynomial},
  author={Witten, E.},
  journal={Communications in Mathematical Physics},
  volume={121},
  number={3},
  pages={351--399},
  year={1989},
  publisher={Springer},
  doi={}
}

@article{alexander1928topological,
  title={Topological invariants of knots and links},
  author={Alexander, J.W.},
  journal={Transactions of the American Mathematical Society},
  volume={30},
  number={2},
  pages={275--306},
  year={1928},
  publisher={JSTOR}
}

@article{leech1970computational,
  title={Computational problems in abstract algebra},
  author={Algebraic Properties, J.H. Conway},
  year={1970},
  journal={John Leech (ed.), Proc.Conf.Oxford, Pergamon Press, Oxford-New York},
  pages={329-358},
  publisher={Citeseer}
}

@article{jones1983invent,
  title={Invent. Math.},
  author={Jones, V.F.R.},
  journal={Index for subfactors},
  volume={72},
  pages={1--25},
  year={1983}
}

@article{freyd1985new,
  title={A new polynomial invariant of knots and links},
  author={Freyd, P. and Yetter, D. and Hoste, J. and Lickorish, W.B.R. and Millett, K. and Ocneanu, A.},
  journal={Bulletin (new series) of the American mathematical society},
  volume={12},
  number={2},
  pages={239--246},
  year={1985},
  publisher={American Mathematical Society}
}

@article{kauffman1987state,
  title={State models and the Jones polynomial},
  author={Kauffman, L.H.},
  journal={Topology},
  volume={26},
  number={3},
  pages={395--407},
  year={1987},
  publisher={Elsevier}
}

@article{przytycki1987kobe,
  title={Kobe J. Math.},
  author={Przytycki, J.H. and Traczyk, K.P.},
  journal={Invariants of links of Conway type},
  volume={4},
  pages={115--139},
  year={1987},
	eprint = {1610.06679},
    archivePrefix = {arXiv},
    primaryClass = {math.GT}
}

@article{Alvarez1994tt,
    author = "Alvarez, M. and Labastida, J.M.F.",
    eprint = "hep-th/9407076",
    archivePrefix = "arXiv",
    primaryClass = {hep-th},
    title = "Numerical knot invariants of finite type from Chern--Simons perturbation theory",
    reportNumber = "US-FT-8-94",
    doi = "",
    journal = "Nucl. Phys. B",
    volume = "433",
    pages = "555--596",
    year = "1995",
}

@article{Alvarez_1997,
   volume={189},
   ISSN={1432-0916},
   DOI={},
   number={3},
    eprint = "q-alg/9604010",
    archivePrefix = "arXiv",
    primaryClass = {math.QA},
   journal={Communications in Mathematical Physics},
    title={Primitive Vassiliev invariants and factorization in Chern--Simons perturbation theory},
   publisher={Springer Science and Business Media LLC},
   author={Alvarez, M. and Labastida, J.M.F.},
   year={1997},
   month={Nov},
   pages={641–654}
}

@article{Alvarez_1993,
   volume={395},
   ISSN={0550-3213},
   DOI={},
   number={1-2},
    title={Analysis of observables in Chern--Simons perturbation theory},
   journal={Nuclear Physics B},
   publisher={Elsevier BV},
   author={Alvarez, M. and Labastida, J.M.F.},
   year={1993},
   month={Apr},
   pages={198–238}
}

@article{Labastida1997uw,
    author = "Labastida, J.M.F. and Perez, Esther",
    archivePrefix = "arXiv",
    doi = "",
    title = "Kontsevich integral for Vassiliev invariants from Chern–Simons perturbation theory in the light-cone gauge",
    eprint = "hep-th/9710176",
    primaryClass = {math.QA},
    journal = "J.Math.Phys.",
    pages = "5183--5198",
    reportNumber = "CERN-TH-97-282, US-FT-32-97",
    volume = "39",
    year = "1998"
}

@article{Guadagnini:1989kr,
    author = "Guadagnini, E. and Martellini, M. and Mintchev, M.",
    reportNumber = "CERN-TH-5234/89, CERN-TH-5324/89, IFUP-TH-6/89",
    doi = "",
    title = "Perturbative aspects of the Chern--Simons field theory",
    journal = "Phys. Lett. B",
    volume = "227",
    pages = "111--117",
    year = "1989"
}

@article{GUADAGNINI1990575,
journal = {Nuclear Physics B},
volume = {330},
number = {2},
pages = {575-607},
year = {1990},
issn = {0550-3213},
doi = {},
title = {Wilson lines in Chern--Simons theory and link invariants},
author = {E. Guadagnini and M. Martellini and M. Mintchev},
}

@article{Reshetikhin,
  title={Invariants of tangles 1},
  author={Reshetikhin, N.Yu.},
  journal={unpublished preprint},
  volume={},
  pages={},
  year={1987},
  publisher={}
}

@article{krasner2008computation,
  title={A computation in Khovanov-Rozansky homology},
  author={K., Daniel},
  journal={arXiv preprint arXiv:0801.4018},
  year={2008},
    eprint = {0801.4018},
    archivePrefix = {arXiv},
    primaryClass = {math.GT}
}

@article{guadagnini1990chern2,
  title={Chern--Simons holonomies and the appearance of quantum groups},
  author={Guadagnini, E. and Martellini, M. and Mintchev, M.},
  journal={Physics Letters B},
  volume={235},
  number={3-4},
  pages={275--281},
  year={1990},
  publisher={Elsevier}
}

@article{reshetikhin1990ribbon,
  title={Ribbon graphs and their invaraints derived from quantum groups},
  author={Reshetikhin, N.Yu. and Turaev, V.G.},
  journal={Communications in Mathematical Physics},
  volume={127},
  number={1},
  pages={1--26},
  year={1990},
  publisher={Springer}
}

@article{turaev1990yang,
  title={The Yang--Baxter equation and invariants of links},
  author={Turaev, V.},
  journal={New Developments in the Theory of Knots},
  volume={11},
  pages={175},
  year={1990},
  publisher={World Scientific}
}

@article{reshetikhin1991invariants,
  title={Invariants of 3-manifolds via link polynomials and quantum groups},
  author={Reshetikhin, N.Yu. and Turaev, V.G.},
  journal={Inventiones mathematicae},
  volume={103},
  number={1},
  pages={547--597},
  year={1991}
}

@inproceedings{mironov2013evolution,
  title={Evolution method and “differential hierarchy” of colored knot polynomials},
  author={Mironov, A. and Morozov, A. and Morozov, And.},
  booktitle={AIP Conference Proceedings},
  volume={1562},
  number={1},
  pages={123--155},
  year={2013},
  organization={American Institute of Physics},
    eprint = {1306.3197},
    archivePrefix = {arXiv},
    primaryClass = {hep-th}
}

@article{artdiff,
	doi = {},
  
	url = {},
  
	year = 2014,
	month = {may},
  
	publisher = {Springer Science and Business Media {LLC}
},
  
	volume = {179},
  
	number = {2},
  
	pages = {509--542},
  
	author = {S. B. Arthamonov and A. D. Mironov and A. Yu. Morozov},
  
	title = {Differential hierarchy and additional grading of knot polynomials},
  
	journal = {Theoretical and Mathematical Physics},
	eprint = {1306.5682},
    archivePrefix = {arXiv},
    primaryClass = {hep-th}
}

@article{morozov2010chern,
  title={Chern--Simons theory in the temporal gauge and knot invariants through the universal quantum R-matrix},
  author={Morozov, A. and Smirnov, A.},
  journal={Nuclear Physics B},
  volume={835},
  number={3},
  pages={284--313},
  year={2010},
  publisher={Elsevier},
    eprint = {1001.2003},
    archivePrefix = {arXiv},
    primaryClass = {hep-th}
}

@article{dasbach2008jones,
  title={The Jones polynomial and graphs on surfaces},
  author={Dasbach, O.T. and Futer, D. and Kalfagianni, E. and Lin, X.-S. and Stoltzfus, N.W.},
  journal={Journal of Combinatorial Theory, Series B},
  volume={98},
  number={2},
  pages={384--399},
  year={2008},
  publisher={Elsevier},
    eprint = {math/0605571},
    archivePrefix = {arXiv},
    primaryClass = {math.GT}
}

@article{lewark2022rasmussen,
  title={Rasmussen invariants of Whitehead doubles and other satellites},
  author={Lewark, L. and Zibrowius, C.},
  journal={arXiv preprint arXiv:2208.13612},
  year={2022},
    eprint = {2208.13612},
    archivePrefix = {arXiv},
    primaryClass = {math.GT}
}

@article{rasmussen2010khovanov,
  title={Khovanov homology and the slice genus},
  author={Rasmussen, J.},
  journal={Inventiones mathematicae},
  volume={182},
  number={2},
  pages={419--447},
  year={2010},
  publisher={Springer},
    eprint = {math/0402131},
    archivePrefix = {arXiv},
    primaryClass = {math.GT}
}

@article{dunin2013superpolynomials,
  title={Superpolynomials for torus knots from evolution induced by cut-and-join operators},
  author={Dunin-Barkowski, P. and Mironov, A. and Morozov, A. and Sleptsov, A. and Smirnov, A.},
  journal={Journal of High Energy Physics},
  volume={2013},
  number={3},
  pages={1--87},
  year={2013},
  publisher={Springer},
    eprint = {1106.4305},
    archivePrefix = {arXiv},
    primaryClass = {hep-th}
}

@article{mironov2016racah,
  title={Racah matrices and hidden integrability in evolution of knots},
  author={Mironov, A. and Morozov, A. and Morozov, And. and Sleptsov, A.},
  journal={Physics Letters B},
  volume={760},
  pages={45--58},
  year={2016},
  publisher={Elsevier},
    eprint = {1605.04881},
    archivePrefix = {arXiv},
    primaryClass = {hep-th}
}

@article{mironov2015colored2,
  title={Colored HOMFLY polynomials for the pretzel knots and links},
  author={Mironov, A and Morozov, A and Sleptsov, A},
  journal={Journal of High Energy Physics},
  volume={2015},
  number={7},
  pages={1--35},
  year={2015},
  publisher={Springer}
}

@article{mironov2014colored,
  title={On colored HOMFLY polynomials for twist knots},
  author={Mironov, A. and Morozov, A. and Morozov, And.},
  journal={Modern Physics Letters A},
  volume={29},
  number={34},
  pages={1450183},
  year={2014},
  publisher={World Scientific},
    eprint = {1408.3076},
    archivePrefix = {arXiv},
    primaryClass = {hep-th}
}

@article{dunin2022evolution,
  title={Evolution for Khovanov polynomials for figure-eight-like family of knots},
  author={Dunin-Barkowski, P. and Popolitov, A. and Popolitova, S.},
  journal={International Journal of Modern Physics A},
  volume={37},
  number={36},
  pages={2250216},
  year={2022},
  publisher={World Scientific},
    eprint = {1812.00858},
    archivePrefix = {arXiv},
    primaryClass = {math-ph}
}

@article{willis2021khovanov,
  title={Khovanov--Rozansky homology for infinite multicolored braids},
  author={Willis, M.},
  journal={Canadian Journal of Mathematics},
  volume={73},
  number={5},
  pages={1239--1277},
  year={2021},
  publisher={Canadian Mathematical Society},
    eprint = {1904.09055},
    archivePrefix = {arXiv},
    primaryClass = {math.GT}
}

@article{nakagane2019action,
  title={The action of full twist on the superpolynomial for torus knots},
  author={Nakagane, K.},
  journal={Topology and its Applications},
  volume={266},
  pages={106841},
  year={2019},
  publisher={Elsevier},
    eprint = {1805.01606},
    archivePrefix = {arXiv},
    primaryClass = {math.GT}
}

@article{anokhina2018khovanov,
  title={Are Khovanov--Rozansky polynomials consistent with evolution in the space of knots?},
  author={Anokhina, A. and Morozov, A.},
  journal={Journal of High Energy Physics},
  volume={2018},
  number={4},
  pages={1--29},
  year={2018},
  publisher={Springer},
    eprint = {1802.09383},
    archivePrefix = {arXiv},
    primaryClass = {hep-th}
}

@article{kaul1998chern,
  title={Chern--Simons theory, knot invariants, vertex models and three-manifold invariants},
  author={Kaul, R.K.},
  journal={arXiv preprint hep-th/9804122},
  year={1998},
    eprint = {hep-th/9804122},
    archivePrefix = {arXiv},
    primaryClass = {hep-th}
}

@article{khovanov2000categorification,
  title={A categorification of the Jones polynomial},
  author={Khovanov, M.},
  journal={Duke Mathematical Journal},
  volume={101},
  number={3},
  pages={359--426},
  year={2000},
  publisher={Duke University Press},
	eprint = {math/9908171},
    archivePrefix = {arXiv},
    primaryClass = {math.QA}
}

@article{bar2002khovanov,
  title={On Khovanov’s categorification of the Jones polynomial},
  author={Bar-Natan, D.},
  journal={Algebraic \& Geometric Topology},
  volume={2},
  number={1},
  pages={337--370},
  year={2002},
  publisher={Mathematical Sciences Publishers},
	eprint = {math/0201043},
    archivePrefix = {arXiv},
    primaryClass = {math.QA}
}

@article{khovanov2004sl,
  title={sl (3) link homology},
  author={Khovanov, M.},
  journal={Algebraic \& Geometric Topology},
  volume={4},
  number={2},
  pages={1045--1081},
  year={2004},
  publisher={Mathematical Sciences Publishers},
	eprint = {math/0304375},
    archivePrefix = {arXiv},
    primaryClass = {math.QA}
}

@article{khovanov2007virtual,
  title={Virtual crossings, convolutions and a categorification of the SO (2N) Kauffman polynomial},
  author={Khovanov, M. and Rozansky, L.},
  journal={arXiv preprint math/0701333},
  year={2007},
	eprint = {math/0701333},
    archivePrefix = {arXiv},
    primaryClass = {math.QA}
}

@article{khovanov2010categorifications,
  title={Categorifications from planar diagrammatics},
  author={Khovanov, M.},
  journal={Japanese Journal of Mathematics},
  volume={5},
  number={2},
  pages={153--181},
  year={2010},
  publisher={Springer},
	eprint = {1008.5084},
    archivePrefix = {arXiv},
    primaryClass = {math.QA}
}

@article{dunfield2006superpolynomial,
  title={The superpolynomial for knot homologies},
  author={Dunfield, N.M. and Gukov, S. and Rasmussen, J.},
  journal={Experimental Mathematics},
  volume={15},
  number={2},
  pages={129--159},
  year={2006},
  publisher={Taylor \& Francis},
    eprint = {math/0505662},
    archivePrefix = {arXiv},
    primaryClass = {math.GT}
}

@article{gukov2005khovanov,
  title={Khovanov--Rozansky homology and topological strings},
  author={Gukov, S. and Schwarz, A. and Vafa, C.},
  journal={Letters in Mathematical Physics},
  volume={74},
  pages={53--74},
  year={2005},
  publisher={Springer},
    eprint = {hep-th/0412243},
    archivePrefix = {arXiv},
    primaryClass = {hep-th}
}

@article{fuji2012volume,
  title={Volume conjecture: refined and categorified},
  author={Fuji, H. and Gukov, S. and Su{\l}kowski, P. and Awata, H.},
  journal={Advances in Theoretical and Mathematical Physics},
  volume={16},
  number={6},
  pages={1669--1777},
  year={2012},
  publisher={International Press},
    eprint = {1203.2182},
    archivePrefix = {arXiv},
    primaryClass = {hep-th}
}

@article{nawata2012super,
  title={Super-A-polynomials for twist knots},
  author={Nawata, S. and Ramadevi, P. and Sun, X.},
  journal={Journal of High Energy Physics},
  volume={2012},
  number={11},
  pages={1--39},
  year={2012},
  publisher={Springer},
    eprint = {1209.1409},
    archivePrefix = {arXiv},
    primaryClass = {hep-th}
}

@misc{katlas, 
title={\href{http://katlas.org}{http://katlas.org}
}
}

@misc{lewark, 
title={\href{http://www.lewark.de/lukas/software.html}{http://www.lewark.de/lukas/software.html}
}
}

@article{anokhina2014cabling,
  title={Cabling procedure for the colored HOMFLY polynomials},
  author={Anokhina, A. and Morozov, And.},
  journal={Theoretical and Mathematical Physics},
  volume={178},
  number={1},
  pages={1--58},
  year={2014},
  publisher={Springer},
    eprint = {1307.2216},
    archivePrefix = {arXiv},
    primaryClass = {hep-th}
}

\newpage

\appendix
\setcounter{equation}{0}
\section{Examples of adjoint Khovanov polynomials}\label{App}
In this section, we provide lists of the adjoint Khovanov polynomilas ${\color{red} \Kh_{\rm adj}^{\cal K}}$ and $\mathfrak{Kh}_{\rm adj}^{\cal K}$ with knots $\cal K$ given by Table~\ref{tab:knots-list}. All boundaries between two different smooth evolutions of  ${\color{red} \Kh_{\rm adj}^{\cal K}}$ are written from the knot diagram independent ${\color{red} \Kh^{{\cal S}_{T_\n}({\cal K})}}$. Namely, jumps occur at the point $\n+2\,{\rm Sh}_{\cal K}=0$. Note that the same adjoint Khovanov polynomials appear both for torus and twist satellites~\eqref{GenKhTwSat},~\eqref{GenMainClaim}.

For any knot $\cal K$, we obtain two polynomials ${\color{red} \Kh_{\rm adj}^{\cal K}}$ which do not differ by a factor. We mark in
red different parts in each pair of polynomials. These red terms give
$\lambda^{-2\,\text{Sh}_{\cal K}}_{\text{adj}}$ which is subtracted from adjoint Khovanov polynomials to form the quantity without jumps $\mathfrak{Kh}_{\rm adj}^{\cal K}$ (see Sections~\ref{sec:KhRed},~\ref{sec:main} for details). Note that sometimes in order to extract $\lambda^{-2\,\text{Sh}_{\cal K}}_{\text{adj}}$ from the adjoint Khovanov polynomial, one should add and subtract the same term from it. These resulting terms are additionally underlined.


\subsection{$\mathcal{K}=tw_k$}
In this section we consider the adjoint Khovanov polynomials for knots ${\cal K}=tw_k$. List the adjoint Khovanov polynomials for $k < 0$ with ${\rm Sh}_{tw_{k<0}}=3\,$:
\begin{equation}\label{Ex1}\nn
 \begin{aligned}
    {\overline{3_1}}=\overline{tw_{-2}}:\quad {\color{red}\Kh^{\overline{3_1}}_{\rm adj}}&{\color{red}(q^{-1},\ST^{-1})}=\begin{cases}
    \ST^{-11} q^{-22}(q^{18} \ST^{11}+q^{12} \ST^7-q^{10} \ST^7{\color{red} +q^{10}
    \ST^5} -q^8 \ST^5+q^6 \ST^3+q^4 \ST^2-q^2 \ST-1),\; \n\geq 6
    \\
    \ST^{-11} q^{-22}(q^{18} \ST^{11}+q^{12} \ST^7-q^{10} \ST^7{\color{red}
    - q^{10} \ST^6} -q^8 \ST^5+q^6 \ST^3+q^4 \ST^2-q^2 \ST-1),\; \n<6
    \end{cases} \\
    3_1=tw_{-2}:\quad {\color{red}\Kh^{3_1}_{\rm adj}}&=\begin{cases}
     \ST^{-11} q^{-22}(q^{18} \ST^{11}+q^{12} \ST^7-q^{10} \ST^7{\color{red}
    - q^{10} \ST^6} -q^8 \ST^5+q^6 \ST^3+q^4 \ST^2-q^2 \ST-1),\quad \n\geq -6 \\
    \ST^{-11} q^{-22}(q^{18} \ST^{11}+q^{12} \ST^7-q^{10} \ST^7{\color{red}
    + q^{10} \ST^5}-q^8 \ST^5+q^6 \ST^3+q^4 \ST^2-q^2 \ST-1), \quad \n<-6
    \end{cases} \\
    \{q^2\ST\}^{-1}\cdot\mathfrak{Kh}_{\rm adj}^{3_1} &= q^{-20} \ST^{-10}(q^{14} \ST^9+q^{10} \ST^7+q^8 \ST^5+q^2 \ST+1)\\
    5_2=tw_{-4}:\quad {\color{red}\Kh^{5_2}_{\rm adj}}&=\begin{cases}
     \ST^{-19}q^{-34}(q^{30} \ST^{19}+q^{28} \ST^{18}+q^{26} \ST^{17}+q^{26}
    \ST^{16}-q^{24} \ST^{16}+2 q^{24} \ST^{15}-q^{22} \ST^{15}\underline{+\,2q^{22} \ST^{14}-}\\ \underline{{\color{red} -q^{22} \ST^{14}}}+q^{20} \ST^{12}-q^{18}
    \ST^{13}-3 q^{18} \ST^{12}+2 q^{18} \ST^{11}-2 q^{16} \ST^{11}+q^{16} \ST^{10}+q^{14} \ST^8-q^{12} \ST^8+\\ +2 q^{12} \ST^7 -2
    q^{10} \ST^7-q^8 \ST^6-2 q^8 \ST^5-q^6 \ST^4+q^6 \ST^3+q^4 \ST^2-q^2 \ST-1),\quad \n\geq -6\\
     \ST^{-19}q^{-34}(q^{30} \ST^{19}+q^{28} \ST^{18}+q^{26} \ST^{17}+q^{26} \ST^{16}-q^{24}
     \ST^{16}+2 q^{24} \ST^{15}-q^{22} \ST^{15}\underline{+2 q^{22} \ST^{14}+}\\ \underline{{\color{red} +q^{22} \ST^{13}}}+
     q^{20} \ST^{12}-q^{18} \ST^{13}-3 q^{18} \ST^{12}+2 q^{18} \ST^{11}-2 q^{16} \ST^{11}+q^{16} \ST^{10}+q^{14} \ST^8-q^{12} \ST^8+\\
     +2 q^{12} \ST^7-2 q^{10} \ST^7-q^8 \ST^6-2 q^8 \ST^5-q^6 \ST^4+q^6 \ST^3+q^4 \ST^2-q^2 \ST-1),\quad \n < -6
    \end{cases}\\
    \{q^2\ST\}^{-1}\cdot \mathfrak{Kh}_{\rm adj}^{5_2}&=q^{-32} \ST^{-18}(q^{26} \ST^{17}+q^{24} \ST^{16}+2 q^{22} \ST^{15}+q^{22} \ST^{14}+2 q^{20} \ST^{13}+q^{18} \ST^{13}+3 q^{18} \ST^{12}+2 q^{16} \ST^{11}+\\ &+q^{16} \ST^{10}+2 q^{14} \ST^9+2 q^{12} \ST^8+2 q^{10} \ST^7+q^{10} \ST^6+q^8 \ST^6+2 q^8 \ST^5+q^6 \ST^4+q^2 \ST+1)  \\
    7_2=tw_{-6}: \quad {\color{red}\Kh^{7_2}_{ \text{adj}}}&=\begin{cases}
    \ST^{-27}q^{-46}(q^{42} \ST^{27}+q^{40} \ST^{26}+q^{38} \ST^{25}+q^{38}
    \ST^{24}+2 q^{36} \ST^{23}\underline{+3q^{34} \ST^{22}{\color{red} -q^{34} \ST^{22}}}-q^{32} \ST^{22}+\\ +q^{32} \ST^{21}
    +q^{32} \ST^{20}-q^{30} \ST^{21}-q^{30} \ST^{20}+3 q^{30} \ST^{19}-q^{28} \ST^{19}+3 q^{28} \ST^{18}-q^{26} \ST^{19}-3 q^{26}
    \ST^{18}+\\ +q^{26} \ST^{16} -2 q^{24} \ST^{17}-2 q^{24} \ST^{16}+3 q^{24} \ST^{15}-q^{22} \ST^{15}+3 q^{22} \ST^{14}-q^{20}
    \ST^{14}+q^{20} \ST^{12}-2 q^{18} \ST^{13}-\\ -3 q^{18} \ST^{12} +2 q^{18} \ST^{11}-q^{16} \ST^{12}-3 q^{16} \ST^{11}+q^{16}
    \ST^{10}-q^{14} \ST^{10}+q^{14} \ST^8-q^{12} \ST^8+2 q^{12} \ST^7-\\ -2 q^{10} \ST^7-q^8 \ST^6 -2 q^8 \ST^5-q^6 \ST^4+q^6 \ST^3+q^4
    \ST^2-q^2 \ST-1),\quad \n\geq -6
    \\
    \ST^{-27}q^{-46}(q^{42} \ST^{27}+q^{40} \ST^{26}+q^{38} \ST^{25}+q^{38} \ST^{24}+2 q^{36}
    \ST^{23}\underline{+3 q^{34} \ST^{22}{\color{red}+q^{34} \ST^{21}}}-q^{32} \ST^{22}+\\ +q^{32} \ST^{21}+q^{32}
    \ST^{20}-q^{30} \ST^{21}-q^{30} \ST^{20}+3 q^{30} \ST^{19}-q^{28} \ST^{19}+3 q^{28} \ST^{18}-q^{26} \ST^{19}-3 q^{26} \ST^{18}+\\
    +q^{26} \ST^{16}-2 q^{24} \ST^{17}-2 q^{24} \ST^{16}+3 q^{24} \ST^{15}-q^{22} \ST^{15}+3 q^{22} \ST^{14}-q^{20} \ST^{14}+q^{20}
    \ST^{12}-2 q^{18} \ST^{13}-\\ -3 q^{18} \ST^{12}+2 q^{18} \ST^{11}-q^{16} \ST^{12}-3 q^{16} \ST^{11}+q^{16} \ST^{10}-q^{14}
    \ST^{10}+q^{14} \ST^8-q^{12} \ST^8+2 q^{12} \ST^7-\\ -2 q^{10} \ST^7-q^8 \ST^6-2 q^8 \ST^5-q^6 \ST^4+q^6 \ST^3+q^4 \ST^2-q^2
    \ST-1),\quad \n<-6
    \end{cases} \\
    \{q^2\ST\}^{-1}\cdot\mathfrak{Kh}_{\rm adj}^{7_2}&=q^{-44} \ST^{-26}(q^{38} \ST^{25}+q^{36} \ST^{24}+2 q^{34} \ST^{23}+q^{34} \ST^{22}+q^{32} \ST^{22}+2 q^{32} \ST^{21}+2 q^{30} \ST^{21}+4 q^{30} \ST^{20}+\\ &+3 q^{28} \ST^{19}+q^{28} \ST^{18}+q^{26} \ST^{19}+3 q^{26} \ST^{18}+3 q^{26} \ST^{17}+2 q^{24} \ST^{17}+4 q^{24} \ST^{16}+3 q^{22} \ST^{15}+q^{22} \ST^{14}+\\ &+2 q^{20} \ST^{14}+3 q^{20} \ST^{13}+2 q^{18} \ST^{13}+4 q^{18} \ST^{12}+q^{16} \ST^{12}+3 q^{16} \ST^{11}+q^{16} \ST^{10}+q^{14} \ST^{10}+2 q^{14} \ST^9+\\ &+2 q^{12} \ST^8+2 q^{10} \ST^7+q^{10} \ST^6+q^8 \ST^6+2 q^8 \ST^5+q^6 \ST^4+q^2 \ST+1)
\end{aligned}
\end{equation}
\begin{equation}\label{Ex2}
 \begin{aligned}
    9_2=tw_{-8}:\quad {\color{red}\Kh^{9_2}_{\rm adj}}&=
    \begin{cases}
    {\ST^{-35}}q^{-58}(q^{54} \ST^{35}+q^{52} \ST^{34}+q^{50} \ST^{33}+q^{50} \ST^{32}+2
    q^{48} \ST^{31}\underline{+3q^{46} \ST^{30} {\color{red}-q^{46} \ST^{30}}}+q^{44} \ST^{29}+q^{44} \ST^{28}+\\ +3 q^{42}
    \ST^{27}-q^{40} \ST^{28}+3 q^{40} \ST^{26}-q^{38} \ST^{27}-q^{38} \ST^{26}+q^{38} \ST^{25}+q^{38} \ST^{24}-q^{36} \ST^{25}+3 q^{36}
    \ST^{23}-\\ -q^{34} \ST^{25}-3 q^{34} \ST^{24}-q^{34} \ST^{23}+3 q^{34} \ST^{22}-2 q^{32} \ST^{23}-2 q^{32} \ST^{22}+q^{32}
    \ST^{21}+q^{32} \ST^{20}-q^{30} \ST^{21}+\\ +3 q^{30} \ST^{19}-q^{28} \ST^{20}-q^{28} \ST^{19}+3 q^{28} \ST^{18}-2 q^{26}
    \ST^{19}-3 q^{26} \ST^{18}+q^{26} \ST^{16}-q^{24} \ST^{18}-3 q^{24} \ST^{17}-\\ -2 q^{24} \ST^{16}+3 q^{24} \ST^{15}-q^{22}
    \ST^{16}-q^{22} \ST^{15}+3 q^{22} \ST^{14}-q^{20} \ST^{14}+q^{20} \ST^{12}-2 q^{18} \ST^{13}-3 q^{18} \ST^{12}+\\ +2 q^{18}
    \ST^{11}-q^{16} \ST^{12}-3 q^{16} \ST^{11}+q^{16} \ST^{10}-q^{14} \ST^{10}+q^{14} \ST^8-q^{12} \ST^8+2 q^{12} \ST^7-2 q^{10} \ST^7-\\
    -q^8 \ST^6-2 q^8 \ST^5-q^6 \ST^4+q^6 \ST^3+q^4 \ST^2-q^2 \ST-1),\quad \n\geq -6
    \\
     \ST^{-35} q^{-58}(q^{54} \ST^{35}+q^{52} \ST^{34}+q^{50} \ST^{33}+q^{50} \ST^{32}+2 q^{48}
     \ST^{31}\underline{+3 q^{46} \ST^{30}{\color{red}+q^{46} \ST^{29}}}+q^{44} \ST^{29}+\\ +q^{44} \ST^{28}+3 q^{42}
     \ST^{27}-q^{40} \ST^{28}+3 q^{40} \ST^{26}-q^{38} \ST^{27}-q^{38} \ST^{26}+q^{38} \ST^{25}+q^{38} \ST^{24}-q^{36} \ST^{25}+\\ +3
     q^{36} \ST^{23}-q^{34} \ST^{25}-3 q^{34} \ST^{24}-q^{34} \ST^{23}+3 q^{34} \ST^{22}-2 q^{32} \ST^{23}-2 q^{32} \ST^{22}+q^{32}
     \ST^{21}+q^{32} \ST^{20}-\\ -q^{30} \ST^{21}+3 q^{30} \ST^{19}-q^{28} \ST^{20}-q^{28} \ST^{19}+3 q^{28} \ST^{18}-2 q^{26}
     \ST^{19}-3 q^{26} \ST^{18}+q^{26} \ST^{16}-q^{24} \ST^{18}-\\ -3 q^{24} \ST^{17}-2 q^{24} \ST^{16}+3 q^{24} \ST^{15}-q^{22}
     \ST^{16}-q^{22} \ST^{15}+3 q^{22} \ST^{14}-q^{20} \ST^{14}+q^{20} \ST^{12}-2 q^{18} \ST^{13}-\\ -3 q^{18} \ST^{12}+2 q^{18}
     \ST^{11}-q^{16} \ST^{12}-3 q^{16} \ST^{11}+q^{16} \ST^{10}-q^{14} \ST^{10}+q^{14} \ST^8-q^{12} \ST^8+2 q^{12} \ST^7-\\ -2 q^{10}
     \ST^7-q^8 \ST^6-2 q^8 \ST^5-q^6 \ST^4+q^6 \ST^3+q^4 \ST^2-q^2 \ST-1),\quad \n< -6
    \end{cases}\\
    \{q^2\ST\}^{-1}\cdot \mathfrak{Kh}_{\rm adj}^{9_2}&=q^{-56} \ST^{-34}(q^{50} \ST^{33}+q^{48} \ST^{32}+2 q^{46} \ST^{31}+q^{46} \ST^{30}+q^{44} \ST^{30}+2 q^{44} \ST^{29}+2 q^{42} \ST^{29}+4 q^{42} \ST^{28}+\\ &+q^{40} \ST^{28}+3 q^{40} \ST^{27}+q^{40} \ST^{26}+2 q^{38} \ST^{27}+4 q^{38} \ST^{26}+3 q^{38} \ST^{25}+3 q^{36} \ST^{25}+4 q^{36} \ST^{24}+q^{34} \ST^{25}+ \\ &+3 q^{34} \ST^{24}+4 q^{34} \ST^{23}+q^{34} \ST^{22}+2 q^{32} \ST^{23}+4 q^{32} \ST^{22}+3 q^{32} \ST^{21}+3 q^{30} \ST^{21}+4 q^{30} \ST^{20}+2 q^{28} \ST^{20}+\\ &+4 q^{28} \ST^{19}+q^{28} \ST^{18}+2 q^{26} \ST^{19}+4 q^{26} \ST^{18}+3 q^{26} \ST^{17}+q^{24} \ST^{18}+3 q^{24} \ST^{17}+4 q^{24} \ST^{16}+q^{22} \ST^{16}+\\ &+3 q^{22} \ST^{15}+q^{22} \ST^{14}+2 q^{20} \ST^{14}+3 q^{20} \ST^{13}+2 q^{18} \ST^{13}+4 q^{18} \ST^{12}+q^{16} \ST^{12}+3 q^{16} \ST^{11}+q^{16} \ST^{10}+\\ &+q^{14} \ST^{10}+2 q^{14} \ST^9+2 q^{12} \ST^8+2 q^{10} \ST^7+q^{10} \ST^6+q^8 \ST^6+2 q^8 \ST^5+q^6 \ST^4+q^2 \ST+1)
\end{aligned}\nn
\end{equation}
The adjoint Khovanov polynomials for $k>0$ with ${\rm Sh}_{tw_{k>0}}=0\,$:
 \begin{equation}\label{KhAdjTwistNeg}
\begin{aligned}
    4_1=tw_{2}:\quad {\color{red}\Kh^{4_1}_{\rm adj}}&=\begin{cases}
        q^{-12} \ST^{-7}(q^{24} \ST^{15}+q^{22} \ST^{14}-q^{20} \ST^{13}-q^{18} \ST^{12}+q^{18} \ST^{11}+2 q^{16} \ST^{10}+q^{16} \ST^9+q^{14} \ST^8-\\ \underline{{\color{red}- q^{12} \ST^8}- q^{12} \ST^8+q^{12} \ST^7}-q^{10} \ST^7-q^8 \ST^6-2 q^8 \ST^5-q^6 \ST^4+q^6 \ST^3+q^4 \ST^2-q^2 \ST-1),\quad \n \geq 0 \\
        q^{-12} \ST^{-7}(q^{24} \ST^{15}+q^{22} \ST^{14}-q^{20} \ST^{13}-q^{18} \ST^{12}+q^{18} \ST^{11}+2 q^{16} \ST^{10}+q^{16} \ST^9+q^{14} \ST^8-\\ \underline{-q^{12} \ST^8+ q^{12} \ST^7{\color{red}+ q^{12} \ST^7}}-q^{10} \ST^7-q^8 \ST^6-2 q^8 \ST^5-q^6 \ST^4+q^6 \ST^3+q^4 \ST^2-q^2 \ST-1), \quad \n < 0
    \end{cases}\\
    \{q^2\ST\}^{-1}\cdot\mathfrak{Kh}_{\rm adj}^{4_1}&=q^{-10} \ST^{-6}(q^{20} \ST^{13}+q^{18} \ST^{12}+q^{14} \ST^9+2 q^{12} \ST^8+q^{12} \ST^7+q^{10} \ST^7+q^{10} \ST^6+q^8 \ST^6+2 q^8 \ST^5+q^6 \ST^4+q^2 \ST+1) \\
    6_1=tw_{4}:\quad {\color{red}\Kh^{6_1}_{\rm adj}}&=\begin{cases}
        q^{-12} \ST^{-7}(q^{36} \ST^{23}+q^{34} \ST^{22}-q^{32} \ST^{21}-q^{30} \ST^{20}+q^{30} \ST^{19}+2 q^{28} \ST^{18}+q^{28} \ST^{17}+\\ +2 q^{26} \ST^{16}-2 q^{24} \ST^{16}+q^{24} \ST^{15}-q^{22} \ST^{15}+q^{22} \ST^{13}-q^{20} \ST^{13}+3 q^{20} \ST^{12}+q^{20} \ST^{11}-\\ -2 q^{18} \ST^{12}+3 q^{18} \ST^{11}+q^{18} \ST^{10}-q^{16} \ST^{11}+q^{16} \ST^{10}+2 q^{16} \ST^9-3 q^{14} \ST^9+q^{14} \ST^8-\\ \underline{{\color{red}- q^{12} \ST^8}- 3q^{12} \ST^8}-q^{10} \ST^7+q^{10} \ST^5-q^8 \ST^6-3 q^8 \ST^5+q^8 \ST^4-2 q^6 \ST^4-q^2 \ST-1),\quad \n\geq 0 \\
        q^{-12} \ST^{-7}(q^{36} \ST^{23}+q^{34} \ST^{22}-q^{32} \ST^{21}-q^{30} \ST^{20}+q^{30} \ST^{19}+2 q^{28} \ST^{18}+q^{28} \ST^{17}+\\ +2 q^{26} \ST^{16}-2 q^{24} \ST^{16}+q^{24} \ST^{15}-q^{22} \ST^{15}+q^{22} \ST^{13}-q^{20} \ST^{13}+3 q^{20} \ST^{12}+q^{20} \ST^{11}-\\ -2 q^{18} \ST^{12}+3 q^{18} \ST^{11}+q^{18} \ST^{10}-q^{16} \ST^{11}+q^{16} \ST^{10}+2 q^{16} \ST^9-3 q^{14} \ST^9+q^{14} \ST^8-\\ \underline{-3 q^{12} \ST^8{\color{red}+q^{12} \ST^7}}-q^{10} \ST^7+q^{10} \ST^5-q^8 \ST^6-3 q^8 \ST^5+q^8 \ST^4-2 q^6 \ST^4-q^2 \ST-1),\quad \n<0
    \end{cases}\\
    \{q^2\ST\}^{-1}\cdot\mathfrak{Kh}_{\rm adj}^{6_1}&= q^{-10} \ST^{-6}(q^{32} \ST^{21}+q^{30} \ST^{20}+q^{26} \ST^{17}+2 q^{24} \ST^{16}+q^{24} \ST^{15}+q^{22} \ST^{15}+2 q^{22} \ST^{14}+2 q^{20} \ST^{13}+\\ &+2 q^{18} \ST^{12}+q^{18} \ST^{11}+q^{16} \ST^{11}+3 q^{16} \ST^{10}+q^{16} \ST^9+4 q^{14} \ST^9+q^{14} \ST^8+4 q^{12} \ST^8+3 q^{12} \ST^7+\\ &+q^{10} \ST^7+2 q^{10} \ST^6+q^8 \ST^6+3 q^8 \ST^5+2 q^6 \ST^4+q^6 \ST^3+q^4 \ST^2+q^2 \ST+1)  \nn
\end{aligned}
\end{equation}
\begin{equation}
\begin{aligned}
    8_1=tw_{6}:\quad {\color{red}\Kh^{8_1}_{\rm adj}}&=\begin{cases}
        q^{-12} \ST^{-7}(q^{48} \ST^{31}+q^{46} \ST^{30}-q^{44} \ST^{29}-q^{42} \ST^{28}+q^{42} \ST^{27}+2 q^{40} \ST^{26}+q^{40} \ST^{25}+2 q^{38} \ST^{24}-\\ -2 q^{36} \ST^{24}+q^{36} \ST^{23}-q^{34} \ST^{23}+q^{34} \ST^{21}-q^{32} \ST^{21}+3 q^{32} \ST^{20}+q^{32} \ST^{19}-2 q^{30} \ST^{20}+\\ +3 q^{30} \ST^{19}+2 q^{30} t
        \ST^{18}-q^{28} \ST^{19}+q^{28} \ST^{17}-3 q^{26} \ST^{17}+q^{26} \ST^{16}+q^{26} \ST^{15}-3 q^{24} \ST^{16}+\\ +2 q^{24} t
        \ST^{15}+3 q^{24} \ST^{14}+q^{24} \ST^{13}-q^{22} \ST^{15}+3 q^{22} \ST^{13}+q^{22} \ST^{12}-3 q^{20} \ST^{13}+2 q^{20} \ST^{12}+\\ +2 q^{20} \ST^{11}-3 q^{18} \ST^{12}+q^{18} \ST^{10}-q^{16} \ST^{11}-q^{16} \ST^{10}+q^{16} \ST^9-3 q^{14} \ST^9+q^{14} \ST^8+q^{14} \ST^7-\\ \underline{-3 q^{12} \ST^8{\color{red}-q^{12} \ST^8}-q^{12} \ST^7}+q^{12} \ST^6-q^{10} \ST^7-q^{10} \ST^6-q^8 \ST^6-3 q^8 \ST^5-2 q^6 \ST^4-q^2 \ST-1),\quad \n \geq 0 \\
        q^{-12} \ST^{-7}(q^{48} \ST^{31}+q^{46} \ST^{30}-q^{44} \ST^{29}-q^{42} \ST^{28}+q^{42} \ST^{27}+2 q^{40} \ST^{26}+q^{40} \ST^{25}+2 q^{38} \ST^{24}-\\ -2 q^{36} \ST^{24}+q^{36} \ST^{23}-q^{34} \ST^{23}+q^{34} \ST^{21}-q^{32} \ST^{21}+3 q^{32} \ST^{20}+q^{32} \ST^{19}-2 q^{30} \ST^{20}+\\ +3 q^{30} \ST^{19}+2 q^{30} \ST^{18}-q^{28} \ST^{19}+q^{28} \ST^{17}-3 q^{26} \ST^{17}+q^{26} \ST^{16}+q^{26} \ST^{15}-3 q^{24} \ST^{16}+\\ +2 q^{24} \ST^{15}+3 q^{24} \ST^{14}+q^{24} \ST^{13}-q^{22} \ST^{15}+3 q^{22} \ST^{13}+q^{22} \ST^{12}-3 q^{20} \ST^{13}+2 q^{20} \ST^{12}+\\ +2 q^{20} \ST^{11}-3 q^{18} \ST^{12}+q^{18} \ST^{10}-q^{16} \ST^{11}-q^{16} \ST^{10}+q^{16} \ST^9-3 q^{14} \ST^9+q^{14} \ST^8+q^{14} \ST^7-\\ \underline{- 3q^{12} \ST^8{\color{red} + q^{12} \ST^7}- q^{12} \ST^7}+q^{12} \ST^6-q^{10} \ST^7-q^{10} \ST^6-q^8 \ST^6-3 q^8 \ST^5-2 q^6 \ST^4-q^2 \ST-1),\quad  \n<0
    \end{cases} \\
    \{q^2 \ST\}^{-1}\cdot\mathfrak{Kh}_{\rm adj}^{8_1}&=q^{-10} \ST^{-6}(q^{44} \ST^{29}+q^{42} \ST^{28}+q^{38} \ST^{25}+2 q^{36} \ST^{24}+q^{36} \ST^{23}+q^{34} \ST^{23}+2 q^{34} \ST^{22}+2 q^{32} \ST^{21}+\\ &+2 q^{30} \ST^{20}+q^{30} \ST^{19}+q^{28} \ST^{19}+3 q^{28} \ST^{18}+q^{28} \ST^{17}+4 q^{26} \ST^{17}+2 q^{26} \ST^{16}+3 q^{24} \ST^{16}+\\ &+2 q^{24} \ST^{15}+q^{22} \ST^{15}+3 q^{22} \ST^{14}+q^{22} \ST^{13}+4 q^{20} \ST^{13}+3 q^{20} \ST^{12}+q^{20} \ST^{11}+3 q^{18} \ST^{12}+4 q^{18} \ST^{11}+\\ &+q^{18} \ST^{10}+q^{16} \ST^{11}+5 q^{16} \ST^{10}+3 q^{16} \ST^9+4 q^{14} \ST^9+2 q^{14} \ST^8+4 q^{12} \ST^8+4 q^{12} \ST^7+q^{10} \ST^7+\\ &+3 q^{10} \ST^6+q^{10} \ST^5+q^8 \ST^6+3 q^8 \ST^5+q^8 \ST^4+2 q^6 \ST^4+q^6 \ST^3+q^4 \ST^2+q^2 \ST+1)  \\
    10_1=tw_{8}:\quad {\color{red}\Kh^{10_1}_{\rm adj}}&=\begin{cases}
        q^{-12} \ST^{-7}(q^{60} \ST^{39}+q^{58} \ST^{38}-q^{56} \ST^{37}-q^{54} \ST^{36}+q^{54} \ST^{35}+2 q^{52} \ST^{34}+q^{52} \ST^{33}+\\ +2 q^{50} \ST^{32}-2 q^{48} \ST^{32}+q^{48} \ST^{31}-q^{46} \ST^{31}+q^{46} \ST^{29}-q^{44} \ST^{29}+3 q^{44} \ST^{28}+q^{44} \ST^{27}-\\ -2 q^{42} \ST^{28}+3 q^{42} \ST^{27}+2 q^{42} \ST^{26}-q^{40} \ST^{27}+q^{40} \ST^{25}-3 q^{38} \ST^{25}+q^{38} \ST^{24}+q^{38} \ST^{23}-\\ -3 q^{36} \ST^{24}+2 q^{36} \ST^{23}+3 q^{36} \ST^{22}+q^{36} \ST^{21}-q^{34} \ST^{23}+3 q^{34} \ST^{21}+2 q^{34} \ST^{20}-3 q^{32} \ST^{21}+\\ +q^{32} \ST^{20}+q^{32} \ST^{19}-3 q^{30} \ST^{20}+q^{30} \ST^{18}+q^{30} \ST^{17}-q^{28} \ST^{19}-q^{28} \ST^{18}+2 q^{28} \ST^{17}+\\ +3 q^{28} \ST^{16}+q^{28} \ST^{15}-3 q^{26} \ST^{17}+q^{26} \ST^{16}+3 q^{26} \ST^{15}+q^{26} \ST^{14}-3 q^{24} \ST^{16}+2 q^{24} \ST^{14}+\\ +2 q^{24} \ST^{13}-q^{22} \ST^{15}-q^{22} \ST^{14}+q^{22} \ST^{12}-3 q^{20} \ST^{13}+q^{20} \ST^{11}-3 q^{18} \ST^{12}+q^{18} \ST^{10}+\\ +q^{18} \ST^9-q^{16} \ST^{11}-q^{16} \ST^{10}+q^{16} \ST^8-3 q^{14}
       \ST^9\underline{-3 q^{12} \ST^8{\color{red}-q^{12} \ST^8}-q^{12} \ST^7}-q^{10} \ST^7-\\ -q^{10} \ST^6 -q^8 \ST^6-3 q^8 \ST^5-2 q^6 \ST^4-q^2 \ST-1),\quad \n\geq 0 \\
        q^{-12} \ST^{-7}(q^{60} \ST^{39}+q^{58} \ST^{38}-q^{56} \ST^{37}-q^{54} \ST^{36}+q^{54} \ST^{35}+2 q^{52} \ST^{34}+q^{52} \ST^{33}+\\ +2 q^{50} \ST^{32}-2 q^{48} \ST^{32}+q^{48} \ST^{31}-q^{46} \ST^{31}+q^{46} \ST^{29}-q^{44} \ST^{29}+3 q^{44} \ST^{28}+q^{44} \ST^{27}-\\ -2 q^{42} \ST^{28}+3 q^{42} \ST^{27}+2 q^{42} \ST^{26}-q^{40} \ST^{27}+q^{40} \ST^{25}-3 q^{38} \ST^{25}+q^{38} \ST^{24}+q^{38} \ST^{23}-\\ -3 q^{36} \ST^{24}+2 q^{36} \ST^{23}+3 q^{36} \ST^{22}+q^{36} \ST^{21}-q^{34} \ST^{23}+3 q^{34} \ST^{21}+2 q^{34} \ST^{20}-3 q^{32} \ST^{21}+\\ +q^{32} \ST^{20}+q^{32} \ST^{19}-3 q^{30} \ST^{20}+q^{30} \ST^{18}+q^{30} \ST^{17}-q^{28} \ST^{19}-q^{28} \ST^{18}+2 q^{28} \ST^{17}+\\ +3 q^{28} \ST^{16}+q^{28} \ST^{15}-3 q^{26} \ST^{17}+q^{26} \ST^{16}+3 q^{26} \ST^{15}+q^{26} \ST^{14}-3 q^{24} \ST^{16}+2 q^{24} \ST^{14}+\\ +2 q^{24} \ST^{13}-q^{22} \ST^{15}-q^{22} \ST^{14}+q^{22} \ST^{12}-3 q^{20} \ST^{13}+q^{20} \ST^{11}-3 q^{18} \ST^{12}+q^{18} \ST^{10}+\\ +q^{18} \ST^9-q^{16} \ST^{11}-q^{16} \ST^{10}+q^{16} \ST^8-3 q^{14} \ST^9\underline{-3q^{12} \ST^8{\color{red} + q^{12} \ST^7}- q^{12} \ST^7}-q^{10} \ST^7-\\ -q^{10} \ST^6-q^8 \ST^6 -3 q^8 \ST^5-2 q^6 \ST^4-q^2 \ST-1),\quad \n<0
    \end{cases} \\
    \{q^2 \ST\}^{-1}\cdot\mathfrak{Kh}_{\rm adj}^{10_1}&=q^{-10} \ST^{-6}(q^{56} \ST^{37}+q^{54} \ST^{36}+q^{50} \ST^{33}+2 q^{48} \ST^{32}+q^{48} \ST^{31}+q^{46} \ST^{31}+2 q^{46} \ST^{30}+2 q^{44} \ST^{29}+\\ &+2 q^{42} \ST^{28}+q^{42} \ST^{27}+q^{40} \ST^{27}+3 q^{40} \ST^{26}+q^{40} \ST^{25}+4 q^{38} \ST^{25}+2 q^{38} \ST^{24}+3 q^{36} \ST^{24}+\\ &+2 q^{36} \ST^{23}+q^{34} \ST^{23}+3 q^{34} \ST^{22}+q^{34} \ST^{21}+4 q^{32} \ST^{21}+3 q^{32} \ST^{20}+q^{32} \ST^{19}+3 q^{30} \ST^{20}+\\ &+4 q^{30} \ST^{19}+2 q^{30} \ST^{18}+q^{28} \ST^{19}+4 q^{28} \ST^{18}+2 q^{28} \ST^{17}+4 q^{26} \ST^{17}+3 q^{26} \ST^{16}+q^{26} \ST^{15}+\\ &+3 q^{24} \ST^{16}+4 q^{24} \ST^{15}+3 q^{24} \ST^{14}+q^{24} \ST^{13}+q^{22} \ST^{15}+4 q^{22} \ST^{14}+4 q^{22} \ST^{13}+q^{22} \ST^{12}+\\ &+4 q^{20} \ST^{13}+5 q^{20} \ST^{12}+3 q^{20} \ST^{11}+3 q^{18} \ST^{12}+4 q^{18} \ST^{11}+2 q^{18} \ST^{10}+q^{16} \ST^{11}+5 q^{16} \ST^{10}+\\ &+4 q^{16} \ST^9+4 q^{14} \ST^9+3 q^{14} \ST^8+q^{14} \ST^7+4 q^{12} \ST^8+4 q^{12} \ST^7+q^{12} \ST^6+q^{10} \ST^7+3 q^{10} \ST^6+\\ &+q^{10} \ST^5+q^8 \ST^6+3 q^8 \ST^5+q^8 \ST^4+2 q^6 \ST^4+q^6 \ST^3+q^4 \ST^2+q^2 \ST+1)
\end{aligned} \nn
\end{equation}

\subsection{$\mathcal{K}=tor_k$}
Calculate $\Kh^{{\cal S}_{tor_n}({\mathcal{K})}}$ for $\mathcal{K}=tor_k$. For $k > 0$, we get $\text{Sh}_{tor_{k>0}}=\frac{3}{2}(k-1)$ and the following adjoint Khovanov
polynomials:
\begin{equation}\label{ExT[2,2p+1]}
\begin{aligned}
    3_1=tor_3:\quad {\color{red} \Kh^{3_1}_{\rm adj}}&=\begin{cases}
        \ST^{-11}q^{-22}(q^{18} \ST^{11}+q^{12} \ST^7-q^{10}
        \ST^7{\color{red}-q^{10}\ST^6}-q^8 \ST^5+q^6 \ST^3+q^4 \ST^2-q^2 \ST-1),\quad \n\geq -6 \\
        \ST^{-11}q^{-22}(q^{18} \ST^{11}+q^{12} \ST^7-q^{10} \ST^7{\color{red}
        + q^{10}\ST^5}-q^8 \ST^5+q^6 \ST^3+q^4 \ST^2-q^2 \ST-1),\quad \n < -6
    \end{cases} \\
    \{q^2\ST\}^{-1}\cdot\mathfrak{Kh}_{\rm adj}^{3_1} &= q^{-20} \ST^{-10}(q^{14} \ST^9+q^{10} \ST^7+q^8 \ST^5+q^2 \ST+1) \\
    5_1=tor_5:\quad {\color{red} \Kh^{5_1}_{\rm adj}}&=\begin{cases}
        \ST^{-19}q^{-38}(q^{30} \ST^{19}+q^{24} \ST^{15}-q^{22} \ST^{15}+q^{22}
        \ST^{13}-q^{20} \ST^{13}+q^{18} \ST^{11}+q^{16} \ST^{10}+q^{16} \ST^9-\\ -2 q^{14} \ST^9{\color{red}
        - q^{14}\ST^8}-q^{12} \ST^8+q^{10} \ST^6+q^{10} \ST^5-q^8 \ST^5+q^8 \ST^4-q^6 \ST^4-q^2 \ST-1),\quad
        \n \geq -12 \\
        \ST^{-19}q^{-38}(q^{30} \ST^{19}+q^{24} \ST^{15}-q^{22} \ST^{15}+q^{22} \ST^{13}-q^{20}
        \ST^{13}+q^{18} \ST^{11}+q^{16} \ST^{10}+q^{16} \ST^9-\\ -2 q^{14} \ST^9{\color{red}
        +q^{14}\ST^7}-q^{12} \ST^8+q^{10} \ST^6+q^{10} \ST^5-q^8 \ST^5+q^8 \ST^4-q^6 \ST^4-q^2 \ST-1),\quad
      \n < -12
    \end{cases}\\
    \{q^2 \ST\}^{-1}\cdot \mathfrak{Kh}_{\rm adj}^{5_1}&=q^{-36} \ST^{-18}(q^{26} \ST^{17}+q^{22} \ST^{15}+q^{20} \ST^{13}+q^{18} \ST^{11}+2 q^{14} \ST^9+q^{12} \ST^8+q^{12} \ST^7+q^8 \ST^5+q^6 \ST^4+\\ &+q^6 \ST^3+q^4 \ST^2+q^2 \ST+1) \\
    7_1=tor_7:\quad {\color{red} \Kh^{7_1}_{\rm adj}}&=\begin{cases}
        \ST^{-27}q^{-54}(q^{42} \ST^{27}+q^{36} \ST^{23}-q^{34} \ST^{23}+q^{34}
        \ST^{21}-q^{32} \ST^{21}+q^{30} \ST^{19}+q^{28} \ST^{18}+q^{28} \ST^{17}-\\ -2 q^{26} \ST^{17}+q^{26} \ST^{15}-q^{24}
        \ST^{16}+q^{22} \ST^{14}+q^{22} \ST^{13}-q^{20} \ST^{13}+q^{20} \ST^{12}+q^{20} \ST^{11}-q^{18} \ST^{12}-\\ -q^{18}
        \ST^{11}{\color{red} -q^{18}\ST^{10}}-q^{14} \ST^9+q^{14} \ST^8+q^{14} \ST^7-q^{12} \ST^8+q^{12}
        \ST^6-q^8 \ST^5-q^6 \ST^4-q^2 \ST-1),\; \n \geq -18 \\
        \ST^{-27}q^{-54}(q^{42} \ST^{27}+q^{36} \ST^{23}-q^{34} \ST^{23}+q^{34} \ST^{21}-q^{32}
        \ST^{21}+q^{30} \ST^{19}+q^{28} \ST^{18}+q^{28} \ST^{17}-\\ -2 q^{26} \ST^{17}+q^{26} \ST^{15}-q^{24} \ST^{16}+q^{22}
        \ST^{14}+q^{22} \ST^{13}-q^{20} \ST^{13}+q^{20} \ST^{12}+q^{20} \ST^{11}-q^{18} \ST^{12}-\\ -q^{18}
        \ST^{11}{\color{red}+q^{18}\ST^9}-q^{14} \ST^9+q^{14} \ST^8+q^{14} \ST^7-q^{12} \ST^8+q^{12} \ST^6-q^8
        \ST^5-q^6 \ST^4-q^2 \ST-1),\; \n < -18
    \end{cases}\\
    \{q^2\ST\}^{-1}\cdot \mathfrak{Kh}_{\rm adj}^{7_1}&=q^{-52} \ST^{-26}(q^{38} \ST^{25}+q^{34} \ST^{23}+q^{32} \ST^{21}+q^{30} \ST^{19}+2 q^{26} \ST^{17}+q^{24} \ST^{16}+q^{24} \ST^{15}+q^{22} \ST^{13}+\\ &+q^{20} \ST^{13}+q^{18} \ST^{12}+2 q^{18} \ST^{11}+q^{16} \ST^{10}+q^{16} \ST^9+q^{14} \ST^9+q^{12} \ST^8+q^{12} \ST^7+q^{10} \ST^6+q^{10} \ST^5+\\ &+q^8 \ST^5+q^8 \ST^4+q^6 \ST^4+q^6 \ST^3+q^4 \ST^2+q^2 \ST+1) \\
    9_1=tor_9:\quad {\color{red}\Kh^{9_1}_{\rm adj}}&=\begin{cases}
        \ST^{-35} q^{-70}(q^{54} \ST^{35}+q^{48} \ST^{31}-q^{46} \ST^{31}+q^{46}
        \ST^{29}-q^{44} \ST^{29}+q^{42} \ST^{27}+q^{40} \ST^{26}+q^{40} \ST^{25}-\\ -2 q^{38} \ST^{25}+q^{38} \ST^{23}-q^{36}
        \ST^{24}+q^{34} \ST^{22}+q^{34} \ST^{21}-q^{32} \ST^{21}+q^{32} \ST^{20}+q^{32} \ST^{19}-q^{30} \ST^{20}-\\ -q^{30}
        \ST^{19}+q^{30} \ST^{17}-q^{26} \ST^{17}+q^{26} \ST^{16}+q^{26} \ST^{15}-q^{24} \ST^{16}+q^{24} \ST^{14}+q^{24} \ST^{13}-q^{22}
        \ST^{13}-\\ {\color{red} -q^{22}\ST^{12}}-q^{20} \ST^{13}-q^{18} \ST^{12}+q^{18} \ST^{10}+q^{18}
        \ST^9+q^{16} \ST^8-q^{14} \ST^9-q^{12} \ST^8-q^8 \ST^5-\\ -q^6 \ST^4-q^2 \ST-1),\quad \n \geq -24 \\
        \ST^{-35} q^{-70}(q^{54} \ST^{35}+q^{48} \ST^{31}-q^{46} \ST^{31}+q^{46} \ST^{29}-q^{44}
        \ST^{29}+q^{42} \ST^{27}+q^{40} \ST^{26}+q^{40} \ST^{25}-\\ -2 q^{38} \ST^{25}+q^{38} \ST^{23}-q^{36} \ST^{24}+q^{34}
        \ST^{22}+q^{34} \ST^{21}-q^{32} \ST^{21}+q^{32} \ST^{20}+q^{32} \ST^{19}-q^{30} \ST^{20}-\\ -q^{30} \ST^{19}+q^{30}
        \ST^{17}-q^{26} \ST^{17}+q^{26} \ST^{16}+q^{26} \ST^{15}-q^{24} \ST^{16}+q^{24} \ST^{14}+q^{24} \ST^{13}-q^{22} \ST^{13}+\\
        {\color{red} +q^{22}\ST^{11}}-q^{20} \ST^{13}-q^{18} \ST^{12}+q^{18} \ST^{10}+q^{18} \ST^9+q^{16}
        \ST^8-q^{14} \ST^9-q^{12} \ST^8-q^8 \ST^5-\\ -q^6 \ST^4-q^2 \ST-1),\quad \n < -24   \nn
    \end{cases}\\
    \{q^2\ST\}^{-1}\cdot \mathfrak{Kh}_{\rm adj}^{9_1}&=q^{-68} \ST^{-34}(q^{50} \ST^{33}+q^{46} \ST^{31}+q^{44} \ST^{29}+q^{42} \ST^{27}+2 q^{38} \ST^{25}+q^{36} \ST^{24}+q^{36} \ST^{23}+q^{34} \ST^{21}+\\ &+q^{32} \ST^{21}+q^{30} \ST^{20}+2 q^{30} \ST^{19}+q^{28} \ST^{18}+q^{28} \ST^{17}+q^{26} \ST^{17}+q^{26} \ST^{15}+q^{24} \ST^{16}+q^{24} \ST^{15}+\\ &+q^{22} \ST^{14}+2 q^{22} \ST^{13}+q^{20} \ST^{13}+q^{20} \ST^{12}+q^{20} \ST^{11}+q^{18} \ST^{12}+q^{18} \ST^{11}+q^{16} \ST^{10}+q^{16} \ST^9+\\ &+q^{14} \ST^9+q^{14} \ST^8+q^{14} \ST^7+q^{12} \ST^8+q^{12} \ST^7+q^{12} \ST^6+q^{10} \ST^6+q^{10} \ST^5+q^8 \ST^5+q^8 \ST^4+q^6 \ST^4+\\ &+q^6 \ST^3+q^4 \ST^2+q^2 \ST+1)
\end{aligned}
\end{equation}
For $k < 0$, we get $\text{Sh}_{tor_{k<0}}=\frac{3}{2}(k+1)$ and the following adjoint Khovanov
polynomials:
\begin{equation}\label{KhAdjTorusNeg}
\begin{aligned}
   \overline{3_1}=tor_{-3}:\quad {\color{red} \Kh^{\overline{3_1}}_{\rm adj}(q,\ST)}&=\begin{cases}
   q^4\ST \left(q^{18} \ST^{11}+q^{16} \ST^{10}-q^{14} \ST^9-q^{12} \ST^8+q^{10} \ST^6{\color{red}-q^8\ST^6}+q^8 \ST^4-q^6 \ST^4-1\right),\, \n \geq 6 \\
    q^4 \ST \left(q^{18} \ST^{11}+q^{16} \ST^{10}-q^{14} \ST^9-q^{12} \ST^8+q^{10} \ST^6{\color{red}+q^8\ST^5}+q^8 \ST^4-q^6 \ST^4-1\right),\, \n < 6 \\
    \end{cases}\\
    \{q^2\ST\}^{-1}\cdot \mathfrak{Kh}_{\rm adj}^{\overline{3_1}}(q,\ST)&=q^6 \ST^2 \left(q^{14} \ST^9+q^{12} \ST^8+q^6 \ST^4+q^4 \ST^2+1\right) \\
    {\color{red} \Kh^{\overline{3_1}}_{\rm adj}(q^{-1},\ST^{-1})}&=\begin{cases}
        \ST^{-11}q^{-22}(q^{18} \ST^{11}+q^{12} \ST^7-q^{10} \ST^7{\color{red}+q^{10}\ST^5}-q^8 \ST^5+q^6 \ST^3+q^4 \ST^2-q^2 \ST-1),\, \n \geq 6 \\
        \ST^{-11}q^{-22}(q^{18} \ST^{11}+q^{12} \ST^7-q^{10} \ST^7{\color{red}-q^{10}\ST^6}-q^8 \ST^5+q^6 \ST^3+q^4 \ST^2-q^2 \ST-1),\, \n < 6
    \end{cases} \\
    \{q^2\ST\}^{-1}\cdot \mathfrak{Kh}_{\rm adj}^{\overline{3_1}}(q^{-1},\ST^{-1})&=q^{-20} \ST^{-11}(q^{14} \ST^9+q^{10} \ST^7+q^8 \ST^5+q^2 \ST+1) \\
    \overline{5_1}=tor_{-5}:\quad {\color{red} \Kh^{\overline{5_1}}_{\rm adj}}&=\begin{cases}
        q^8\ST (q^{30} \ST^{19}+q^{28} \ST^{18}+q^{24} \ST^{15}-q^{22} \ST^{15}+q^{22} \ST^{14}-q^{20} \ST^{14}-q^{20} \ST^{13}+q^{18} \ST^{11}-\\ {\color{red}-q^{16}\ST^{12}}+2 q^{16} \ST^{10}-q^{14} \ST^{10}-q^{14} \ST^9-q^{12} \ST^8+q^{10} \ST^6-q^8 \ST^6+q^8 \ST^4-q^6 \ST^4-1),\; \n \geq 12 \\
        q^8\ST (q^{30} \ST^{19}+q^{28} \ST^{18}+q^{24} \ST^{15}-q^{22} \ST^{15}+q^{22} \ST^{14}-q^{20} \ST^{14}-q^{20} \ST^{13}+q^{18} \ST^{11}+ \\ {\color{red}+q^{16}\ST^{11}}+2 q^{16} \ST^{10}-q^{14} \ST^{10}-q^{14} \ST^9-q^{12} \ST^8+q^{10} \ST^6-q^8 \ST^6+q^8 \ST^4-q^6 \ST^4-1),\; \n < 12
    \end{cases} \\
    \{q^2\ST\}^{-1}\cdot \mathfrak{Kh}_{\rm adj}^{\overline{5_1}}&=q^{10} \ST^2 (q^{26} \ST^{17}+q^{24} \ST^{16}+q^{22} \ST^{15}+q^{20} \ST^{14}+q^{20} \ST^{13}+q^{18} \ST^{12}+q^{14} \ST^{10}+q^{14} \ST^9+\\ &+2 q^{12} \ST^8+q^8 \ST^6+q^6 \ST^4+q^4 \ST^2+1) \\
    \overline{7_1}=tor_{-7}:\quad {\color{red} \Kh^{\overline{7_1}}_{\rm adj}}&=\begin{cases}
        q^{12}\ST (q^{42} \ST^{27}+q^{40} \ST^{26}+q^{36} \ST^{23}+q^{34} \ST^{22}-q^{30} \ST^{21}+q^{30} \ST^{19}-q^{28} \ST^{20}-q^{28} \ST^{19}+\\ +q^{28} \ST^{18}{\color{red}-q^{24}\ST^{18}}+q^{24} \ST^{16}+q^{24} \ST^{15}-q^{22} \ST^{16}-q^{22} \ST^{15}+q^{22} \ST^{14}-q^{20} \ST^{14}-\\ -q^{20} \ST^{13}+q^{18} \ST^{11}-q^{16} \ST^{12}+2 q^{16} \ST^{10}-q^{14} \ST^{10}-q^{14} \ST^9-q^{12} \ST^8+q^{10} \ST^6-\\ -q^8 \ST^6+q^8 \ST^4-q^6 \ST^4-1),\quad \n \geq 18 \\
        q^{12}\ST (q^{42} \ST^{27}+q^{40} \ST^{26}+q^{36} \ST^{23}+q^{34} \ST^{22}-q^{30} \ST^{21}+q^{30} \ST^{19}-q^{28} \ST^{20}-q^{28} \ST^{19}+\\ +q^{28} \ST^{18}{\color{red}+q^{24}\ST^{17}} +q^{24} \ST^{16}+q^{24} \ST^{15}-q^{22} \ST^{16}-q^{22} \ST^{15}+q^{22} \ST^{14}-q^{20} \ST^{14}-\\ -q^{20} \ST^{13}+q^{18} \ST^{11}-q^{16} \ST^{12}+2 q^{16} \ST^{10}-q^{14} \ST^{10}-q^{14} \ST^9-q^{12} \ST^8+q^{10} \ST^6-\\ -q^8 \ST^6+q^8 \ST^4-q^6 \ST^4-1), \quad \n < 18
    \end{cases}\\
    \{q^2\ST\}^{-1}\cdot \mathfrak{Kh}_{\rm adj}^{\overline{7_1}}&=q^{14} \ST^2 (q^{38} \ST^{25}+q^{36} \ST^{24}+q^{34} \ST^{23}+q^{32} \ST^{22}+q^{32} \ST^{21}+q^{30} \ST^{21}+q^{30} \ST^{20}+q^{28} \ST^{20}+\\ &+q^{28} \ST^{19}+q^{26} \ST^{18}+q^{26} \ST^{17}+q^{24} \ST^{16}+q^{22} \ST^{16}+q^{22} \ST^{15}+2 q^{20} \ST^{14}+q^{20} \ST^{13}+\\ &+q^{18} \ST^{12}+q^{16} \ST^{12}+q^{14} \ST^{10}+q^{14} \ST^9+2 q^{12} \ST^8+q^8 \ST^6+q^6 \ST^4+q^4 \ST^2+1) \\
    \overline{9_1}=tor_{-9}:\quad {\color{red} \Kh^{\overline{9_1}}_{\rm adj}}&=\begin{cases}
        q^{16}\ST (q^{54} \ST^{35}+q^{52} \ST^{34}+q^{48} \ST^{31}+q^{46} \ST^{30}+q^{42} \ST^{27}+q^{40} \ST^{26}-q^{38} \ST^{27}-q^{36} \ST^{26}-\\ -q^{36} \ST^{25}+q^{36} \ST^{23}+q^{34} \ST^{22}{\color{red}-q^{32}\ST^{24}} +q^{32} \ST^{22}-q^{30} \ST^{22}-q^{30} \ST^{21}+q^{30} \ST^{19}-\\ -q^{28} \ST^{20}-q^{28} \ST^{19}+q^{28} \ST^{18}-q^{24} \ST^{18}+q^{24} \ST^{16}+q^{24} \ST^{15}-q^{22} \ST^{16}-q^{22} \ST^{15}+\\ +q^{22} \ST^{14}-q^{20} \ST^{14}-q^{20} \ST^{13}+q^{18} \ST^{11}-q^{16} \ST^{12}+2 q^{16} \ST^{10}-q^{14} \ST^{10}-q^{14} \ST^9-\\ -q^{12} \ST^8+q^{10} \ST^6-q^8 \ST^6+q^8 \ST^4-q^6 \ST^4-1), \quad \n \geq 24 \\ \nn
        q^{16}\ST (q^{54} \ST^{35}+q^{52} \ST^{34}+q^{48} \ST^{31}+q^{46} \ST^{30}+q^{42} \ST^{27}+q^{40} \ST^{26}-q^{38} \ST^{27}-q^{36} \ST^{26}-\\ -q^{36} \ST^{25}+q^{36} \ST^{23}+q^{34} \ST^{22}{\color{red}+q^{32}\ST^{23}}+q^{32} \ST^{22}-q^{30} \ST^{22}-q^{30} \ST^{21}+q^{30} \ST^{19}-\\ -q^{28} \ST^{20}-q^{28} \ST^{19}+q^{28} \ST^{18}-q^{24} \ST^{18}+q^{24} \ST^{16}+q^{24} \ST^{15}-q^{22} \ST^{16}-q^{22} \ST^{15}+\\ +q^{22} \ST^{14}-q^{20} \ST^{14}-q^{20} \ST^{13}+q^{18} \ST^{11}-q^{16} \ST^{12}+2 q^{16} \ST^{10}-q^{14} \ST^{10}-q^{14} \ST^9-\\ -q^{12} \ST^8+q^{10} \ST^6-q^8 \ST^6+q^8 \ST^4-q^6 \ST^4-1), \quad \n < 24
    \end{cases}\\
    \{q^2\ST\}^{-1}\cdot \mathfrak{Kh}_{\rm adj}^{\overline{9_1}}&=q^{18} \ST^2 (q^{50} \ST^{33}+q^{48} \ST^{32}+q^{46} \ST^{31}+q^{44} \ST^{30}+q^{44} \ST^{29}+q^{42} \ST^{29}+q^{42} \ST^{28}+q^{40} \ST^{28}+\\ &+q^{40} \ST^{27}+q^{38} \ST^{27}+q^{38} \ST^{26}+q^{38} \ST^{25}+q^{36} \ST^{26}+q^{36} \ST^{25}+q^{36} \ST^{24}+q^{34} \ST^{24}+q^{34} \ST^{23}+\\ &+q^{32} \ST^{22}+q^{32} \ST^{21}+q^{30} \ST^{22}+q^{30} \ST^{21}+q^{30} \ST^{20}+2 q^{28} \ST^{20}+q^{28} \ST^{19}+q^{26} \ST^{18}+q^{26} \ST^{17}+\\ &+q^{24} \ST^{18}+q^{24} \ST^{16}+q^{22} \ST^{16}+q^{22} \ST^{15}+2 q^{20} \ST^{14}+q^{20} \ST^{13}+q^{18} \ST^{12}+q^{16} \ST^{12}+\\ &+q^{14} \ST^{10}+q^{14} \ST^9+2 q^{12} \ST^8+q^8 \ST^6+q^6 \ST^4+q^4 \ST^2+1)
\end{aligned}
\end{equation}

\subsection{Other knots}
For non-torus and non-twist knots, we get the following adjoint Khovanov
polynomials:
\begin{equation}
\begin{aligned}
    6_2:\quad {\color{red} \Kh_{\rm adj}^{6_2}}&=\begin{cases}
        q^{-28} \ST^{-15}(q^{36} \ST^{23}+q^{34} \ST^{22}-q^{32} \ST^{21}-q^{30} \ST^{20}+2 q^{30} \ST^{19}+3 q^{28} \ST^{18}+2 q^{28} \ST^{17}-\\ -q^{26} \ST^{17}+2 q^{26} \ST^{16}-2 q^{24} \ST^{16}+3 q^{24} \ST^{15}-q^{22} \ST^{15}+2 q^{22} \ST^{14}+2 q^{22} \ST^{13}-q^{20} \ST^{14}-\\ -3 q^{20} \ST^{13}+4 q^{20} \ST^{12}+q^{20} \ST^{11}-4 q^{18} \ST^{12}+3 q^{18} \ST^{11}+q^{18} \ST^{10}-2 q^{16} \ST^{11}\underline{+ 2 q^{16} \ST^9-}\\ \underline{{\color{red} -q^{16} \ST^{10}}+q^{16} \ST^{10}}-3 q^{14} \ST^9+3 q^{14} \ST^8-5 q^{12} \ST^8+q^{12} \ST^7-2 q^{10} \ST^7-q^{10} \ST^6+q^{10} \ST^5-3 q^8 \ST^5+2 q^8 \ST^4-\\ -3 q^6 \ST^4+q^6 \ST^3-q^4 \ST^3-q^4 \ST^2-2 q^2 \ST-1),\quad \n \geq -6 \\
        q^{-28} \ST^{-15}(q^{36} \ST^{23}+q^{34} \ST^{22}-q^{32} \ST^{21}-q^{30} \ST^{20}+2 q^{30} \ST^{19}+3 q^{28} \ST^{18}+2 q^{28} \ST^{17}-\\ -q^{26} \ST^{17}+2 q^{26} \ST^{16}-2 q^{24} \ST^{16}+3 q^{24} \ST^{15}-q^{22} \ST^{15}+2 q^{22} \ST^{14}+2 q^{22} \ST^{13}-q^{20} \ST^{14}-\\ -3 q^{20} \ST^{13}+4 q^{20} \ST^{12}+q^{20} \ST^{11}-4 q^{18} \ST^{12}+3 q^{18} \ST^{11}+q^{18} \ST^{10}-2 q^{16} \ST^{11}\underline{{\color{red}+ q^{16} \ST^{9}}+}\\ \underline{+2 q^{16} \ST^9+ q^{16} \ST^{10}}-3 q^{14} \ST^9+3 q^{14} \ST^8-5 q^{12} \ST^8+q^{12} \ST^7-2 q^{10} \ST^7-q^{10} \ST^6+q^{10} \ST^5-3 q^8 \ST^5+\\ +2 q^8 \ST^4-3 q^6 \ST^4+q^6 \ST^3-q^4 \ST^3-q^4 \ST^2-2 q^2 \ST-1),\quad \n < -6
    \end{cases}\\
    \{q^2\ST\}^{-1}\cdot \mathfrak{Kh}_{\rm adj}^{6_2}&=q^{-26} \ST^{-14}(q^{32} \ST^{21}+q^{30} \ST^{20}+2 q^{26} \ST^{17}+3 q^{24} \ST^{16}+2 q^{24} \ST^{15}+q^{22} \ST^{15}+2 q^{22} \ST^{14}+q^{20} \ST^{14}+\\ &+5 q^{20} \ST^{13}+4 q^{18} \ST^{12}+2 q^{18} \ST^{11}+2 q^{16} \ST^{11}+4 q^{16} \ST^{10}+q^{16} \ST^9+5 q^{14} \ST^9+q^{14} \ST^8+5 q^{12} \ST^8+\\ &+3 q^{12} \ST^7+2 q^{10} \ST^7+4 q^{10} \ST^6+4 q^8 \ST^5+3 q^6 \ST^4+q^6 \ST^3+q^4 \ST^3+2 q^4 \ST^2+2 q^2 \ST+1) \\
    6_3:\quad {\color{red} \Kh_{\rm adj}^{6_3}}&=\begin{cases}
        q^{-18} \ST^{-11}(q^{36} \ST^{23}+2 q^{34} \ST^{22}+q^{32} \ST^{20}-2 q^{30} \ST^{20}+3 q^{30} \ST^{19}-q^{28} \ST^{19}+4 q^{28} \ST^{18}+\\ +q^{28} \ST^{17}+q^{26} \ST^{17}+4 q^{26} \ST^{16}-4 q^{24} \ST^{16}+5 q^{24} \ST^{15}-3 q^{22} \ST^{15}+4 q^{22} \ST^{14}+2 q^{22} \ST^{13}-\\ -q^{20} \ST^{14}-q^{20} \ST^{13}+5 q^{20} \ST^{12}-q^{18} \ST^{13}\underline{-6 q^{18} \ST^{12}{\color{red}-q^{18} \ST^{12}}+6 q^{18} \ST^{11}}+q^{18} \ST^{10}-5 q^{16} \ST^{11}+\\ +q^{16} \ST^{10}+q^{16} \ST^9-2 q^{14} \ST^{10}-4 q^{14} \ST^9+3 q^{14} \ST^8-5 q^{12} \ST^8+4 q^{12} \ST^7-4 q^{10} \ST^7-q^{10} \ST^6-\\ -q^8 \ST^6-4 q^8 \ST^5+q^8 \ST^4-3 q^6 \ST^4+2 q^6 \ST^3-q^4 \ST^3-2 q^2 \ST-1),\quad \n \geq 0 \\
        q^{-18} \ST^{-11}(q^{36} \ST^{23}+2 q^{34} \ST^{22}+q^{32} \ST^{20}-2 q^{30} \ST^{20}+3 q^{30} \ST^{19}-q^{28} \ST^{19}+4 q^{28} \ST^{18}+\\ +q^{28} \ST^{17}+q^{26} \ST^{17}+4 q^{26} \ST^{16}-4 q^{24} \ST^{16}+5 q^{24} \ST^{15}-3 q^{22} \ST^{15}+4 q^{22} \ST^{14}+2 q^{22} \ST^{13}-\\ -q^{20} \ST^{14}-q^{20} \ST^{13}+5 q^{20} \ST^{12}-q^{18} \ST^{13}\underline{-6 q^{18} \ST^{12}{\color{red}+q^{18} \ST^{11}}+6 q^{18} \ST^{11}}+q^{18} \ST^{10}-5 q^{16} \ST^{11}+\\ +q^{16} \ST^{10}+q^{16} \ST^9-2 q^{14} \ST^{10}-4 q^{14} \ST^9+3 q^{14} \ST^8-5 q^{12} \ST^8+4 q^{12} \ST^7-4 q^{10} \ST^7-\\ -q^{10} \ST^6-q^8 \ST^6-4 q^8 \ST^5+q^8 \ST^4-3 q^6 \ST^4+2 q^6 \ST^3-q^4 \ST^3-2 q^2 \ST-1),\quad \n < 0
    \end{cases}\\
    \{q^2\ST\}^{-1}\cdot \mathfrak{Kh}_{\rm adj}^{6_3}&=q^{-16} \ST^{-10}\left(q^8 \ST^6+q^4 \ST^3+1\right) (q^{24} \ST^{15}+2 q^{22} \ST^{14}+q^{20} \ST^{13}+q^{18} \ST^{11}+4 q^{16} \ST^{10}+\\ &+4 q^{14} \ST^9+q^{14} \ST^8+q^{12} \ST^8+q^{12} \ST^7+q^{10} \ST^7+4 q^{10} \ST^6+4 q^8 \ST^5+q^6 \ST^4+q^4 \ST^2+2 q^2 \ST+1) \\
    \overline{7_3}:\quad {\color{red} \Kh_{\rm adj}^{\overline{7_3}}}&=\begin{cases}
        q^8\ST (q^{42} \ST^{27}+q^{40} \ST^{26}+2 q^{36} \ST^{23}-q^{34} \ST^{23}+3 q^{34} \ST^{22}+q^{34} \ST^{21}-q^{32} \ST^{22}+2 q^{32} \ST^{20}-\\ -q^{30} \ST^{20}+4 q^{30} \ST^{19}-q^{28} \ST^{19}+5 q^{28} \ST^{18}+q^{28} \ST^{17}-2 q^{26} \ST^{18}-q^{26} \ST^{17}+3 q^{26} \ST^{16}-\\ -q^{24} \ST^{17}-5 q^{24} \ST^{16}+4 q^{24} \ST^{15}-3 q^{22} \ST^{15}+4 q^{22} \ST^{14}+q^{22} \ST^{13}-2 q^{20} \ST^{14}+4 q^{20} \ST^{12}-\\ -q^{18} \ST^{13}-5 q^{18} \ST^{12}+3 q^{18} \ST^{11}\underline{{\color{red}-q^{16} \ST^{12}}-4 q^{16} \ST^{11}}+q^{16} \ST^{10}-2 q^{14} \ST^{10}-2 q^{14} \ST^9+\\ +2 q^{14} \ST^8-q^{12} \ST^9-4 q^{12} \ST^8+3 q^{12} \ST^7+q^{12} \ST^6-2 q^{10} \ST^7-q^8 \ST^6-2 q^8 \ST^5+q^8 \ST^4-\\ -2 q^6 \ST^4+q^6 \ST^3-q^4 \ST^3-q^4 \ST^2-q^2 \ST-1),\quad \n \geq 12 \\
        q^8\ST (q^{42} \ST^{27}+q^{40} \ST^{26}+2 q^{36} \ST^{23}-q^{34} \ST^{23}+3 q^{34} \ST^{22}+q^{34} \ST^{21}-q^{32} \ST^{22}+2 q^{32} \ST^{20}-\\ -q^{30} \ST^{20}+4 q^{30} \ST^{19}-q^{28} \ST^{19}+5 q^{28} \ST^{18}+q^{28} \ST^{17}-2 q^{26} \ST^{18}-q^{26} \ST^{17}+3 q^{26} \ST^{16}-\\ -q^{24} \ST^{17}-5 q^{24} \ST^{16}+4 q^{24} \ST^{15}-3 q^{22} \ST^{15}+4 q^{22} \ST^{14}+q^{22} \ST^{13}-2 q^{20} \ST^{14}+4 q^{20} \ST^{12}-\\ -q^{18} \ST^{13}-5 q^{18} \ST^{12}+3 q^{18} \ST^{11}\underline{{\color{red}+ q^{16} \ST^{11}}-4q^{16} \ST^{11}}+q^{16} \ST^{10}-2 q^{14} \ST^{10}-2 q^{14} \ST^9+2 q^{14} \ST^8-\\ -q^{12} \ST^9-4 q^{12} \ST^8+3 q^{12} \ST^7+q^{12} \ST^6-2 q^{10} \ST^7-q^8 \ST^6-2 q^8 \ST^5+q^8 \ST^4-2 q^6 \ST^4+\\ +q^6 \ST^3-q^4 \ST^3-q^4 \ST^2-q^2 \ST-1),\quad \n < 12
    \end{cases} \\ \nn
    \{q^2\ST\}^{-1}\cdot \mathfrak{Kh}_{\rm adj}^{\overline{7_3}}&=q^{10} \ST^2 (q^{38} \ST^{25}+q^{36} \ST^{24}+q^{34} \ST^{23}+q^{32} \ST^{22}+2 q^{32} \ST^{21}+3 q^{30} \ST^{20}+q^{30} \ST^{19}+2 q^{28} \ST^{19}+\\ &+2 q^{28} \ST^{18}+2 q^{26} \ST^{18}+5 q^{26} \ST^{17}+q^{24} \ST^{17}+7 q^{24} \ST^{16}+q^{24} \ST^{15}+4 q^{22} \ST^{15}+3 q^{22} \ST^{14}+2 q^{20} \ST^{14}+\\ &+5 q^{20} \ST^{13}+q^{18} \ST^{13}+7 q^{18} \ST^{12}+q^{18} \ST^{11}+5 q^{16} \ST^{11}+4 q^{16} \ST^{10}+2 q^{14} \ST^{10}+4 q^{14} \ST^9+q^{12} \ST^9+\\ &+5 q^{12} \ST^8+2 q^{10} \ST^7+2 q^{10} \ST^6+q^8 \ST^6+3 q^8 \ST^5+q^8 \ST^4+2 q^6 \ST^4+q^4 \ST^3+2 q^4 \ST^2+q^2 \ST+1) \\
\end{aligned}
\end{equation}
\begin{equation}
\begin{aligned}
\overline{7_4}:\quad {\color{red} \Kh_{\rm adj}^{\overline{7_4}}}&=\begin{cases}
        q^4\ST (q^{42} \ST^{27}+q^{40} \ST^{26}-q^{38} \ST^{25}-q^{36} \ST^{24}+2 q^{36} \ST^{23}+3 q^{34} \ST^{22}+2 q^{34} \ST^{21}-q^{32} \ST^{21}+\\ +3 q^{32} \ST^{20}-3 q^{30} \ST^{20}+2 q^{30} \ST^{19}-q^{28} \ST^{19}+2 q^{28} \ST^{18}+2 q^{28} \ST^{17}-2 q^{26} \ST^{17}+\\ +6 q^{26} \ST^{16}+q^{26} \ST^{15}-5 q^{24} \ST^{16}+5 q^{24} \ST^{15}+3 q^{24} \ST^{14}-2 q^{22} \ST^{15}+3 q^{22} \ST^{13}-4 q^{20} \ST^{13}+\\ +4 q^{20} \ST^{12}-6 q^{18} \ST^{12}+4 q^{18} \ST^{11}+3 q^{18} \ST^{10}-3 q^{16} \ST^{11}-2 q^{16} \ST^{10}+5 q^{16} \ST^9+q^{16} \ST^8-\\ -6 q^{14} \ST^9+3 q^{14} \ST^8-5 q^{12} \ST^8+3 q^{12} \ST^7+2 q^{12} \ST^6-2 q^{10} \ST^7-2 q^{10} \ST^6+2 q^{10} \ST^5\underline{{\color{red}-q^8 \ST^6}-}\\ \underline{-6 q^8 \ST^5}+q^8 \ST^4-4 q^6 \ST^4-2 q^4 \ST^3-3 q^4 \ST^2-2 q^2 \ST-1),\quad \n \geq 6 \\
        q^4\ST (q^{42} \ST^{27}+q^{40} \ST^{26}-q^{38} \ST^{25}-q^{36} \ST^{24}+2 q^{36} \ST^{23}+3 q^{34} \ST^{22}+2 q^{34} \ST^{21}-q^{32} \ST^{21}+\\ +3 q^{32} \ST^{20}-3 q^{30} \ST^{20}+2 q^{30} \ST^{19}-q^{28} \ST^{19}+2 q^{28} \ST^{18}+2 q^{28} \ST^{17}-2 q^{26} \ST^{17}+6 q^{26} \ST^{16}+\\ +q^{26} \ST^{15}-5 q^{24} \ST^{16}+5 q^{24} \ST^{15}+3 q^{24} \ST^{14}-2 q^{22} \ST^{15}+3 q^{22} \ST^{13}-4 q^{20} \ST^{13}+4 q^{20} \ST^{12}-\\ -6 q^{18} \ST^{12}+4 q^{18} \ST^{11}+3 q^{18} \ST^{10}-3 q^{16} \ST^{11}-2 q^{16} \ST^{10}+5 q^{16} \ST^9+q^{16} \ST^8-6 q^{14} \ST^9+\\ +3 q^{14} \ST^8-5 q^{12} \ST^8+3 q^{12} \ST^7+2 q^{12} \ST^6-2 q^{10} \ST^7-2 q^{10} \ST^6+2 q^{10} \ST^5\underline{{\color{red}+ q^8 \ST^5}-6 q^8 \ST^5}+q^8 \ST^4-\\ -4 q^6 \ST^4-2 q^4 \ST^3-3 q^4 \ST^2-2 q^2 \ST-1),\quad \n < 6
    \end{cases}\\
    \{q^2\ST\}^{-1}\cdot \mathfrak{Kh}_{\rm adj}^{\overline{7_4}}&=q^6 \ST^2 (q^{38} \ST^{25}+q^{36} \ST^{24}+2 q^{32} \ST^{21}+3 q^{30} \ST^{20}+2 q^{30} \ST^{19}+q^{28} \ST^{19}+3 q^{28} \ST^{18}+4 q^{26} \ST^{17}+\\ &+5 q^{24} \ST^{16}+2 q^{24} \ST^{15}+2 q^{22} \ST^{15}+6 q^{22} \ST^{14}+q^{22} \ST^{13}+7 q^{20} \ST^{13}+3 q^{20} \ST^{12}+6 q^{18} \ST^{12}+4 q^{18} \ST^{11}+\\ &+3 q^{16} \ST^{11}+7 q^{16} \ST^{10}+8 q^{14} \ST^9+3 q^{14} \ST^8+5 q^{12} \ST^8+5 q^{12} \ST^7+q^{12} \ST^6+2 q^{10} \ST^7+6 q^{10} \ST^6+\\ &+8 q^8 \ST^5+3 q^8 \ST^4+4 q^6 \ST^4+2 q^6 \ST^3+2 q^4 \ST^3+4 q^4 \ST^2+2 q^2 \ST+1) \\
    7_5:\quad {\color{red} \Kh_{\rm adj}^{7_5}}&=\begin{cases}
        q^{-50} \ST^{-27}(q^{42} \ST^{27}+q^{40} \ST^{26}+q^{38} \ST^{25}+q^{38} \ST^{24}-q^{36} \ST^{24}+3 q^{36} \ST^{23}-q^{34} \ST^{23}+4 q^{34} \ST^{22}+\\ +2 q^{34} \ST^{21}+3 q^{32} \ST^{20}-q^{30} \ST^{21}-4 q^{30} \ST^{20}+7 q^{30} \ST^{19}+q^{30} \ST^{18}-3 q^{28} \ST^{19}+5 q^{28} \ST^{18}+\\ +3 q^{28} \ST^{17}-q^{26} \ST^{18}-3 q^{26} \ST^{17}\underline{+6 q^{26} \ST^{16}{\color{red} -q^{26} \ST^{16}}}-6 q^{24} \ST^{16}+8 q^{24} \ST^{15}+q^{24} \ST^{14}-6 q^{22} \ST^{15}+\\ +3 q^{22} \ST^{14}+4 q^{22} \ST^{13}-2 q^{20} \ST^{14}-5 q^{20} \ST^{13}+7 q^{20} \ST^{12}-8 q^{18} \ST^{12}+7 q^{18} \ST^{11}+q^{18} \ST^{10}-\\ -6 q^{16} \ST^{11}+q^{16} \ST^{10}+3 q^{16} \ST^9-2 q^{14} \ST^{10}-7 q^{14} \ST^9+4 q^{14} \ST^8-8 q^{12} \ST^8+3 q^{12} \ST^7-4 q^{10} \ST^7-\\ -q^{10} \ST^6+q^{10} \ST^5-q^8 \ST^6-5 q^8 \ST^5+2 q^8 \ST^4-4 q^6 \ST^4+q^6 \ST^3-q^4 \ST^3-q^4 \ST^2-2 q^2 \ST-1),\quad \n \geq -12 \\
        q^{-50} \ST^{-27}(q^{42} \ST^{27}+q^{40} \ST^{26}+q^{38} \ST^{25}+q^{38} \ST^{24}-q^{36} \ST^{24}+3 q^{36} \ST^{23}-q^{34} \ST^{23}+4 q^{34} \ST^{22}+\\ +2 q^{34} \ST^{21}+3 q^{32} \ST^{20}-q^{30} \ST^{21}-4 q^{30} \ST^{20}+7 q^{30} \ST^{19}+q^{30} \ST^{18}-3 q^{28} \ST^{19}+5 q^{28} \ST^{18}+\\ +3 q^{28} \ST^{17}-q^{26} \ST^{18}-3 q^{26} \ST^{17}\underline{+ 6 q^{26} \ST^{16}{\color{red}+q^{26} \ST^{15}}}-6 q^{24} \ST^{16}+8 q^{24} \ST^{15}+q^{24} \ST^{14} -6 q^{22} \ST^{15}+\\ +3 q^{22} \ST^{14}+4 q^{22} \ST^{13}-2 q^{20} \ST^{14}-5 q^{20} \ST^{13}+7 q^{20} \ST^{12}-8 q^{18} \ST^{12}+7 q^{18} \ST^{11} +q^{18} \ST^{10}-\\ -6 q^{16} \ST^{11}+q^{16} \ST^{10}+3 q^{16} \ST^9-2 q^{14} \ST^{10}-7 q^{14} \ST^9+4 q^{14} \ST^8-8 q^{12} \ST^8+3 q^{12} \ST^7-4 q^{10} \ST^7-q^{10} \ST^6+\\ +q^{10} \ST^5-q^8 \ST^6-5 q^8 \ST^5+2 q^8 \ST^4-4 q^6 \ST^4+q^6 \ST^3-q^4 \ST^3-q^4 \ST^2 -2 q^2 \ST-1),\quad \n < -12
    \end{cases}\\
    \{q^2\ST\}^{-1}\cdot\mathfrak{Kh}_{\rm adj}^{7_5}&=q^{-48} \ST^{-26}(q^{38} \ST^{25}+q^{36} \ST^{24}+2 q^{34} \ST^{23}+q^{34} \ST^{22}+3 q^{32} \ST^{21}+q^{30} \ST^{21}+5 q^{30} \ST^{20}+2 q^{30} \ST^{19}+\\ &+3 q^{28} \ST^{19}+3 q^{28} \ST^{18}+q^{26} \ST^{18}+9 q^{26} \ST^{17}+q^{26} \ST^{16}+8 q^{24} \ST^{16}+3 q^{24} \ST^{15}+6 q^{22} \ST^{15}+\\ &+7 q^{22} \ST^{14}+2 q^{20} \ST^{14}+11 q^{20} \ST^{13}+q^{20} \ST^{12}+10 q^{18} \ST^{12}+4 q^{18} \ST^{11}+6 q^{16} \ST^{11}+8 q^{16} \ST^{10}+\\ &+2 q^{14} \ST^{10}+11 q^{14} \ST^9+q^{14} \ST^8+9 q^{12} \ST^8+3 q^{12} \ST^7+4 q^{10} \ST^7+5 q^{10} \ST^6+q^8 \ST^6+6 q^8 \ST^5+4 q^6 \ST^4+\\ &+q^6 \ST^3+q^4 \ST^3+2 q^4 \ST^2+2 q^2 \ST+1) \\
    7_6:\quad {\color{red} \Kh_{\rm adj}^{7_6}}&=\begin{cases}
        q^{-34} \ST^{-19}(q^{42} \ST^{27}+2 q^{40} \ST^{26}+q^{38} \ST^{24}-2 q^{36} \ST^{24}+4 q^{36} \ST^{23}-q^{34} \ST^{23}+5 q^{34} \ST^{22}+\\ +2 q^{34} \ST^{21}+5 q^{32} \ST^{20}-4 q^{30} \ST^{20}+9 q^{30} \ST^{19}+q^{30} \ST^{18}-4 q^{28} \ST^{19}+5 q^{28} \ST^{18}+4 q^{28} \ST^{17}-\\ -2 q^{26} \ST^{18}-3 q^{26} \ST^{17}+10 q^{26} \ST^{16}+q^{26} \ST^{15}-8 q^{24} \ST^{16}+10 q^{24} \ST^{15}+2 q^{24} \ST^{14}-7 q^{22} \ST^{15}+\\ \underline{+3 q^{22} \ST^{14}{\color{red}-q^{22} \ST^{14}}+5 q^{22} \ST^{13}}-2 q^{20} \ST^{14}-7 q^{20} \ST^{13}+8 q^{20} \ST^{12}-q^{18} \ST^{13}-13 q^{18} \ST^{12}+6 q^{18} \ST^{11}+\\ +q^{18} \ST^{10}-7 q^{16} \ST^{11}+4 q^{16} \ST^9-q^{14} \ST^{10}-7 q^{14} \ST^9+6 q^{14} \ST^8-9 q^{12} \ST^8+3 q^{12} \ST^7-5 q^{10} \ST^7-\\ -3 q^{10} \ST^6+q^{10} \ST^5-q^8 \ST^6-6 q^8 \ST^5+2 q^8 \ST^4-4 q^6 \ST^4+q^6 \ST^3-q^4 \ST^3-q^4 \ST^2-2 q^2 \ST-1),\quad \n \geq -6 \\
        q^{-34} \ST^{-19}(q^{42} \ST^{27}+2 q^{40} \ST^{26}+q^{38} \ST^{24}-2 q^{36} \ST^{24}+4 q^{36} \ST^{23}-q^{34} \ST^{23}+5 q^{34} \ST^{22}+\\ +2 q^{34} \ST^{21}+5 q^{32} \ST^{20}-4 q^{30} \ST^{20}+9 q^{30} \ST^{19}+q^{30} \ST^{18}-4 q^{28} \ST^{19}+5 q^{28} \ST^{18}+4 q^{28} \ST^{17}-\\ -2 q^{26} \ST^{18}-3 q^{26} \ST^{17}+10 q^{26} \ST^{16}+q^{26} \ST^{15}-8 q^{24} \ST^{16}+10 q^{24} \ST^{15}+2 q^{24} \ST^{14}-7 q^{22} \ST^{15}+\\ \underline{+3 q^{22} \ST^{14}{\color{red} +q^{22} \ST^{13}}+5 q^{22} \ST^{13}}-2 q^{20} \ST^{14}-7 q^{20} \ST^{13}+8 q^{20} \ST^{12}-q^{18} \ST^{13}-13 q^{18} \ST^{12}+6 q^{18} \ST^{11}+\\ +q^{18} \ST^{10}-7 q^{16} \ST^{11}+4 q^{16} \ST^9-q^{14} \ST^{10}-7 q^{14} \ST^9+6 q^{14} \ST^8-9 q^{12} \ST^8+3 q^{12} \ST^7-5 q^{10} \ST^7-\\ -3 q^{10} \ST^6+q^{10} \ST^5-q^8 \ST^6-6 q^8 \ST^5+2 q^8 \ST^4-4 q^6 \ST^4+q^6 \ST^3-q^4 \ST^3-q^4 \ST^2-2 q^2 \ST-1),\quad \n < -6
    \end{cases}\\ \nn
    \{q^2\ST\}^{-1}\cdot \mathfrak{Kh}_{\rm adj}^{7_6}&=q^{-32} \ST^{-18}(q^{38} \ST^{25}+2 q^{36} \ST^{24}+q^{34} \ST^{23}+q^{34} \ST^{22}+4 q^{32} \ST^{21}+6 q^{30} \ST^{20}+2 q^{30} \ST^{19}+4 q^{28} \ST^{19}+\\ &+5 q^{28} \ST^{18}+2 q^{26} \ST^{18}+11 q^{26} \ST^{17}+q^{26} \ST^{16}+10 q^{24} \ST^{16}+4 q^{24} \ST^{15}+8 q^{22} \ST^{15}+11 q^{22} \ST^{14}+\\ &+q^{22} \ST^{13}+2 q^{20} \ST^{14}+14 q^{20} \ST^{13}+2 q^{20} \ST^{12}+q^{18} \ST^{13}+14 q^{18} \ST^{12}+6 q^{18} \ST^{11}+7 q^{16} \ST^{11}+10 q^{16} \ST^{10}+\\ &+q^{14} \ST^{10}+12 q^{14} \ST^9+q^{14} \ST^8+10 q^{12} \ST^8+4 q^{12} \ST^7+5 q^{10} \ST^7+7 q^{10} \ST^6+q^8 \ST^6+7 q^8 \ST^5+4 q^6 \ST^4+\\ &+q^6 \ST^3+q^4 \ST^3+2 q^4 \ST^2+2 q^2 \ST+1) \\
\end{aligned}
\end{equation}
\begin{equation}
\begin{aligned}
7_7:\quad {\color{red} \Kh_{\rm adj}^{7_7}}&=\begin{cases}
        q^{-18} \ST^{-11}(q^{42} \ST^{27}+2 q^{40} \ST^{26}+q^{38} \ST^{24}-2 q^{36} \ST^{24}+4 q^{36} \ST^{23}-q^{34} \ST^{23}+6 q^{34} \ST^{22}+\\ +2 q^{34} \ST^{21}+q^{32} \ST^{21}+6 q^{32} \ST^{20}-6 q^{30} \ST^{20}+8 q^{30} \ST^{19}-5 q^{28} \ST^{19}+6 q^{28} \ST^{18}+4 q^{28} \ST^{17}-\\ -q^{26} \ST^{18}-q^{26} \ST^{17}+12 q^{26} \ST^{16}+q^{26} \ST^{15}-9 q^{24} \ST^{16}+13 q^{24} \ST^{15}+2 q^{24} \ST^{14}-8 q^{22} \ST^{15}+\\ +6 q^{22} \ST^{14}+7 q^{22} \ST^{13}-3 q^{20} \ST^{14}-8 q^{20} \ST^{13}+10 q^{20} \ST^{12}-q^{18} \ST^{13}\underline{-14 q^{18} \ST^{12}{\color{red}-q^{18} \ST^{12}}+9 q^{18} \ST^{11}}+\\ +2 q^{18} \ST^{10}-9 q^{16} \ST^{11}+q^{16} \ST^{10}+5 q^{16} \ST^9-4 q^{14} \ST^{10}-11 q^{14} \ST^9+7 q^{14} \ST^8-12 q^{12} \ST^8+\\ +4 q^{12} \ST^7-6 q^{10} \ST^7-3 q^{10} \ST^6+q^{10} \ST^5-q^8 \ST^6-7 q^8 \ST^5+3 q^8 \ST^4-6 q^6 \ST^4+2 q^6 \ST^3-\\ -2 q^4 \ST^3-2 q^4 \ST^2-3 q^2 \ST-1),\quad \n \geq 0 \\
        q^{-18} \ST^{-11}(q^{42} \ST^{27}+2 q^{40} \ST^{26}+q^{38} \ST^{24}-2 q^{36} \ST^{24}+4 q^{36} \ST^{23}-q^{34} \ST^{23}+6 q^{34} \ST^{22}+\\ +2 q^{34} \ST^{21}+q^{32} \ST^{21}+6 q^{32} \ST^{20}-6 q^{30} \ST^{20}+8 q^{30} \ST^{19}-5 q^{28} \ST^{19}+6 q^{28} \ST^{18}+4 q^{28} \ST^{17}-\\ -q^{26} \ST^{18}-q^{26} \ST^{17}+12 q^{26} \ST^{16}+q^{26} \ST^{15}-9 q^{24} \ST^{16}+13 q^{24} \ST^{15}+2 q^{24} \ST^{14}-8 q^{22} \ST^{15}+\\ +6 q^{22} \ST^{14}+7 q^{22} \ST^{13}-3 q^{20} \ST^{14}-8 q^{20} \ST^{13}+10 q^{20} \ST^{12}-q^{18} \ST^{13}\underline{-14 q^{18} \ST^{12}+}\\ \underline{{\color{red}+q^{18} \ST^{11}}+9 q^{18} \ST^{11}}+2 q^{18} \ST^{10}-9 q^{16} \ST^{11}+q^{16} \ST^{10}+5 q^{16} \ST^9-4 q^{14} \ST^{10}-11 q^{14} \ST^9+7 q^{14} \ST^8-\\ -12 q^{12} \ST^8+4 q^{12} \ST^7-6 q^{10} \ST^7-3 q^{10} \ST^6+q^{10} \ST^5-q^8 \ST^6-7 q^8 \ST^5+3 q^8 \ST^4-6 q^6 \ST^4+\\ +2 q^6 \ST^3-2 q^4 \ST^3-2 q^4 \ST^2-3 q^2 \ST-1),\quad \n < 0
    \end{cases} \\
    \{q^2\ST\}^{-1}\cdot \mathfrak{Kh}_{\rm adj}^{7_7}&=q^{-16} \ST^{-10}(q^{38} \ST^{25}+2 q^{36} \ST^{24}+q^{34} \ST^{23}+q^{34} \ST^{22}+4 q^{32} \ST^{21}+7 q^{30} \ST^{20}+2 q^{30} \ST^{19}+\\ &+5 q^{28} \ST^{19}+6 q^{28} \ST^{18}+q^{26} \ST^{18}+10 q^{26} \ST^{17}+12 q^{24} \ST^{16}+4 q^{24} \ST^{15}+9 q^{22} \ST^{15}+12 q^{22} \ST^{14}+\\ &+q^{22} \ST^{13}+3 q^{20} \ST^{14}+17 q^{20} \ST^{13}+2 q^{20} \ST^{12}+q^{18} \ST^{13}+18 q^{18} \ST^{12}+8 q^{18} \ST^{11}+9 q^{16} \ST^{11}+\\ &+12 q^{16} \ST^{10}+4 q^{14} \ST^{10}+17 q^{14} \ST^9+2 q^{14} \ST^8+13 q^{12} \ST^8+5 q^{12} \ST^7+6 q^{10} \ST^7+9 q^{10} \ST^6+\\ &+q^8 \ST^6+9 q^8 \ST^5+6 q^6 \ST^4+q^6 \ST^3+2 q^4 \ST^3+3 q^4 \ST^2+3 q^2 \ST+1) \\
{\color{blue} \mathbf{\overline{8_{19}}}}:\quad {\color{red} \Kh_{\rm adj}^{\overline{8_{19}}}}&=\begin{cases}
        q^{12}\ST (q^{34} \ST^{23}+q^{32} \ST^{22}+q^{28} \ST^{19}-q^{26} \ST^{19}+q^{26} \ST^{18}+q^{26} \ST^{17}-q^{24} \ST^{18}-q^{24} \ST^{17}+\\ +q^{24} \ST^{16}+q^{24} \ST^{15}{\color{red}-q^{20} \ST^{16}}+2 q^{20} \ST^{14}-q^{18} \ST^{14}-q^{18} \ST^{13}+q^{18} \ST^{12}+q^{18} \ST^{11}-\\ -2 q^{16} \ST^{12}-q^{16} \ST^{11}+q^{16} \ST^{10}-q^{14} \ST^9-q^{12} \ST^{10}-q^{10} \ST^8+q^{10} \ST^6-q^8 \ST^6+q^8 \ST^4-\\ -q^6 \ST^4-1),\; \n \geq {\color{blue} \mathbf{16}} \\
        q^{12}\ST (q^{34} \ST^{23}+q^{32} \ST^{22}+q^{28} \ST^{19}-q^{26} \ST^{19}+q^{26} \ST^{18}+q^{26} \ST^{17}-q^{24} \ST^{18}-q^{24} \ST^{17}+\\ +q^{24} \ST^{16}+q^{24} \ST^{15}{\color{red}+q^{20} \ST^{15}}+2 q^{20} \ST^{14}-q^{18} \ST^{14}-q^{18} \ST^{13}+q^{18} \ST^{12}+q^{18} \ST^{11}-\\ -2 q^{16} \ST^{12}-q^{16} \ST^{11}+q^{16} \ST^{10}-q^{14} \ST^9-q^{12} \ST^{10}-q^{10} \ST^8+q^{10} \ST^6-q^8 \ST^6+q^8 \ST^4-\\ -q^6 \ST^4-1),\; \n < {\color{blue} \mathbf{16}}
    \end{cases} \\
    \{q^2\ST\}^{-1}\cdot \mathfrak{Kh}_{\rm adj}^{\overline{8_{19}}}&=q^{14} \ST^2 (q^{30} \ST^{21}+q^{28} \ST^{20}+q^{26} \ST^{19}+q^{24} \ST^{18}+q^{24} \ST^{17}+q^{22} \ST^{16}+q^{22} \ST^{15}+q^{20} \ST^{14}+\\ &+q^{20} \ST^{13}+q^{18} \ST^{14}+q^{18} \ST^{13}+3 q^{16} \ST^{12}+q^{16} \ST^{11}+q^{14} \ST^{10}+q^{14} \ST^9+q^{12} \ST^{10}+\\ &+q^{12} \ST^8+q^{10} \ST^8+q^8 \ST^6+q^6 \ST^4+q^4 \ST^2+1) \\ \nn
    8_{20}:\quad {\color{red} \Kh^{8_{20}}_{\rm adj}}&=\begin{cases}
        q^{-30} \ST^{-19}(q^{36} \ST^{24}+q^{36} \ST^{23}+q^{34} \ST^{23}+2 q^{34} \ST^{22}-q^{32} \ST^{22}\underline{{\color{red}-q^{30} \ST^{20}}+q^{30} \ST^{20}+q^{30} \ST^{19}}+ \\ +q^{28} \ST^{19}
        +3 q^{28} \ST^{18}-q^{26} \ST^{19}-3 q^{26} \ST^{18}+q^{26} \ST^{17}+q^{26} \ST^{16}-2 q^{24} \ST^{17}-2 q^{24} \ST^{16}+\\ +q^{24} \ST^{15}+q^{22} \ST^{15}+3 q^{22} \ST^{14}+2 q^{20} \ST^{13}+q^{20} \ST^{12}-3 q^{18} \ST^{13}-4 q^{18} \ST^{12}+\\ +q^{18} \ST^{11}-q^{16} \ST^{12}-3 q^{16} \ST^{11}+q^{14} \ST^9+q^{14} \ST^8+2 q^{12} \ST^7-2 q^{10} \ST^7-q^8 \ST^6-\\ -2 q^8 \ST^5-q^6 \ST^4+q^6 \ST^3+q^4 \ST^2-q^2 \ST-1)\,,\quad n\geq 0 \\
        q^{-30} \ST^{-19}(q^{36} \ST^{24}+q^{36} \ST^{23}+q^{34} \ST^{23}+2 q^{34} \ST^{22}-q^{32} \ST^{22}+\underline{q^{30} \ST^{20}+q^{30} \ST^{19}{\color{red}+q^{30} \ST^{19}}}+\\ +q^{28} \ST^{19}+3 q^{28} \ST^{18}-q^{26} \ST^{19}-3 q^{26} \ST^{18}+q^{26} \ST^{17}+q^{26} \ST^{16}-2 q^{24} \ST^{17}-\\ -2 q^{24} \ST^{16}+q^{24} \ST^{15}+q^{22} \ST^{15}+3 q^{22} \ST^{14}+2 q^{20} \ST^{13}+q^{20} \ST^{12}-3 q^{18} \ST^{13}-\\ -4 q^{18} \ST^{12}+q^{18} \ST^{11}-q^{16} \ST^{12}-3 q^{16} \ST^{11}+q^{14} \ST^9+q^{14} \ST^8+2 q^{12} \ST^7-2 q^{10} \ST^7-\\ -q^8 \ST^6-2 q^8 \ST^5-q^6 \ST^4+q^6 \ST^3+q^4 \ST^2-q^2 \ST-1)\,,\quad n<0
    \end{cases} \\
    \{q^2\ST\}^{-1}\cdot \mathfrak{Kh}_{\rm adj}^{8_{20}}&=q^{-28} \ST^{-18}(q^{32} \ST^{22}+q^{32} \ST^{21}+q^{30} \ST^{21}+2 q^{30} \ST^{20}+q^{28} \ST^{19}+q^{26} \ST^{19}+3 q^{26} \ST^{18}+\\ &+q^{26} \ST^{17}+2 q^{24} \ST^{17}+3 q^{24} \ST^{16}+2 q^{22} \ST^{15}+q^{22} \ST^{14}+q^{20} \ST^{14}+q^{20} \ST^{13}+3 q^{18} \ST^{13}+\\ &+4 q^{18} \ST^{12}+q^{16} \ST^{12}+3 q^{16} \ST^{11}+q^{16} \ST^{10}+q^{14} \ST^9+q^{12} \ST^8+2 q^{10} \ST^7+q^{10} \ST^6+\\ &+q^8 \ST^6+2 q^8 \ST^5+q^6 \ST^4+q^2 \ST+1) 
    \end{aligned}
\end{equation}
\begin{equation}
\begin{aligned}
8_{21}:\quad {\color{red} \Kh_{\rm adj}^{8_{21}}}&=\begin{cases}
        q^{-40} \ST^{-23}(q^{38} \ST^{25}+q^{36} \ST^{24}+3 q^{36} \ST^{23}-q^{34} \ST^{23}+3 q^{34} \ST^{22}+3 q^{32} \ST^{21}+\\ +2 q^{32} \ST^{20}+6 q^{30} \ST^{19}-q^{28} \ST^{20}-5 q^{28} \ST^{19}\underline{{\color{red}-q^{28} \ST^{18}}+6 q^{28} \ST^{18}+q^{28} \ST^{17}}-\\ -3 q^{26} \ST^{18} +3 q^{26} \ST^{16}-q^{24} \ST^{17}-3 q^{24} \ST^{16}+8 q^{24} \ST^{15}-4 q^{22} \ST^{15}+7 q^{22} \ST^{14}+\\ +q^{22} \ST^{13} -5 q^{20} \ST^{14}-3 q^{20} \ST^{13}+3 q^{20} \ST^{12}-2 q^{18} \ST^{13}-7 q^{18} \ST^{12}+5 q^{18} \ST^{11}-\\ -5 q^{16} \ST^{11} +4 q^{16} \ST^{10}+q^{16} \ST^9-3 q^{14} \ST^{10}-2 q^{14} \ST^9+3 q^{14} \ST^8-q^{12} \ST^9-6 q^{12} \ST^8+\\ +3 q^{12} \ST^7-4 q^{10} \ST^7-q^8 \ST^6-3 q^8 \ST^5+q^8 \ST^4-3 q^6 \ST^4+2 q^6 \ST^3-q^4 \ST^3-2 q^2 \ST-1)\,,\quad n\geq -6 \\
        q^{-40} \ST^{-23}(q^{38} \ST^{25}+q^{36} \ST^{24}+3 q^{36} \ST^{23}-q^{34} \ST^{23}+3 q^{34} \ST^{22}+3 q^{32} \ST^{21}+\\ +2 q^{32} \ST^{20}+6 q^{30} \ST^{19}-q^{28} \ST^{20}-5 q^{28} \ST^{19}\underline{+6 q^{28} \ST^{18}+ q^{28} \ST^{17}{\color{red}+q^{28} \ST^{17}}}-\\ -3 q^{26} \ST^{18}+3 q^{26} \ST^{16}-q^{24} \ST^{17}-3 q^{24} \ST^{16}+8 q^{24} \ST^{15}-4 q^{22} \ST^{15}+\\ +7 q^{22} \ST^{14}+q^{22} \ST^{13}-5 q^{20} \ST^{14}-3 q^{20} \ST^{13}+3 q^{20} \ST^{12}-2 q^{18} \ST^{13}-\\ -7 q^{18} \ST^{12}+5 q^{18} \ST^{11}-5 q^{16} \ST^{11}+4 q^{16} \ST^{10}+q^{16} \ST^9-3 q^{14} \ST^{10}-\\ -2 q^{14} \ST^9+3 q^{14} \ST^8-q^{12} \ST^9-6 q^{12} \ST^8+3 q^{12} \ST^7-4 q^{10} \ST^7-q^8 \ST^6-\\ -3 q^8 \ST^5+q^8 \ST^4-3 q^6 \ST^4+2 q^6 \ST^3-q^4 \ST^3-2 q^2 \ST-1)\,,\quad n < -6 
        \end{cases} \\
        \{q^2\ST\}^{-1}\cdot \mathfrak{Kh}_{\rm adj}^{8_{21}}&=q^{-38} \ST^{-22}(q^{34} \ST^{23}+q^{32} \ST^{22}+3 q^{32} \ST^{21}+3 q^{30} \ST^{20}+q^{28} \ST^{20}+6 q^{28} \ST^{19}+\\ &+2 q^{28} \ST^{18}+3 q^{26} \ST^{18}+6 q^{26} \ST^{17}+q^{24} \ST^{17}+8 q^{24} \ST^{16}+q^{24} \ST^{15}+\\ &+6 q^{22} \ST^{15}+3 q^{22} \ST^{14}+5 q^{20} \ST^{14}+9 q^{20} \ST^{13}+2 q^{18} \ST^{13}+10 q^{18} \ST^{12}+\\ &+q^{18} \ST^{11}+6 q^{16} \ST^{11}+3 q^{16} \ST^{10}+3 q^{14} \ST^{10}+6 q^{14} \ST^9+q^{12} \ST^9+7 q^{12} \ST^8+\\ &+q^{12} \ST^7+4 q^{10} \ST^7+3 q^{10} \ST^6+q^8 \ST^6+4 q^8 \ST^5+3 q^6 \ST^4+q^4 \ST^3+q^4 \ST^2+2 q^2 \ST+1) \\
        {\color{blue} \mathbf{9_{42}}}:\quad {\color{red} \Kh_{\rm adj}^{9_{42}}}&=\begin{cases}
            q^{-20} \ST^{-15}(q^{40} \ST^{27}+q^{38} \ST^{26}-q^{36} \ST^{25}-q^{34} \ST^{24}+q^{34} \ST^{23}+\\ +2 q^{32} \ST^{22}+q^{32} \ST^{21}+q^{30} \ST^{21}+3 q^{30} \ST^{20}+q^{30} \ST^{19}-q^{28} \ST^{20}+\\ +2 q^{28} \ST^{19}+q^{28} \ST^{18}-2 q^{26} \ST^{19}-2 q^{26} \ST^{18}-q^{26} \ST^{17}-q^{24} \ST^{18}-\\ -2 q^{24} \ST^{17}+q^{24} \ST^{15}+4 q^{22} \ST^{15}+3 q^{22} \ST^{14}\underline{{\color{red}-q^{20} \ST^{16}}-q^{20} \ST^{15}}+\\ +2 q^{20} \ST^{14}+2 q^{20} \ST^{13}+q^{20} \ST^{12}-q^{18} \ST^{14}-4 q^{18} \ST^{13}-3 q^{18} \ST^{12}+\\ +q^{18} \ST^{11}-2 q^{16} \ST^{12}-2 q^{16} \ST^{11}+q^{14} \ST^{10}+q^{14} \ST^9+q^{14} \ST^8-q^{12} \ST^{10}-\\ -q^{12} \ST^9+2 q^{12} \ST^7-q^{10} \ST^8-2 q^{10} \ST^7-q^8 \ST^6-2 q^8 \ST^5-q^6 \ST^4+\\ +q^6 \ST^3+q^4 \ST^2-q^2 \ST-1)\,,\quad n\geq 0 \\
            q^{-20} \ST^{-15}(q^{40} \ST^{27}+q^{38} \ST^{26}-q^{36} \ST^{25}-q^{34} \ST^{24}+q^{34} \ST^{23}+2 q^{32} \ST^{22}+\\ +q^{32} \ST^{21}+q^{30} \ST^{21}+3 q^{30} \ST^{20}+q^{30} \ST^{19}-q^{28} \ST^{20}+2 q^{28} \ST^{19}+q^{28} \ST^{18}-\\ -2 q^{26} \ST^{19}-2 q^{26} \ST^{18}-q^{26} \ST^{17}-q^{24} \ST^{18}-2 q^{24} \ST^{17}+q^{24} \ST^{15}+\\ +4 q^{22} \ST^{15}+3 q^{22} \ST^{14}\underline{{\color{red}+q^{20}\ST^{15}}-q^{20}\ST^{15}}+2 q^{20} \ST^{14}+2 q^{20} \ST^{13}+q^{20} \ST^{12}-q^{18} \ST^{14}-\\ -4 q^{18} \ST^{13}-3 q^{18} \ST^{12}+q^{18} \ST^{11}-2 q^{16} \ST^{12}-2 q^{16} \ST^{11}+q^{14} \ST^{10}+\\ +q^{14} \ST^9+q^{14} \ST^8-q^{12} \ST^{10}-q^{12} \ST^9+2 q^{12} \ST^7-q^{10} \ST^8-2 q^{10} \ST^7-\\ -q^8 \ST^6-2 q^8 \ST^5-q^6 \ST^4+q^6 \ST^3+q^4 \ST^2-q^2 \ST-1)\,,\quad n < 0
        \end{cases} \\
        \{q^2\ST\}^{-1}\cdot \mathfrak{Kh}_{\rm adj}^{9_{42}}&=q^{-18} \ST^{-14}(q^{36} \ST^{25}+q^{34} \ST^{24}+q^{30} \ST^{21}+2 q^{28} \ST^{20}+q^{28} \ST^{19}+2 q^{26} \ST^{19}+\\ &+3 q^{26} \ST^{18}+q^{26} \ST^{17}+q^{24} \ST^{18}+3 q^{24} \ST^{17}+q^{24} \ST^{16}+q^{22} \ST^{16}+q^{20} \ST^{15}+q^{20} \ST^{14}+\\
        &+q^{20} \ST^{13}+q^{18} \ST^{14}+4 q^{18} \ST^{13}+3 q^{18} \ST^{12}+3 q^{16} \ST^{12}+3 q^{16} \ST^{11}+q^{16} \ST^{10}+\\ &+q^{14} \ST^9+q^{12} \ST^{10}+q^{12} \ST^9+q^{12} \ST^8+q^{10} \ST^8+2 q^{10} \ST^7+q^{10} \ST^6+q^8 \ST^6+2 q^8 \ST^5+\\ 
        &+q^6 \ST^4+q^2 \ST+1) \nn
    \end{aligned}
    \end{equation}
    \begin{equation}
    \begin{aligned}
    \overline{9_{43}}:\quad {\color{red} \Kh_{\rm adj}^{\overline{9_{43}}}}&=\begin{cases}
           q^{-4} \ST^{-7}(q^{42} \ST^{28}+q^{42} \ST^{27}+2 q^{40} \ST^{27}+2 q^{40} \ST^{26}+q^{38} \ST^{25}-2 q^{36} \ST^{25}+\\ +q^{36} \ST^{23}-q^{34} \ST^{24}+q^{34} \ST^{23}+3 q^{34} \ST^{22}+q^{32} \ST^{22}+4 q^{32} \ST^{21}+q^{32} \ST^{20}-\\ -2 q^{30} \ST^{21}+q^{30} \ST^{20}+2 q^{30} \ST^{19}\underline{-2 q^{28} \ST^{20}{\color{red}-q^{28} \ST^{20}}-q^{28} \ST^{19}}+4 q^{28} \ST^{18}-q^{26} \ST^{19}-\\ -2 q^{26} \ST^{18}+2 q^{26} \ST^{17}+q^{26} \ST^{16}-q^{24} \ST^{18}-4 q^{24} \ST^{17}-q^{24} \ST^{16}+3 q^{24} \ST^{15}-\\ -2 q^{22} \ST^{16}-2 q^{22} \ST^{15}+3 q^{22} \ST^{14}-2 q^{20} \ST^{14}+q^{20} \ST^{13}+2 q^{20} \ST^{12}-\\ -2 q^{18} \ST^{13}-2 q^{18} \ST^{12}+3 q^{18} \ST^{11}-2 q^{16} \ST^{12}-4 q^{16} \ST^{11}+2 q^{16} \ST^{10}+\\ +q^{16} \ST^9-2 q^{14} \ST^{10}-q^{14} \ST^9+q^{14} \ST^8-2 q^{12} \ST^8+2 q^{12} \ST^7-2 q^{10} \ST^7+\\ +q^{10} \ST^6-2 q^8 \ST^6-3 q^8 \ST^5-2 q^6 \ST^4+q^6 \ST^3+q^4 \ST^2-q^2 \ST-1)\,,\quad n \geq 12 \\
           q^{-4} \ST^{-7}(q^{42} \ST^{28}+q^{42} \ST^{27}+2 q^{40} \ST^{27}+2 q^{40} \ST^{26}+q^{38} \ST^{25}-2 q^{36} \ST^{25}+\\ +q^{36} \ST^{23}-q^{34} \ST^{24}+q^{34} \ST^{23}+3 q^{34} \ST^{22}+q^{32} \ST^{22}+4 q^{32} \ST^{21}+q^{32} \ST^{20}-\\ -2 q^{30} \ST^{21}+q^{30} \ST^{20}+2 q^{30} \ST^{19}\underline{-2 q^{28} \ST^{20}{\color{red} +q^{28} \ST^{19}}-q^{28} \ST^{19}}+4 q^{28} \ST^{18}-q^{26} \ST^{19}-\\ -2 q^{26} \ST^{18} +2 q^{26} \ST^{17}+q^{26} \ST^{16}-q^{24} \ST^{18}-4 q^{24} \ST^{17}-q^{24} \ST^{16}+3 q^{24} \ST^{15}-\\ -2 q^{22} \ST^{16} -2 q^{22} \ST^{15}+3 q^{22} \ST^{14}-2 q^{20} \ST^{14}+q^{20} \ST^{13}+2 q^{20} \ST^{12}-2 q^{18} \ST^{13}-\\ -2 q^{18} \ST^{12}+3 q^{18} \ST^{11}-2 q^{16} \ST^{12}-4 q^{16} \ST^{11}+2 q^{16} \ST^{10}+q^{16} \ST^9-2 q^{14} \ST^{10}-\\ -q^{14} \ST^9+q^{14} \ST^8-2 q^{12} \ST^8+2 q^{12} \ST^7-2 q^{10} \ST^7+q^{10} \ST^6-2 q^8 \ST^6-3 q^8 \ST^5-\\ -2 q^6 \ST^4+q^6 \ST^3+q^4 \ST^2-q^2 \ST-1)\,,\quad n < 12
        \end{cases} \\
        \{q^2\ST\}^{-1}\cdot \mathfrak{Kh}_{\rm adj}^{\overline{9_{43}}}&=q^{-2} \ST^{-6}(q^{38} \ST^{26}+q^{38} \ST^{25}+2 q^{36} \ST^{25}+2 q^{36} \ST^{24}+q^{34} \ST^{24}+2 q^{34} \ST^{23}+\\ &+2 q^{32} \ST^{22}+q^{32} \ST^{21}+3 q^{30} \ST^{21}+3 q^{30} \ST^{20}+3 q^{28} \ST^{20}+5 q^{28} \ST^{19}+q^{28} \ST^{18}+\\ &+q^{26} \ST^{19}+4 q^{26} \ST^{18}+2 q^{26} \ST^{17}+q^{24} \ST^{18}+4 q^{24} \ST^{17}+5 q^{24} \ST^{16}+2 q^{22} \ST^{16}+\\ &+4 q^{22} \ST^{15}+q^{22} \ST^{14}+4 q^{20} \ST^{14}+3 q^{20} \ST^{13}+2 q^{18} \ST^{13}+4 q^{18} \ST^{12}+2 q^{16} \ST^{12}+\\ &+4 q^{16} \ST^{11}+2 q^{16} \ST^{10}+2 q^{14} \ST^{10}+3 q^{14} \ST^9+4 q^{12} \ST^8+q^{12} \ST^7+2 q^{10} \ST^7+\\ &+q^{10} \ST^6+2 q^8 \ST^6+3 q^8 \ST^5+2 q^6 \ST^4+q^2 \ST+1) \\ \nn
    {\color{blue} \mathbf{\overline{10_{124}}}}:\quad {\color{red} \Kh_{\rm adj}^{\overline{10_{124}}}}&=\begin{cases}
        q^{16}\ST (q^{42} \ST^{29}+q^{40} \ST^{28}+q^{38} \ST^{27}+q^{38} \ST^{26}+q^{36} \ST^{26}+2 q^{36} \ST^{25}-2 q^{34} \ST^{25}+\\ +q^{34} \ST^{23}-2 q^{32} \ST^{24}-2 q^{32} \ST^{23}+q^{30} \ST^{22}+2 q^{30} \ST^{21}{\color{red}-q^{28} \ST^{22}}+4 q^{28} \ST^{20}+\\ +2 q^{28} \ST^{19}-2 q^{26} \ST^{20}-3 q^{26} \ST^{19}+q^{26} \ST^{18}+q^{26} \ST^{17}-2 q^{24} \ST^{18}-2 q^{24} \ST^{17}+\\ +q^{24} \ST^{15}+q^{22} \ST^{16}-2 q^{20} \ST^{16}+3 q^{20} \ST^{14}-2 q^{18} \ST^{14}-q^{18} \ST^{13}+q^{18} \ST^{12}+\\ +q^{18} \ST^{11}-2 q^{16} \ST^{12}-q^{16} \ST^{11}+q^{16} \ST^{10}-q^{14} \ST^9-q^{12} \ST^{10}-q^{10} \ST^8+q^{10} \ST^6-\\ -q^8 \ST^6+q^8 \ST^4-q^6 \ST^4-1),\quad \n \geq {\color{blue} \mathbf{22}} \\
        q^{16}\ST (q^{42} \ST^{29}+q^{40} \ST^{28}+q^{38} \ST^{27}+q^{38} \ST^{26}+q^{36} \ST^{26}+2 q^{36} \ST^{25}-2 q^{34} \ST^{25}+\\ +q^{34} \ST^{23}-2 q^{32} \ST^{24}-2 q^{32} \ST^{23}+q^{30} \ST^{22}+2 q^{30} \ST^{21}{\color{red}+q^{28} \ST^{21}}+4 q^{28} \ST^{20}+\\ +2 q^{28} \ST^{19}-2 q^{26} \ST^{20}-3 q^{26} \ST^{19}+q^{26} \ST^{18}+q^{26} \ST^{17}-2 q^{24} \ST^{18}-2 q^{24} \ST^{17}+\\ +q^{24} \ST^{15}+q^{22} \ST^{16}-2 q^{20} \ST^{16}+3 q^{20} \ST^{14}-2 q^{18} \ST^{14}-q^{18} \ST^{13}+q^{18} \ST^{12}+\\ +q^{18} \ST^{11}-2 q^{16} \ST^{12}-q^{16} \ST^{11}+q^{16} \ST^{10}-q^{14} \ST^9-q^{12} \ST^{10}-q^{10} \ST^8+q^{10} \ST^6-\\ -q^8 \ST^6+q^8 \ST^4-q^6 \ST^4-1),\quad \n < {\color{blue} \mathbf{22}}
    \end{cases}\\
    \{q^2\ST\}^{-1}\cdot \mathfrak{Kh}_{\rm adj}^{\overline{10_{124}}}&=q^{18} \ST^2 (q^{38} \ST^{27}+q^{36} \ST^{26}+2 q^{34} \ST^{25}+q^{34} \ST^{24}+2 q^{32} \ST^{24}+2 q^{32} \ST^{23}+q^{30} \ST^{22}+\\ &+q^{30} \ST^{21}+2 q^{26} \ST^{20}+3 q^{26} \ST^{19}+4 q^{24} \ST^{18}+2 q^{24} \ST^{17}+q^{22} \ST^{16}+q^{22} \ST^{15}+2 q^{20} \ST^{16}+\\ &+q^{20} \ST^{13}+2 q^{18} \ST^{14}+q^{18} \ST^{13}+3 q^{16} \ST^{12}+q^{16} \ST^{11}+q^{14} \ST^{10}+q^{14} \ST^9+q^{12} \ST^{10}+\\ &+q^{12} \ST^8+q^{10} \ST^8+q^8 \ST^6+q^6 \ST^4+q^4 \ST^2+1)
    \end{aligned}
    \end{equation}
    \begin{equation}
    \begin{aligned}
    {\color{blue} \mathbf{\overline{10_{128}}}}:\quad {\color{red} \Kh^{\overline{10_{128}}}_{\rm adj}}&=\begin{cases}
        q^{12}\ST (q^{46} \ST^{31}+q^{44} \ST^{30}-q^{42} \ST^{29}-q^{40} \ST^{28}+q^{40} \ST^{27}+2 q^{38} \ST^{26}+2 q^{38} \ST^{25}+\\ +3 q^{36} \ST^{24}+q^{36} \ST^{23}-2 q^{34} \ST^{24}+q^{34} \ST^{23}+q^{34} \ST^{22}-q^{32} \ST^{23}+q^{32} \ST^{21}-\\ -q^{30} \ST^{21}+4 q^{30} \ST^{20}+3 q^{30} \ST^{19}-2 q^{28} \ST^{20}+3 q^{28} \ST^{19}+5 q^{28} \ST^{18}+q^{28} \ST^{17}-\\ -2 q^{26} \ST^{19}-q^{26} \ST^{18}+q^{26} \ST^{17}+q^{26} \ST^{16}-q^{24} \ST^{18}-3 q^{24} \ST^{17}+q^{24} \ST^{16}+\\ +3 q^{24} \ST^{15}-3 q^{22} \ST^{16}+4 q^{22} \ST^{14}+q^{22} \ST^{13}\underline{{\color{red}-q^{20} \ST^{16}}-2 q^{20} \ST^{15}}-3 q^{20} \ST^{14}+\\ +3 q^{20} \ST^{12}-q^{18} \ST^{14}-3 q^{18} \ST^{13}-2 q^{18} \ST^{12}+2 q^{18} \ST^{11}-2 q^{16} \ST^{12}-2 q^{16} \ST^{11}+\\ +q^{16} \ST^{10}-q^{14} \ST^{11}-2 q^{14} \ST^{10}-q^{14} \ST^9+2 q^{14} \ST^8-q^{12} \ST^{10}-2 q^{12} \ST^9-\\ -3 q^{12} \ST^8+3 q^{12} \ST^7+q^{12} \ST^6-q^{10} \ST^8-2 q^{10} \ST^7-q^8 \ST^6-2 q^8 \ST^5+q^8 \ST^4-\\ -2 q^6 \ST^4+q^6 \ST^3-q^4 \ST^3-q^4 \ST^2-q^2 \ST-1)\,,\quad n\geq {\color{blue} \mathbf{16}} \\
        q^{12}\ST (q^{46} \ST^{31}+q^{44} \ST^{30}-q^{42} \ST^{29}-q^{40} \ST^{28}+q^{40} \ST^{27}+2 q^{38} \ST^{26}+2 q^{38} \ST^{25}+\\ +3 q^{36} \ST^{24}+q^{36} \ST^{23}-2 q^{34} \ST^{24}+q^{34} \ST^{23}+q^{34} \ST^{22}-q^{32} \ST^{23}+q^{32} \ST^{21}-\\ -q^{30} \ST^{21}+4 q^{30} \ST^{20}+3 q^{30} \ST^{19}-2 q^{28} \ST^{20}+3 q^{28} \ST^{19}+5 q^{28} \ST^{18}+q^{28} \ST^{17}-\\ -2 q^{26} \ST^{19}-q^{26} \ST^{18}+q^{26} \ST^{17}+q^{26} \ST^{16}-q^{24} \ST^{18}-3 q^{24} \ST^{17}+q^{24} \ST^{16}+\\ +3 q^{24} \ST^{15}-3 q^{22} \ST^{16}+4 q^{22} \ST^{14}+q^{22} \ST^{13}\underline{-2q^{20} \ST^{15}{\color{red}+q^{20} \ST^{15}}}-3 q^{20} \ST^{14}+3 q^{20} \ST^{12}-\\ -q^{18} \ST^{14}-3 q^{18} \ST^{13}-2 q^{18} \ST^{12}+2 q^{18} \ST^{11}-2 q^{16} \ST^{12}-2 q^{16} \ST^{11}+q^{16} \ST^{10}-\\ -q^{14} \ST^{11}-2 q^{14} \ST^{10}-q^{14} \ST^9+2 q^{14} \ST^8-q^{12} \ST^{10}-2 q^{12} \ST^9-3 q^{12} \ST^8+\\ +3 q^{12} \ST^7+q^{12} \ST^6-q^{10} \ST^8-2 q^{10} \ST^7-q^8 \ST^6-2 q^8 \ST^5+q^8 \ST^4-2 q^6 \ST^4+\\ +q^6 \ST^3-q^4 \ST^3-q^4 \ST^2-q^2 \ST-1)\,,\quad n < {\color{blue} \mathbf{16}}
    \end{cases} \\ \nn
    \{q^2\ST\}^{-1}\cdot \mathfrak{Kh}_{\rm adj}^{\overline{10_{128}}}&=q^{14} \ST^2 (q^{42} \ST^{29}+q^{40} \ST^{28}+q^{36} \ST^{25}+2 q^{34} \ST^{24}+2 q^{34} \ST^{23}+q^{32} \ST^{23}+3 q^{32} \ST^{22}+\\ &+q^{32} \ST^{21}+3 q^{30} \ST^{21}+q^{30} \ST^{20}+3 q^{28} \ST^{20}+2 q^{28} \ST^{19}+2 q^{26} \ST^{19}+5 q^{26} \ST^{18}+\\ &+3 q^{26} \ST^{17}+q^{24} \ST^{18}+5 q^{24} \ST^{17}+5 q^{24} \ST^{16}+q^{24} \ST^{15}+4 q^{22} \ST^{16}+4 q^{22} \ST^{15}+\\ &+q^{22} \ST^{14}+2 q^{20} \ST^{15}+6 q^{20} \ST^{14}+4 q^{20} \ST^{13}+q^{18} \ST^{14}+4 q^{18} \ST^{13}+5 q^{18} \ST^{12}+\\ &+q^{18} \ST^{11}+3 q^{16} \ST^{12}+4 q^{16} \ST^{11}+3 q^{16} \ST^{10}+q^{14} \ST^{11}+3 q^{14} \ST^{10}+3 q^{14} \ST^9+\\ &+q^{12} \ST^{10}+2 q^{12} \ST^9+4 q^{12} \ST^8+q^{10} \ST^8+2 q^{10} \ST^7+2 q^{10} \ST^6+q^8 \ST^6+3 q^8 \ST^5+\\ &+q^8 \ST^4+2 q^6 \ST^4+q^4 \ST^3+2 q^4 \ST^2+q^2 \ST+1)\\
    {\color{blue} \mathbf{10_{132}}}:\quad {\color{red} \Kh_{\rm adj}^{10_{132}}}&=\begin{cases}
        q^{-42} \ST^{-27}(q^{44} \ST^{30}+q^{42} \ST^{29}+2 q^{40} \ST^{27}+q^{38} \ST^{27}+4 q^{38} \ST^{26}-q^{36} \ST^{26}-\\ -q^{34} \ST^{25}-3 q^{34} \ST^{24}+q^{34} \ST^{23}+5 q^{32} \ST^{22}+q^{32} \ST^{21}-q^{30} \ST^{23}\underline{- q^{30} \ST^{22}{\color{red} - q^{30} \ST^{22}}+}\\ \underline{+4 q^{30} \ST^{21}}+4 q^{30} \ST^{20}+q^{30} \ST^{19}-2 q^{28} \ST^{21}-4 q^{28} \ST^{20}+q^{28} \ST^{19}+q^{28} \ST^{18}-\\ -3 q^{26} \ST^{19}-q^{26} \ST^{18}-q^{26} \ST^{17}+2 q^{24} \ST^{17}+q^{24} \ST^{16}+q^{24} \ST^{15}-2 q^{22} \ST^{17}-\\ -2 q^{22} \ST^{16}+4 q^{22} \ST^{15}+3 q^{22} \ST^{14}-q^{20} \ST^{16}-4 q^{20} \ST^{15}+q^{20} \ST^{13}+q^{20} \ST^{12}-\\ -q^{18} \ST^{14}-3 q^{18} \ST^{13}-3 q^{18} \ST^{12}+q^{18} \ST^{11}-q^{16} \ST^{12}-q^{16} \ST^{11}-q^{14} \ST^{11}+\\ +q^{14} \ST^{10}+q^{14} \ST^9+q^{14} \ST^8-q^{12} \ST^{10}-q^{12} \ST^9+2 q^{12} \ST^7-q^{10} \ST^8-\\ -2 q^{10} \ST^7-q^8 \ST^6-2 q^8 \ST^5-q^6 \ST^4+q^6 \ST^3+q^4 \ST^2-q^2 \ST-1)\,,\quad n \geq -6 \\
        q^{-42} \ST^{-27}(q^{44} \ST^{30}+q^{42} \ST^{29}+2 q^{40} \ST^{27}+q^{38} \ST^{27}+4 q^{38} \ST^{26}-q^{36} \ST^{26}-\\ -q^{34} \ST^{25}-3 q^{34} \ST^{24}+q^{34} \ST^{23}+5 q^{32} \ST^{22}+q^{32} \ST^{21}-q^{30} \ST^{23}\underline{-q^{30} \ST^{22}+}\\ \underline{{\color{red} +q^{30} \ST^{21}}+4 q^{30} \ST^{21}}+4 q^{30} \ST^{20}+q^{30} \ST^{19}-2 q^{28} \ST^{21}-4 q^{28} \ST^{20}+q^{28} \ST^{19}+q^{28} \ST^{18}-\\ -3 q^{26} \ST^{19}-q^{26} \ST^{18}-q^{26} \ST^{17}+2 q^{24} \ST^{17}+q^{24} \ST^{16}+q^{24} \ST^{15}-2 q^{22} \ST^{17}-\\ -2 q^{22} \ST^{16}+4 q^{22} \ST^{15}+3 q^{22} \ST^{14}-q^{20} \ST^{16}-4 q^{20} \ST^{15}+q^{20} \ST^{13}+q^{20} \ST^{12}-\\ -q^{18} \ST^{14}-3 q^{18} \ST^{13}-3 q^{18} \ST^{12}+q^{18} \ST^{11}-q^{16} \ST^{12}-q^{16} \ST^{11}-q^{14} \ST^{11}+\\ +q^{14} \ST^{10}+q^{14} \ST^9+q^{14} \ST^8-q^{12} \ST^{10}-q^{12} \ST^9+2 q^{12} \ST^7-q^{10} \ST^8-\\ -2 q^{10} \ST^7-q^8 \ST^6-2 q^8 \ST^5-q^6 \ST^4+q^6 \ST^3+q^4 \ST^2-q^2 \ST-1)\,,\quad n < -6
    \end{cases}\\
    \{q^2\ST\}^{-1}\cdot \mathfrak{Kh}_{\rm adj}^{10_{132}}&=q^{-40} \ST^{-26}(q^{40} \ST^{28}+q^{38} \ST^{27}+q^{36} \ST^{26}+2 q^{36} \ST^{25}+2 q^{34} \ST^{25}+4 q^{34} \ST^{24}+\\ &+2 q^{32} \ST^{23}+q^{30} \ST^{23}+q^{30} \ST^{22}+q^{30} \ST^{21}+2 q^{28} \ST^{21}+5 q^{28} \ST^{20}+q^{28} \ST^{19}+\\ &+5 q^{26} \ST^{19}+4 q^{26} \ST^{18}+q^{26} \ST^{17}+q^{24} \ST^{18}+2 q^{24} \ST^{17}+q^{24} \ST^{16}+2 q^{22} \ST^{17}+\\ &+3 q^{22} \ST^{16}+q^{20} \ST^{16}+4 q^{20} \ST^{15}+2 q^{20} \ST^{14}+q^{20} \ST^{13}+q^{18} \ST^{14}+4 q^{18} \ST^{13}+\\ &+3 q^{18} \ST^{12}+2 q^{16} \ST^{12}+2 q^{16} \ST^{11}+q^{16} \ST^{10}+q^{14} \ST^{11}+q^{14} \ST^9+q^{12} \ST^{10}+\\ &+q^{12} \ST^9+q^{12} \ST^8+q^{10} \ST^8+2 q^{10} \ST^7+q^{10} \ST^6+q^8 \ST^6+2 q^8 \ST^5+q^6 \ST^4+q^2 \ST+1)
    \end{aligned}
    \end{equation}
    \begin{equation}
    \begin{aligned}
    {\color{blue} \mathbf{10_{136}}}:\quad {\color{red} \Kh_{\rm adj}^{10_{136}}}&=\begin{cases}
        q^{-20} \ST^{-15}(q^{46} \ST^{31}+q^{44} \ST^{30}-q^{42} \ST^{29}+q^{42} \ST^{28}-q^{40} \ST^{28}+3 q^{40} \ST^{27}+\\ +q^{40} \ST^{26}+2 q^{38} \ST^{26}+2 q^{38} \ST^{25}-2 q^{36} \ST^{25}+3 q^{36} \ST^{24}+q^{36} \ST^{23} -3 q^{34} \ST^{24}+\\ +3 q^{34} \ST^{23}+3 q^{34} \ST^{22}-q^{32} \ST^{23}+4 q^{32} \ST^{21}+q^{32} \ST^{20} -2 q^{30} \ST^{21}+\\ +5 q^{30} \ST^{20}+4 q^{30} \ST^{19}-3 q^{28} \ST^{20}+4 q^{28} \ST^{19}+6 q^{28} \ST^{18} -3 q^{26} \ST^{19}-3 q^{26} \ST^{18}+\\ +3 q^{26} \ST^{17}+2 q^{26} \ST^{16}-q^{24} \ST^{18}-6 q^{24} \ST^{17} +q^{24} \ST^{16}+6 q^{24} \ST^{15}-4 q^{22} \ST^{16}+\\ +6 q^{22} \ST^{14}+q^{22} \ST^{13}\underline{{\color{red} -q^{20} \ST^{16}} -3 q^{20} \ST^{15}}-5 q^{20} \ST^{14}+3 q^{20} \ST^{12}-q^{18} \ST^{14}-\\ -6 q^{18} \ST^{13}-4 q^{18} \ST^{12} +4 q^{18} \ST^{11}-2 q^{16} \ST^{12}-3 q^{16} \ST^{11}+3 q^{16} \ST^{10}+q^{16} \ST^9-\\ -q^{14} \ST^{11} -3 q^{14} \ST^{10}-q^{14} \ST^9+3 q^{14} \ST^8-q^{12} \ST^{10}-3 q^{12} \ST^9-5 q^{12} \ST^8+3 q^{12} \ST^7-\\ -q^{10} \ST^8-4 q^{10} \ST^7-q^8 \ST^6-3 q^8 \ST^5+q^8 \ST^4-3 q^6 \ST^4+2 q^6 \ST^3-q^4 \ST^3-2 q^2 \ST-1)\,,\quad n\geq 0 \\
        q^{-20} \ST^{-15}(q^{46} \ST^{31}+q^{44} \ST^{30}-q^{42} \ST^{29}+q^{42} \ST^{28}-q^{40} \ST^{28}+3 q^{40} \ST^{27}+\\ +q^{40} \ST^{26}+2 q^{38} \ST^{26}+2 q^{38} \ST^{25}-2 q^{36} \ST^{25}+3 q^{36} \ST^{24}+q^{36} \ST^{23} -3 q^{34} \ST^{24}+\\ +3 q^{34} \ST^{23}+3 q^{34} \ST^{22}-q^{32} \ST^{23}+4 q^{32} \ST^{21}+q^{32} \ST^{20} -2 q^{30} \ST^{21}+5 q^{30} \ST^{20}+\\ +4 q^{30} \ST^{19}-3 q^{28} \ST^{20}+4 q^{28} \ST^{19}+6 q^{28} \ST^{18}
        -3 q^{26} \ST^{19} -3 q^{26} \ST^{18}+3 q^{26} \ST^{17}+\\ +2 q^{26} \ST^{16}-q^{24} \ST^{18}-6 q^{24} \ST^{17} +q^{24} \ST^{16}+6 q^{24} \ST^{15}-4 q^{22} \ST^{16}+6 q^{22} \ST^{14}+\\ +q^{22} \ST^{13}\underline{-3 q^{20} \ST^{15}{\color{red} +q^{20} \ST^{15}}} -5 q^{20} \ST^{14}+3 q^{20} \ST^{12}-q^{18} \ST^{14}-6 q^{18} \ST^{13}-4 q^{18} \ST^{12}+\\ +4 q^{18} \ST^{11}-2 q^{16} \ST^{12}-3 q^{16} \ST^{11}+3 q^{16} \ST^{10}+q^{16} \ST^9 -q^{14} \ST^{11}-3 q^{14} \ST^{10}-\\ -q^{14} \ST^9+3 q^{14} \ST^8-q^{12} \ST^{10}-3 q^{12} \ST^9-5 q^{12} \ST^8+3 q^{12} \ST^7-q^{10} \ST^8-\\ -4 q^{10} \ST^7-q^8 \ST^6-3 q^8 \ST^5+q^8 \ST^4-3 q^6 \ST^4+2 q^6 \ST^3-q^4 \ST^3-2 q^2 \ST-1)\,,\quad n < 0 \nn
    \end{cases} \\
    \{q^2\ST\}^{-1}\cdot \mathfrak{Kh}_{\rm adj}^{10_{136}}&=q^{-18} \ST^{-14}(q^{42} \ST^{29}+q^{40} \ST^{28}+q^{38} \ST^{26}+3 q^{36} \ST^{25}+q^{36} \ST^{24}+3 q^{34} \ST^{24}+2 q^{34} \ST^{23}+\\ &+q^{32} \ST^{23}+4 q^{32} \ST^{22}+q^{32} \ST^{21}+5 q^{30} \ST^{21}+3 q^{30} \ST^{20}+4 q^{28} \ST^{20}+5 q^{28} \ST^{19}+\\ &+q^{28} \ST^{18}+3 q^{26} \ST^{19}+8 q^{26} \ST^{18}+4 q^{26} \ST^{17}+q^{24} \ST^{18}+9 q^{24} \ST^{17}+7 q^{24} \ST^{16}+\\ &+5 q^{22} \ST^{16}+7 q^{22} \ST^{15}+2 q^{22} \ST^{14}+3 q^{20} \ST^{15}+8 q^{20} \ST^{14}+6 q^{20} \ST^{13}+q^{18} \ST^{14}+\\ &+7 q^{18} \ST^{13}+8 q^{18} \ST^{12}+q^{18} \ST^{11}+3 q^{16} \ST^{12}+6 q^{16} \ST^{11}+3 q^{16} \ST^{10}+q^{14} \ST^{11}+\\ &+4 q^{14} \ST^{10}+5 q^{14} \ST^9+q^{12} \ST^{10}+3 q^{12} \ST^9+6 q^{12} \ST^8+q^{12} \ST^7+q^{10} \ST^8+\\ &+4 q^{10} \ST^7+3 q^{10} \ST^6+q^8 \ST^6+4 q^8 \ST^5+3 q^6 \ST^4+q^4 \ST^3+q^4 \ST^2+2 q^2 \ST+1) \\
    {\color{blue} \mathbf{\overline{10_{139}}}}:\quad {\color{red} \Kh_{\rm adj}^{\overline{10_{139}}}}&=\begin{cases}
        q^{16}\ST (q^{50} \ST^{35}+q^{48} \ST^{34}-q^{46} \ST^{33}-q^{44} \ST^{32}+q^{44} \ST^{31}+2 q^{42} \ST^{30}+\\ +3 q^{42} \ST^{29}+4 q^{40} \ST^{28}+q^{40} \ST^{27}-2 q^{38} \ST^{28}+q^{38} \ST^{26}-q^{36} \ST^{27}+\\ +3 q^{36} \ST^{25}-2 q^{34} \ST^{25}+4 q^{34} \ST^{24}+2 q^{34} \ST^{23}-4 q^{32} \ST^{24}-q^{32} \ST^{23}+\\ +2 q^{32} \ST^{22}-q^{30} \ST^{23}-2 q^{30} \ST^{22}+q^{30} \ST^{21}\underline{{\color{red} -q^{28} \ST^{22}}-2 q^{28} \ST^{21}}+\\ +4 q^{28} \ST^{20}+2 q^{28} \ST^{19}-3 q^{26} \ST^{20}-q^{26} \ST^{19}+3 q^{26} \ST^{18}+q^{26} \ST^{17}-\\ -q^{24} \ST^{19}-4 q^{24} \ST^{18}+q^{24} \ST^{16}+q^{24} \ST^{15}-2 q^{22} \ST^{17}-q^{22} \ST^{16}-\\ -2 q^{20} \ST^{16}-q^{20} \ST^{15}+3 q^{20} \ST^{14}-2 q^{18} \ST^{14}+q^{18} \ST^{12}+q^{18} \ST^{11}-\\ -q^{16} \ST^{13}-3 q^{16} \ST^{12}-q^{16} \ST^{11}+q^{16} \ST^{10}-q^{14} \ST^{11}-q^{14} \ST^9-\\ -q^{12} \ST^{10}-q^{10} \ST^8+q^{10} \ST^6-q^8 \ST^6+q^8 \ST^4-q^6 \ST^4-1)\,,\quad n \geq {\color{blue} \mathbf{22}} \\
        q^{16}\ST (q^{50} \ST^{35}+q^{48} \ST^{34}-q^{46} \ST^{33}-q^{44} \ST^{32}+q^{44} \ST^{31}+2 q^{42} \ST^{30}+\\ +3 q^{42} \ST^{29}+4 q^{40} \ST^{28}+q^{40} \ST^{27}-2 q^{38} \ST^{28}+q^{38} \ST^{26}-q^{36} \ST^{27}+\\ +3 q^{36} \ST^{25}-2 q^{34} \ST^{25}+4 q^{34} \ST^{24}+2 q^{34} \ST^{23}-4 q^{32} \ST^{24}-q^{32} \ST^{23}+\\ +2 q^{32} \ST^{22}-q^{30} \ST^{23}-2 q^{30} \ST^{22}+q^{30} \ST^{21}\underline{{\color{red}+q^{28} \ST^{21}}-2q^{28} \ST^{21}}+4 q^{28} \ST^{20}+\\ +2 q^{28} \ST^{19}-3 q^{26} \ST^{20}-q^{26} \ST^{19}+3 q^{26} \ST^{18}+q^{26} \ST^{17}-q^{24} \ST^{19}-\\ -4 q^{24} \ST^{18}+q^{24} \ST^{16}+q^{24} \ST^{15}-2 q^{22} \ST^{17}-q^{22} \ST^{16}-2 q^{20} \ST^{16}-\\ -q^{20} \ST^{15}+3 q^{20} \ST^{14}-2 q^{18} \ST^{14}+q^{18} \ST^{12}+q^{18} \ST^{11}-q^{16} \ST^{13}-\\ -3 q^{16} \ST^{12}-q^{16} \ST^{11}+q^{16} \ST^{10}-q^{14} \ST^{11}-q^{14} \ST^9-q^{12} \ST^{10}-\\ -q^{10} \ST^8+q^{10} \ST^6-q^8 \ST^6+q^8 \ST^4-q^6 \ST^4-1)\,,\quad n < {\color{blue} \mathbf{22}}
    \end{cases} \\
    \{q^2\ST\}^{-1}\cdot \mathfrak{Kh}_{\rm adj}^{\overline{10_{139}}}&=q^{18} \ST^2 (q^{46} \ST^{33}+q^{44} \ST^{32}+q^{40} \ST^{29}+2 q^{38} \ST^{28}+3 q^{38} \ST^{27}+q^{36} \ST^{27}+\\ &+4 q^{36} \ST^{26}+q^{36} \ST^{25}+3 q^{34} \ST^{25}+q^{34} \ST^{24}+4 q^{32} \ST^{24}+4 q^{32} \ST^{23}+q^{30} \ST^{23}+\\ &+5 q^{30} \ST^{22}+2 q^{30} \ST^{21}+3 q^{28} \ST^{21}+2 q^{28} \ST^{20}+3 q^{26} \ST^{20}+3 q^{26} \ST^{19}+\\ &+q^{24} \ST^{19}+6 q^{24} \ST^{18}+2 q^{24} \ST^{17}+2 q^{22} \ST^{17}+3 q^{22} \ST^{16}+q^{22} \ST^{15}+2 q^{20} \ST^{16}+\\ &+2 q^{20} \ST^{15}+q^{20} \ST^{14}+q^{20} \ST^{13}+2 q^{18} \ST^{14}+q^{18} \ST^{13}+q^{16} \ST^{13}+4 q^{16} \ST^{12}+\\ &+q^{16} \ST^{11}+q^{14} \ST^{11}+q^{14} \ST^{10}+q^{14} \ST^9+q^{12} \ST^{10}+q^{12} \ST^8+q^{10} \ST^8+q^8 \ST^6+q^6 \ST^4+q^4 \ST^2+1)
\end{aligned}
\end{equation}
\begin{equation}
\begin{aligned}
    {\color{blue} \mathbf{10_{145}}}:\quad {\color{red} \Kh_{\rm adj}^{10_{145}}}&=\begin{cases}
        q^{-58} \ST^{-35}(q^{50} \ST^{35}+q^{44} \ST^{31}+2 q^{44} \ST^{30}-q^{42} \ST^{31}+3 q^{42} \ST^{29}+q^{42} \ST^{28} -q^{40} \ST^{29}-q^{40} \ST^{28}+\\ +2 q^{40} \ST^{27}+q^{38} \ST^{27}+5 q^{38} \ST^{26}+3 q^{36} \ST^{25} +q^{36} \ST^{24}-4 q^{34} \ST^{25}\underline{-5 q^{34} \ST^{24}{\color{red}-q^{34} \ST^{24}}+2 q^{34} \ST^{23}}-\\ -q^{32} \ST^{24}-3 q^{32} \ST^{23} +4 q^{32} \ST^{22}+q^{32} \ST^{21}+5 q^{30} \ST^{21}+4 q^{30} \ST^{20}+q^{30} \ST^{19}-2 q^{28} \ST^{21}-\\ -2 q^{28} \ST^{20}+3 q^{28} \ST^{19}+q^{28} \ST^{18}-q^{26} \ST^{20}-6 q^{26} \ST^{19}-q^{26} \ST^{18} -q^{26} \ST^{17}-2 q^{24} \ST^{18}+\\ +q^{24} \ST^{16}+q^{24} \ST^{15}-q^{22} \ST^{17}-q^{22} \ST^{16}+5 q^{22} \ST^{15} +3 q^{22} \ST^{14}-q^{20} \ST^{16}-4 q^{20} \ST^{15}+q^{20} \ST^{14}+\\ +q^{20} \ST^{13}+q^{20} \ST^{12} -2 q^{18} \ST^{14}-4 q^{18} \ST^{13}-3 q^{18} \ST^{12}+q^{18} \ST^{11}-2 q^{16} \ST^{12}-q^{16} \ST^{11} -q^{14} \ST^{11}+\\ +q^{14} \ST^{10} +q^{14} \ST^9+q^{14} \ST^8-q^{12} \ST^{10}-q^{12} \ST^9+2 q^{12} \ST^7 -q^{10} \ST^8-2 q^{10} \ST^7-q^8 \ST^6-2 q^8 \ST^5-\\ -q^6 \ST^4+q^6 \ST^3+q^4 \ST^2-q^2 \ST-1)\,,\quad n \geq -12 \\
        q^{-58} \ST^{-35}(q^{50} \ST^{35}+q^{44} \ST^{31}+2 q^{44} \ST^{30}-q^{42} \ST^{31}+3 q^{42} \ST^{29}+q^{42} \ST^{28} -q^{40} \ST^{29}-q^{40} \ST^{28}+\\ +2 q^{40} \ST^{27}+q^{38} \ST^{27}+5 q^{38} \ST^{26}+3 q^{36} \ST^{25} +q^{36} \ST^{24}-4 q^{34} \ST^{25}\underline{-5 q^{34} \ST^{24}{\color{red} +q^{34} \ST^{23}}+2 q^{34} \ST^{23}}-\\ -q^{32} \ST^{24}-3 q^{32} \ST^{23} +4 q^{32} \ST^{22}+q^{32} \ST^{21}+5 q^{30} \ST^{21}+4 q^{30} \ST^{20}+q^{30} \ST^{19}-2 q^{28} \ST^{21} -2 q^{28} \ST^{20}+\\ +3 q^{28} \ST^{19} +q^{28} \ST^{18}-q^{26} \ST^{20}-6 q^{26} \ST^{19}-q^{26} \ST^{18} -q^{26} \ST^{17}-2 q^{24} \ST^{18}+q^{24} \ST^{16}+q^{24} \ST^{15}-\\ -q^{22} \ST^{17}-q^{22} \ST^{16}+5 q^{22} \ST^{15} +3 q^{22} \ST^{14}-q^{20} \ST^{16}-4 q^{20} \ST^{15}+q^{20} \ST^{14}+q^{20} \ST^{13}+q^{20} \ST^{12}-\\ -2 q^{18} \ST^{14}-4 q^{18} \ST^{13}-3 q^{18} \ST^{12}+q^{18} \ST^{11}-2 q^{16} \ST^{12}-q^{16} \ST^{11} -q^{14} \ST^{11}+q^{14} \ST^{10}+\\ +q^{14} \ST^9+q^{14} \ST^8-q^{12} \ST^{10}-q^{12} \ST^9+2 q^{12} \ST^7 -q^{10} \ST^8-2 q^{10} \ST^7-q^8 \ST^6-2 q^8 \ST^5-\\ -q^6 \ST^4+q^6 \ST^3+q^4 \ST^2-q^2 \ST-1)\,,\quad n < -12 \nn
    \end{cases}  \\
    \{q^2\ST\}^{-1}\cdot \mathfrak{Kh}_{\rm adj}^{10_{145}}&=q^{-56} \ST^{-34}(q^{46} \ST^{33}+q^{42} \ST^{31}+q^{40} \ST^{29}+2 q^{40} \ST^{28}+3 q^{38} \ST^{27}+q^{38} \ST^{26}+q^{36} \ST^{26}+\\ &+2 q^{36} \ST^{25}+4 q^{34} \ST^{25}+6 q^{34} \ST^{24}+q^{32} \ST^{24}+5 q^{32} \ST^{23}+q^{32} \ST^{22}+q^{30} \ST^{22}+2 q^{30} \ST^{21}+\\ &+2 q^{28} \ST^{21}+5 q^{28} \ST^{20}+q^{28} \ST^{19}+q^{26} \ST^{20}+7 q^{26} \ST^{19}+4 q^{26} \ST^{18}+q^{26} \ST^{17}+3 q^{24} \ST^{18}+\\ &+4 q^{24} \ST^{17}+q^{24} \ST^{16}+q^{22} \ST^{17}+3 q^{22} \ST^{16}+q^{20} \ST^{16}+4 q^{20} \ST^{15}+2 q^{20} \ST^{14}+q^{20} \ST^{13}+\\ &+2 q^{18} \ST^{14}+5 q^{18} \ST^{13}+3 q^{18} \ST^{12}+3 q^{16} \ST^{12}+2 q^{16} \ST^{11}+q^{16} \ST^{10}+q^{14} \ST^{11}+q^{14} \ST^9+\\ &+q^{12} \ST^{10}+q^{12} \ST^9+q^{12} \ST^8+q^{10} \ST^8+2 q^{10} \ST^7+q^{10} \ST^6+q^8 \ST^6+2 q^8 \ST^5+q^6 \ST^4+q^2 \ST+1)\,,\quad n < -12 \\
    {\color{blue} \mathbf{10_{152}}}:\quad {\color{red} \Kh_{\rm adj}^{10_{152}}}&=\begin{cases}
        q^{-72} \ST^{-39}(q^{56} \ST^{39}+q^{50} \ST^{35}-q^{48} \ST^{35}+q^{48} \ST^{33}-q^{46} \ST^{33}+2 q^{46} \ST^{31} +q^{44} \ST^{30}+2 q^{44} \ST^{29}+\\ +q^{42} \ST^{30}-q^{42} \ST^{29}+q^{42} \ST^{28}-q^{40} \ST^{29}+q^{40} \ST^{28} +4 q^{40} \ST^{27}+q^{40} \ST^{26}-q^{38} \ST^{28}-q^{38} \ST^{27}+\\ +2 q^{38} \ST^{26}+5 q^{38} \ST^{25}-q^{36} \ST^{26} -5 q^{36} \ST^{25}+4 q^{36} \ST^{24}+5 q^{36} \ST^{23}-2 q^{34} \ST^{24}-2 q^{34} \ST^{23}+\\ +4 q^{34} \ST^{22} -q^{32} \ST^{24}-q^{32} \ST^{23}+7 q^{32} \ST^{21}+2 q^{32} \ST^{20}-q^{30} \ST^{22}-2 q^{30} \ST^{21} +6 q^{30} \ST^{20}+\\ +8 q^{30} \ST^{19}-5 q^{28} \ST^{20}-8 q^{28} \ST^{19}\underline{+8 q^{28} \ST^{18}{\color{red} -q^{28} \ST^{18}}+q^{28} \ST^{17}} -q^{26} \ST^{19}-8 q^{26} \ST^{18} +q^{26} \ST^{17}+\\ +3 q^{26} \ST^{16}-3 q^{24} \ST^{17}+q^{24} \ST^{16} +10 q^{24} \ST^{15}-2 q^{22} \ST^{16}-4 q^{22} \ST^{15}+10 q^{22} \ST^{14} +q^{22} \ST^{13}-\\ -q^{20} \ST^{15} -10 q^{20} \ST^{14}-q^{20} \ST^{13}+4 q^{20} \ST^{12}-5 q^{18} \ST^{13}-8 q^{18} \ST^{12}+5 q^{18} \ST^{11} -q^{16} \ST^{12}-\\ -8 q^{16} \ST^{11}+4 q^{16} \ST^{10}+q^{16} \ST^9-5 q^{14} \ST^{10}-q^{14} \ST^9+3 q^{14} \ST^8 -2 q^{12} \ST^9 -7 q^{12} \ST^8+\\ +3 q^{12} \ST^7-5 q^{10} \ST^7-q^8 \ST^6-3 q^8 \ST^5+q^8 \ST^4-3 q^6 \ST^4 +2 q^6 \ST^3-q^4 \ST^3-2 q^2 \ST-1)\,,\quad n \geq {\color{blue} \mathbf{-22}} \\
        q^{-72} \ST^{-39}(q^{56} \ST^{39}+q^{50} \ST^{35}-q^{48} \ST^{35}+q^{48} \ST^{33}-q^{46} \ST^{33}+2 q^{46} \ST^{31} +q^{44} \ST^{30}+2 q^{44} \ST^{29}+\\ +q^{42} \ST^{30}-q^{42} \ST^{29}+q^{42} \ST^{28}-q^{40} \ST^{29}+q^{40} \ST^{28} +4 q^{40} \ST^{27}+q^{40} \ST^{26}-q^{38} \ST^{28}-q^{38} \ST^{27}+\\ +2 q^{38} \ST^{26}+5 q^{38} \ST^{25}-q^{36} \ST^{26} -5 q^{36} \ST^{25}+4 q^{36} \ST^{24}+5 q^{36} \ST^{23}-2 q^{34} \ST^{24}-2 q^{34} \ST^{23}+\\ +4 q^{34} \ST^{22} -q^{32} \ST^{24}-q^{32} \ST^{23}+7 q^{32} \ST^{21}+2 q^{32} \ST^{20}-q^{30} \ST^{22}-2 q^{30} \ST^{21} +6 q^{30} \ST^{20}+8 q^{30} \ST^{19}-\\ -5 q^{28} \ST^{20}-8 q^{28} \ST^{19}\underline{+8 q^{28} \ST^{18} {\color{red} + q^{28} \ST^{17}}+ q^{28} \ST^{17}} -q^{26} \ST^{19}-8 q^{26} \ST^{18}+q^{26} \ST^{17} +3 q^{26} \ST^{16}-\\ -3 q^{24} \ST^{17}+q^{24} \ST^{16}+10 q^{24} \ST^{15} -2 q^{22} \ST^{16}-4 q^{22} \ST^{15} +10 q^{22} \ST^{14}+q^{22} \ST^{13} -q^{20} \ST^{15}-\\ -10 q^{20} \ST^{14} -q^{20} \ST^{13}+4 q^{20} \ST^{12}-5 q^{18} \ST^{13}-8 q^{18} \ST^{12} +5 q^{18} \ST^{11}-q^{16} \ST^{12}-8 q^{16} \ST^{11} +\\ +4 q^{16} \ST^{10}+q^{16} \ST^9-5 q^{14} \ST^{10}-q^{14} \ST^9+3 q^{14} \ST^8-2 q^{12} \ST^9-7 q^{12} \ST^8 +3 q^{12} \ST^7-\\ -5 q^{10} \ST^7-q^8 \ST^6-3 q^8 \ST^5+q^8 \ST^4-3 q^6 \ST^4+2 q^6 \ST^3-q^4 \ST^3-2 q^2 \ST-1)\,,\quad n < {\color{blue} \mathbf{-22}}
    \end{cases}\\
    \mathfrak{Kh}_{\rm adj}^{10_{152}}&=q^{-70} \ST^{-38}(q^{52} \ST^{37}+q^{48} \ST^{35}+q^{46} \ST^{33}+q^{44} \ST^{31}+2 q^{42} \ST^{29}+q^{40} \ST^{29}+q^{40} \ST^{28}+\\ &+2 q^{40} \ST^{27}+q^{38} \ST^{28}+q^{38} \ST^{27}+q^{38} \ST^{26}+2 q^{36} \ST^{26}+6 q^{36} \ST^{25}+q^{36} \ST^{24}+3 q^{34} \ST^{24}+\\ &+5 q^{34} \ST^{23}+q^{32} \ST^{24}+q^{32} \ST^{23}+5 q^{32} \ST^{22}+5 q^{32} \ST^{21}+q^{30} \ST^{22}+3 q^{30} \ST^{21}+4 q^{30} \ST^{20}+\\ &+5 q^{28} \ST^{20}+12 q^{28} \ST^{19}+2 q^{28} \ST^{18}+q^{26} \ST^{19}+10 q^{26} \ST^{18}+8 q^{26} \ST^{17}+4 q^{24} \ST^{17}+\\ &+10 q^{24} \ST^{16}+q^{24} \ST^{15}+2 q^{22} \ST^{16}+9 q^{22} \ST^{15}+3 q^{22} \ST^{14}+q^{20} \ST^{15}+11 q^{20} \ST^{14}+11 q^{20} \ST^{13}+\\ &+5 q^{18} \ST^{13}+13 q^{18} \ST^{12}+q^{18} \ST^{11}+q^{16} \ST^{12}+10 q^{16} \ST^{11}+4 q^{16} \ST^{10}+5 q^{14} \ST^{10}+6 q^{14} \ST^9+\\ &+2 q^{12} \ST^9+8 q^{12} \ST^8+q^{12} \ST^7+5 q^{10} \ST^7+3 q^{10} \ST^6+q^8 \ST^6+4 q^8 \ST^5+3 q^6 \ST^4+q^4 \ST^3+q^4 \ST^2+2 q^2 \ST+1) 
\end{aligned}
\end{equation}
\begin{equation} \nn
\begin{aligned}
    {\color{blue} \mathbf{10_{153}}}:\quad {\color{red} \Kh_{\rm adj}^{10_{153}}}&=\begin{cases}
        q^{-30} \ST^{-19}(q^{56} \ST^{39}+q^{54} \ST^{38}-q^{52} \ST^{37}-q^{50} \ST^{36}+q^{50} \ST^{35}+2 q^{48} \ST^{34}+q^{48} \ST^{33} +2 q^{46} \ST^{32}+\\ +q^{46} \ST^{31} -2 q^{44} \ST^{32}+2 q^{44} \ST^{30}+q^{44} \ST^{29}-q^{42} \ST^{31}-q^{42} \ST^{30} +2 q^{42} \ST^{28}+2 q^{40} \ST^{27}-q^{38} \ST^{28}+\\ +3 q^{38} \ST^{27} +5 q^{38} \ST^{26}+2 q^{38} \ST^{25}-q^{36} \ST^{27} -q^{36} \ST^{26}+q^{36} \ST^{25}+7 q^{36} \ST^{24}+q^{36} \ST^{23}-4 q^{34} \ST^{25}-\\ -7 q^{34} \ST^{24} +4 q^{34} \ST^{23} +3 q^{34} \ST^{22}-q^{32} \ST^{24}-3 q^{32} \ST^{23}-3 q^{32} \ST^{22}+2 q^{32} \ST^{21}+q^{30} \ST^{22}+\\ +q^{30} \ST^{21} \underline{+8 q^{30} \ST^{20}{\color{red} -q^{30} \ST^{20}}+q^{30} \ST^{19}}-q^{28} \ST^{21}-2 q^{28} \ST^{20}+8 q^{28} \ST^{19}+5 q^{28} \ST^{18}-q^{26} \ST^{20} -7 q^{26} \ST^{19}- \\ -7 q^{26} \ST^{18}+4 q^{26} \ST^{17}+q^{26} \ST^{16}-2 q^{24} \ST^{18}-10 q^{24} \ST^{17}+q^{24} \ST^{15} -3 q^{22} \ST^{16}+6 q^{22} \ST^{15}+\\ +5 q^{22} \ST^{14} -q^{20} \ST^{15}+5 q^{20} \ST^{13}+q^{20} \ST^{12}-q^{18} \ST^{14} -11 q^{18} \ST^{13}-4 q^{18} \ST^{12}+q^{18} \ST^{11}-\\ -5 q^{16} \ST^{12}-4 q^{16} \ST^{11}-q^{14} \ST^{10}+2 q^{14} \ST^9 +q^{14} \ST^8-2 q^{12} \ST^9+q^{12} \ST^8+2 q^{12} \ST^7-\\ -q^{10} \ST^8-3 q^{10} \ST^7-2 q^8 \ST^6-2 q^8 \ST^5 -q^6 \ST^4+q^6 \ST^3+q^4 \ST^2-q^2 \ST-1)\,,\quad n \geq 0 \\
        q^{-30} \ST^{-19}(q^{56} \ST^{39}+q^{54} \ST^{38}-q^{52} \ST^{37}-q^{50} \ST^{36}+q^{50} \ST^{35}+2 q^{48} \ST^{34}+q^{48} \ST^{33} +2 q^{46} \ST^{32}+\\ +q^{46} \ST^{31}-2 q^{44} \ST^{32}+2 q^{44} \ST^{30}+q^{44} \ST^{29}-q^{42} \ST^{31}-q^{42} \ST^{30} +2 q^{42} \ST^{28}+2 q^{40} \ST^{27} -\\ -q^{38} \ST^{28}+3 q^{38} \ST^{27}+5 q^{38} \ST^{26} +2 q^{38} \ST^{25}-q^{36} \ST^{27} -q^{36} \ST^{26}+q^{36} \ST^{25}+7 q^{36} \ST^{24}+\\ +q^{36} \ST^{23}-4 q^{34} \ST^{25}-7 q^{34} \ST^{24}+4 q^{34} \ST^{23} +3 q^{34} \ST^{22}-q^{32} \ST^{24}-3 q^{32} \ST^{23}-3 q^{32} \ST^{22}+\\ +2 q^{32} \ST^{21}+q^{30} \ST^{22}+q^{30} \ST^{21} \underline{+8 q^{30} \ST^{20}{\color{red} + q^{30} \ST^{19}}+ q^{30} \ST^{19}}-q^{28} \ST^{21}-2 q^{28} \ST^{20}+8 q^{28} \ST^{19}+\\ +5 q^{28} \ST^{18}-q^{26} \ST^{20} -7 q^{26} \ST^{19}-7 q^{26} \ST^{18}+4 q^{26} \ST^{17}+q^{26} \ST^{16}-2 q^{24} \ST^{18}-10 q^{24} \ST^{17}+\\ +q^{24} \ST^{15} -3 q^{22} \ST^{16}+6 q^{22} \ST^{15}+5 q^{22} \ST^{14}-q^{20} \ST^{15}+5 q^{20} \ST^{13}+q^{20} \ST^{12}-q^{18} \ST^{14} -11 q^{18} \ST^{13}-\\ -4 q^{18} \ST^{12}+q^{18} \ST^{11} -5 q^{16} \ST^{12}-4 q^{16} \ST^{11}-q^{14} \ST^{10}+2 q^{14} \ST^9 +q^{14} \ST^8 -2 q^{12} \ST^9 +q^{12} \ST^8+\\ +2 q^{12} \ST^7-q^{10} \ST^8-3 q^{10} \ST^7-2 q^8 \ST^6-2 q^8 \ST^5 -q^6 \ST^4+q^6 \ST^3+q^4 \ST^2-q^2 \ST-1)\,,\quad n < 0
    \end{cases} \\
    \{q^2\ST\}^{-1}\cdot \mathfrak{Kh}_{\rm adj}^{10_{153}}&=q^{-28} \ST^{-18}(q^{52} \ST^{37}+q^{50} \ST^{36}+q^{46} \ST^{33}+2 q^{44} \ST^{32}+q^{44} \ST^{31}+q^{42} \ST^{31}+2 q^{42} \ST^{30}+q^{42} \ST^{29}+\\ &+q^{40} \ST^{29}+2 q^{40} \ST^{28}+q^{40} \ST^{27}+q^{38} \ST^{28}+q^{38} \ST^{27}+2 q^{38} \ST^{26}+q^{36} \ST^{27}+2 q^{36} \ST^{26}+3 q^{36} \ST^{25}+\\ &+4 q^{34} \ST^{25}+7 q^{34} \ST^{24}+2 q^{34} \ST^{23}+q^{32} \ST^{24}+4 q^{32} \ST^{23}+ 7q^{32} \ST^{22}+q^{32} \ST^{21}+6 q^{30} \ST^{21}+\\ &+3 q^{30} \ST^{20}+q^{28} \ST^{21}+4 q^{28} \ST^{20}+3 q^{28} \ST^{19}+q^{26} \ST^{20}+7 q^{26} \ST^{19}+11 q^{26} \ST^{18}+q^{26} \ST^{17}+\\ &+2 q^{24} \ST^{18}+11 q^{24} \ST^{17}+5 q^{24} \ST^{16}+4 q^{22} \ST^{16}+5 q^{22} \ST^{15}+q^{22} \ST^{14}+q^{20} \ST^{15}+5 q^{20} \ST^{14}+\\ &+q^{20} \ST^{13}+q^{18} \ST^{14}+11 q^{18} \ST^{13}+6 q^{18} \ST^{12}+5 q^{16} \ST^{12}+6 q^{16} \ST^{11}+q^{16} \ST^{10}+2 q^{14} \ST^{10}+\\ &+q^{14} \ST^9+2 q^{12} \ST^9+q^{12} \ST^8+q^{10} \ST^8+3 q^{10} \ST^7+q^{10} \ST^6+2 q^8 \ST^6+2 q^8 \ST^5+q^6 \ST^4+q^2 \ST+1) \\
    {\color{blue} \mathbf{\overline{14n_{21881}}}}:\quad {\color{red} \Kh_{\rm adj}^{\overline{14n_{21881}}}}&=\begin{cases}
        q^{24}\ST (q^{58} \ST^{41}+q^{56} \ST^{40}+q^{54} \ST^{39}+q^{54} \ST^{38}+q^{52} \ST^{38}+2 q^{52} \ST^{37}-q^{50} \ST^{37}-q^{48} \ST^{36}+\\ +q^{48} \ST^{34}-q^{46} \ST^{35}+q^{46} \ST^{34}+4 q^{46} \ST^{33}+q^{46} \ST^{32}-q^{44} \ST^{34}-2 q^{44} \ST^{33} +2 q^{44} \ST^{32}+\\ +2 q^{44} \ST^{31}-q^{42} \ST^{32}-2 q^{42} \ST^{31}-q^{42} \ST^{30}{\color{red} -q^{40} \ST^{32}}+q^{40} \ST^{30}+q^{40} \ST^{29}-q^{38} \ST^{30}-\\ -2 q^{38} \ST^{29}+3 q^{38} \ST^{28}+4 q^{38} \ST^{27}+q^{38} \ST^{26}-3 q^{36} \ST^{28}-3 q^{36} \ST^{27}+3 q^{36} \ST^{26} +3 q^{36} \ST^{25}-\\ -q^{34} \ST^{26}-3 q^{34} \ST^{25}-q^{34} \ST^{24}+q^{34} \ST^{23}-q^{32} \ST^{26}+q^{32} \ST^{24}-q^{32} \ST^{23} -2 q^{30} \ST^{24}-\\ -q^{30} \ST^{23}+3 q^{30} \ST^{22}+2 q^{30} \ST^{21}-3 q^{28} \ST^{22}-2 q^{28} \ST^{21}+3 q^{28} \ST^{20} +2 q^{28} \ST^{19} -2 q^{26} \ST^{20}-\\ -3 q^{26} \ST^{19}+q^{26} \ST^{18}+q^{26} \ST^{17}-q^{24} \ST^{20}-q^{24} \ST^{18}-2 q^{24} \ST^{17} +q^{24} \ST^{15} -q^{22} \ST^{18}+\\ +q^{22} \ST^{16}-2 q^{20} \ST^{16}+3 q^{20} \ST^{14}-2 q^{18} \ST^{14}-q^{18} \ST^{13}+q^{18} \ST^{12} +q^{18} \ST^{11} -2 q^{16} \ST^{12}-\\ -q^{16} \ST^{11}+q^{16} \ST^{10}-q^{14} \ST^9-q^{12} \ST^{10}-q^{10} \ST^8+q^{10} \ST^6-q^8 \ST^6+q^8 \ST^4-q^6 \ST^4-1)\,,\quad n \geq {\color{blue} \mathbf{32}} \\
        q^{24}\ST (q^{58} \ST^{41}+q^{56} \ST^{40}+q^{54} \ST^{39}+q^{54} \ST^{38}+q^{52} \ST^{38}+2 q^{52} \ST^{37}-q^{50} \ST^{37}-q^{48} \ST^{36}+\\ +q^{48} \ST^{34}-q^{46} \ST^{35}+q^{46} \ST^{34}+4 q^{46} \ST^{33}+q^{46} \ST^{32}-q^{44} \ST^{34}-2 q^{44} \ST^{33}+2 q^{44} \ST^{32}+\\ +2 q^{44} \ST^{31}-q^{42} \ST^{32}-2 q^{42} \ST^{31}-q^{42} \ST^{30}{\color{red} +q^{40} \ST^{31}}+q^{40} \ST^{30}+q^{40} \ST^{29}-q^{38} \ST^{30}-\\ -2 q^{38} \ST^{29}+3 q^{38} \ST^{28}+4 q^{38} \ST^{27}+q^{38} \ST^{26}-3 q^{36} \ST^{28}-3 q^{36} \ST^{27}+3 q^{36} \ST^{26} +3 q^{36} \ST^{25}-\\ -q^{34} \ST^{26}-3 q^{34} \ST^{25}-q^{34} \ST^{24}+q^{34} \ST^{23}-q^{32} \ST^{26}+q^{32} \ST^{24}-q^{32} \ST^{23} -2 q^{30} \ST^{24}-\\ -q^{30} \ST^{23} +3 q^{30} \ST^{22}+2 q^{30} \ST^{21}-3 q^{28} \ST^{22}-2 q^{28} \ST^{21}+3 q^{28} \ST^{20} +2 q^{28} \ST^{19}-2 q^{26} \ST^{20}-\\ -3 q^{26} \ST^{19}+q^{26} \ST^{18}+q^{26} \ST^{17}-q^{24} \ST^{20}-q^{24} \ST^{18}-2 q^{24} \ST^{17} +q^{24} \ST^{15}-q^{22} \ST^{18}+\\ +q^{22} \ST^{16}-2 q^{20} \ST^{16}+3 q^{20} \ST^{14}-2 q^{18} \ST^{14}-q^{18} \ST^{13}+q^{18} \ST^{12} +q^{18} \ST^{11}-2 q^{16} \ST^{12}-\\ -q^{16} \ST^{11}+q^{16} \ST^{10}-q^{14} \ST^9-q^{12} \ST^{10}-q^{10} \ST^8+q^{10} \ST^6-q^8 \ST^6+q^8 \ST^4-q^6 \ST^4-1)\,,\quad n < {\color{blue} \mathbf{32}}
    \end{cases}\\
    \{q^2\ST\}^{-1}\cdot \mathfrak{Kh}_{\rm adj}^{\overline{14n_{21881}}}&=q^{26} \ST^2 (q^{54} \ST^{39}+q^{52} \ST^{38}+2 q^{50} \ST^{37}+q^{50} \ST^{36}+2 q^{48} \ST^{36}+2 q^{48} \ST^{35}+q^{46} \ST^{35}+q^{46} \ST^{34}+q^{44} \ST^{34}+\\ &+2 q^{44} \ST^{33}+q^{44} \ST^{32}+2 q^{42} \ST^{32}+4 q^{42} \ST^{31}+q^{42} \ST^{30}+3 q^{40} \ST^{30}+2 q^{40} \ST^{29}+q^{38} \ST^{30}+2 q^{38} \ST^{29}+\\ &4 q^{36} \ST^{28}+3 q^{36} \ST^{27}+3 q^{34} \ST^{26}+4 q^{34} \ST^{25}+q^{34} \ST^{24}+q^{32} \ST^{26}+3 q^{32} \ST^{24}+3 q^{32} \ST^{23}+2 q^{30} \ST^{24}+\\ &+q^{30} \ST^{23}+q^{30} \ST^{21}+4 q^{28} \ST^{22}+2 q^{28} \ST^{21}+3 q^{26} \ST^{20}+3 q^{26} \ST^{19}+q^{24} \ST^{20}+3 q^{24} \ST^{18}+2 q^{24} \ST^{17}+q^{22} \ST^{18}+\\ &+q^{22} \ST^{16}+q^{22} \ST^{15}+2 q^{20} \ST^{16}+q^{20} \ST^{13}+2 q^{18} \ST^{14}+q^{18} \ST^{13}+3 q^{16} \ST^{12}+q^{16} \ST^{11}+q^{14} \ST^{10}+q^{14} \ST^9+\\ &+q^{12} \ST^{10}+q^{12} \ST^8+q^{10} \ST^8+q^8 \ST^6+q^6 \ST^4+q^4 \ST^2+1)
\end{aligned}
\end{equation}

\end{document}